\documentclass{ar-1col}

\usepackage{url}
\usepackage{amsmath}
\usepackage{amssymb}
\usepackage{mathrsfs}
\usepackage[comma]{natbib}

\jname{Annu. Rev. Astron. Astrophys.} 
\jvol{AA}
\jyear{2021}
\doi{10.1146/((please add article doi))}

\newcommand{\be}{\begin{equation}}
\newcommand{\ee}{\end{equation}}
\newcommand{\bea}{\begin{eqnarray}}
\newcommand{\eea}{\end{eqnarray}}
\newcommand{\beas}{\begin{eqnarray*}}
\newcommand{\eeas}{\end{eqnarray*}}

\def\gsim{ \lower .75ex \hbox{$\sim$} \llap{\raise .27ex \hbox{$>$}} }
\def\lsim{ \lower .75ex \hbox{$\sim$} \llap{\raise .27ex \hbox{$<$}} }

\newcommand{\ba}{\begin{eqnarray}}
\newcommand{\ea}{\end{eqnarray}}

\begin{document}

\hoffset=-1.0in
\voffset=-1.0in

\markboth{Hui}{Wave Dark Matter}

\title{Wave Dark Matter}

\author{Lam Hui
\affil{Center for Theoretical Physics, Department of Physics, Columbia
  University, New York, NY 10027, USA; email: lh399@columbia.edu}}

\begin{abstract}
We review the physics and phenomenology of wave dark matter: a bosonic
dark matter candidate lighter than about $30$ eV. Such particles have a de Broglie wavelength
exceeding the average inter-particle separation in a galaxy like the Milky Way, thus well 
described as a set of classical waves. We outline the particle physics
motivations for them, including the QCD axion as well as ultra-light
axion-like-particles such as fuzzy dark matter. 
The wave nature of the dark matter implies a rich phenomenology:
\begin{itemize}
\setlength{\leftskip}{-5.5mm}
\item Wave interference gives rise to order unity density fluctuations
  on \\ de Broglie scale in halos. One manifestation is
  vortices 
  where the \\ density vanishes and around which the velocity 
  circulates. There is \\ one vortex ring per de Broglie 
  volume on average.
\item For sufficiently low masses, soliton condensation occurs at
  centers \\ of halos. The soliton oscillates and random walks, another
  \\ manifestation of wave interference. The halo and subhalo \\ abundance
  is expected to be suppressed at small masses, but the \\
  precise prediction from numerical wave simulations remains to be \\ determined.
\item For ultra-light $\sim 10^{-22}$ eV dark matter, the wave
  interference \\ substructures can be probed by tidal
  streams/gravitational \\ 
  lensing. The signal can be distinguished from that due
  to subhalos \\ by the dependence on stream orbital radius/image separation.
\item Axion detection experiments are sensitive to interference \\
  substructures for wave dark matter that is moderately light. The \\
  stochastic nature of the waves affects the interpretation of \\
  experimental constraints and motivates the measurement of \\
  correlation functions.
\end{itemize}
Current constraints and open questions, covering detection experiments
and cosmological/galactic/black-hole observations, are discussed.
\end{abstract}

\maketitle

\tableofcontents

\section{INTRODUCTION}
\label{intro}

The astronomical evidence for the existence of dark matter,
accumulated over decades, is rich and compelling
\citep[e.g.,][]{Zwicky:1933gu,Smith:1936mlg,
  Rubin1970,Freeman:1970mx,OstrikerPeebles1973,Hoekstra:2003pn,Clowe2006, WMAP2013,Aghanim:2018eyx}.
Yet, the identity and basic properties of dark matter remain shrouded in mystery.
An example is the constituent's mass:
proposals range from ultra-light $\sim 10^{-22}$ eV
\citep*{Hu:2000ke} to astronomical $\sim 10 {\,\rm M_\odot}$
\citep{Bird:2016dcv,Garcia-Bellido:2017mdw,Sasaki:2018dmp,Jedamzik:2020ypm}.
In this vast spectrum, there is nonetheless a useful demarcation
point. Dynamical measurements tell us the dark matter mass density
in the solar neighborhood is about $0.4  {\,\rm GeV \, cm^{-3}}$. 
\footnote{
A range of local dark matter density values have been reported in the
literature: e.g. $0.008 {\,\rm M_\odot / pc^3} = 0.3 {\,\rm GeV / cm^3}$
\citep{Bovy:2012tw}, $0.0122 {\,\rm M_\odot / pc^3} = 0.46 {\,\rm GeV /
  cm^3}$ \citep{Sivertsson:2017rkp}, $0.013  {\,\rm M_\odot / pc^3} = 0.49 {\,\rm GeV /
  cm^3}$ \citep{McKee2015}.
}
From this, one can deduce the average inter-particle separation, {\it
  given} a dark matter particle mass.
We can compare it against the de Broglie wavelength of the particle:
\begin{equation}
\label{lambdadB}
 \lambda_{\rm dB}  \equiv {2\pi \over {m v}} = 0.48 {\,\rm kpc}  \left(
  {10^{-22} {\,\rm eV} \over m} \right) \left( {250 {\,\rm km/s} \over
  v} \right) = 1.49  {\,\rm km} \left(
  {10^{-6} {\,\rm eV} \over m} \right) \left( {250 {\,\rm km/s} \over
  v} \right) \, ,
\end{equation}
where $v$ is the velocity dispersion of the galactic halo, and $m$ is the dark matter particle mass, for which 
two representative values are chosen for illustration.
\footnote{In this article, $\hbar$ and $c$ are set to unity. In most cases, restoring $\hbar$ is a matter of
replacing $m$ by $m/\hbar$. For instance, the de Broglie wavelength is
$\lambda_{\rm dB} = 2\pi \hbar/(mv) = h/(mv)$. The Compton wavelength
is $\lambda_{\rm Compton} = 2\pi \hbar/(mc)$. 
}
It can be shown that the de Broglie wavelength exceeds
the inter-particle separation if $m \, \lsim \, 30$ eV. In other
words, in a Milky-Way-like environment, the average number of particles in a 
de Broglie volume $\lambda_{\rm dB}^3$ is:
\begin{equation}
N_{\rm dB} \sim \left( {34 {\,\rm eV} \over m} \right)^4 \left( {250 {\,\rm km/s} \over v} \right)^3 \, .
\end{equation}
For $m \ll 30$ eV, the occupancy $N_{\rm dB}$ is so large that the set of particles is best described by
classical waves, much as in electromagnetism, a state with a 
large number of photons 
is well described by the classical electric and magnetic fields.
\footnote{A more precise statement is that a coherent state of photons has negligible 
quantum fluctuations if the {\it average} occupation number is large.
See e.g. the classic paper by
\cite{Glauber:1963fi}.}
The associated wave phenomena is the subject of this review. We emphasize
{\it classical}, for large occupancy implies negligible quantum
fluctuations. 
The question of how the classical description relates to the
underlying quantum one is a fascinating
subject. We unfortunately do not have the space to explore it here 
\citep[see][]{Sikivie:2009qn,Guth:2014hsa,Dvali:2017ruz,Lentz:2018mpf,Allali:2020ttz}.

Such a light dark matter particle is necessarily bosonic, for the 
Pauli exclusion principle precludes multiple occupancies 
for fermions---this is the essence of the bound by \citet{Tremaine:1979we}.
For concreteness, we focus on a spin zero (scalar) particle, 
although much of the wave phenomenology
applies to higher spin cases as well
\citep{Graham:2015rva,Kolb:2020fwh,Aoki:2016zgp}. 
There is a long history of investigations of dark matter as a scalar
field \citep[e.g.,][]{Baldeschi:1983mq,Turner:1983he,Press:1989id,Sin:1992bg,Peebles:2000yy,Goodman:2000tg,Lesgourgues:2002hk, Amendola:2005ad,Chavanis:2011zi,Suarez:2011yf,RindlerDaller:2011kx,Berezhiani:2015bqa,Fan:2016rda,Alexander:2016glq}.
Perhaps the most well motivated example is the Quantum Chromodynamics 
(QCD) axion
\citep{Peccei:1977hh,Kim:1979if,Weinberg:1977ma,Wilczek:1977pj,
Shifman:1979if,Zhitnitsky:1980tq,Dine:1981rt,Preskill:1982cy,Abbott:1982af,Dine:1982ah}. Its
possible mass spans a large range---experimental detection has focused 
on masses around $10^{-6}$ eV, with newer experiments reaching down to much lower
values. For recent reviews, see
\cite{Graham:2015ouw,Marsh:2015xka,Sikivie:2020zpn}. String
theory also predicts a large number of axion-like-particles (ALP), 
one or some of which could
be dark matter
\citep{Svrcek:2006yi,Arvanitaki:2009fg,Halverson:2017deq,Bachlechner:2018gew}. 
At the extreme end of the spectrum is the possibility of an ALP with
mass around $10^{-22} - 10^{-20}$ eV, with a relic abundance that
naturally matches the observed dark matter density (see
Section \ref{motivations}). More generally, ultra-light dark matter in
this mass range is often referred to as fuzzy dark matter (FDM).
It was proposed by \citet*{Hu:2000ke} 
to address small scale structure issues thought to be associated with
conventional cold dark mater (CDM) \citep{Spergel:1999mh}.
This is a large subject we will not discuss in depth, though it
will be touched upon in Section \ref{implications}. It remains unclear
whether the small scale structure issues point to novelty in the
dark matter sector, or can be resolved by baryonic physics, once
the complexities of galaxy formation are properly understood
\citep[for a recent review, see][]{Weinberg:2013aya}.

In this article, we take a broad perspective on wave dark
matter ($m \, \lsim \, 30$ eV), and discuss novel features that
distinguish it from particle dark matter ($m \,\gsim \, 30$ eV). 
The underlying wave dynamics is the same whether the dark matter
is ultra-light like 
fuzzy dark matter, or merely light like the QCD axion. The length scale of
the wave phenomena (i.e. the de Broglie wavelength) depends of course on the mass.
For the higher masses, the length scales are small, which can be
probed by laboratory detection experiments.
(The higher masses can have astrophysical consequences too,
despite the short de Broglie wavelength, for instance around black
holes or in solitons, as we will see.)
For the ultra-light end of the spectrum, fuzzy dark matter ($m \sim 10^{-22} -
10^{-20}$ eV), the length scales are long and 
there can be striking astrophysical signatures,
which we will highlight.\footnote{There is a recent flurry of activities
  on this front, starting from the paper by \citet*{Schive:2014dra}:
\citet{Schive:2014hza,
Veltmaat:2016rxo,Schwabe:2016rze,Hui:2016ltb,Mocz:2017wlg,Nori:2018hud,Levkov:2018kau,Bar-Or:2018pxz,Bar:2018acw,Church:2018sro,Li:2018kyk,Marsh:2018zyw,Schive:2019rrw,Mocz:2019pyf,Lancaster:2019mde,Chan:2020exg,Hui:2020hbq}. A
recent review can be found in \cite{Niemeyer:2019aqm}.}
A mass $m \, < \, 10^{-22}$ eV is possible, but only
if the particle constitutes a small fraction of dark matter, for the
simple reason that an excessively large $\lambda_{\rm dB}$
precludes the existence of dark matter dominated dwarf galaxies \citep{Hu:2000ke}.
When the mass approaches the
size of the Hubble constant today $m \sim 10^{-33}$ eV, the scalar field is so slowly
rolling that it is essentially a form of dark energy
\citep{Hlozek:2014lca}.
(The distinction between a slowly rolling scalar field as dark energy,
and oscillating scalar field as dark matter, is discussed in Section \ref{motivations}.)

An outline of the article is as follows. 
Particle physics motivations for considering wave dark matter are
discussed in Section \ref{motivations}. 
The bulk of this review is devoted to elucidating the dynamics
and phenomenology of wave dark matter,
in Section \ref{dynamics}. The observational/experimental implications and constraints are summarized in Section \ref{implications}. 
We conclude in Section \ref{conclude} with a discussion of open questions
and directions for further research.
This article is intended to be pedagogical: we emphasize
results that can be understood in an intuitive way, while providing
ample references. We devote
more space to elucidating the physics than to summarizing the current
constraints, which evolve, sometimes rapidly.

\begin{textbox}[t]
\section{Terminology}
We use the term axion to loosely refer to
both the QCD axion, and an axion-like-particle (Section \ref{motivations}). 
The term fuzzy dark
matter (FDM) is reserved for the ultra-light part of the mass spectrum $m
\sim 10^{-22} - 10^{-20}$ eV. Wave dark matter is the more general
term, $m \, \lsim \, 30$ eV, for which dark matter exhibits wave
phenomena. Wave dark matter, such as the axion, 
is in fact one form of cold dark matter (CDM), assuming it is {\it not}
produced by thermal freeze-out (see Section \ref{motivations}). 
We use the term
particle dark matter for cases where $m \,\gsim \, 30 {\,\rm
  eV}$, the primary example of which is Weakly Interacting Massive
Particle (WIMP). We sometimes refer to it as {\it conventional}
CDM.
\end{textbox}

\section{Particle physics motivations}
\label{motivations}

In this section, we describe the axion---the QCD axion or
an axion-like-particle---as a concrete example of wave dark
matter: (1) how it is motivated by high energy physics considerations
independent of the dark matter problem; (2) how a relic abundance that
matches the observed dark matter density can be naturally obtained;
(3) how it is weakly interacting and cold.
Readers not interested in the details can skip to Section \ref{dynamics} without
loss of continuity.

We are interested in a scalar field $\phi$ that has a small mass $m$.
A natural starting point is a massless Goldstone boson, associated with the
spontaneous breaking of some symmetry. Non-perturbative 
quantum effects can generate a small
mass---hence, a {\it pseudo} Goldstone boson---or more generally a potential $V(\phi)$, giving a Lagrangian density of the
form: \footnote{\label{instanton} By non-perturbative effects, we mean something that is
exponentially suppressed in the $\hbar \rightarrow 0$ limit, analogous
to how the tunneling amplitude in quantum mechanics is exponentially suppressed
$\sim e^{-S_{\rm instanton} /\hbar}$. A moderate value for $S_{\rm
  instanton}/\hbar$ could yield a small mass, starting from some high
energy scale. See \cite{Marsh:2015xka} for examples.
}
\begin{equation}
\label{phiLag}
{\cal L} = - {1\over 2} \partial_\mu \phi \, \partial^\mu \phi  -
V(\phi) \, .
\end{equation}
A concrete realization is the axion, which is a real angular field, in the
sense that $\phi$ and $\phi + 2\pi f$ are identified i.e. $\phi/f$ is
effectively an angle. The periodicity scale $f$, an energy scale, is 
often referred to as the axion decay constant. 

The classic example is the QCD axion, a particle that couples to the
gluon field strength and derives its mass from 
the presence of this coupling (and confinement). It was introduced to address 
the strong CP (charge-conjugation parity)
problem: that a certain parameter in the standard model, the angle
$\theta_{\rm QCD}$, is constrained to be less
than $10^{-9}$ from experimental bounds on the neutron 
electric dipole moment. 
\footnote{
\label{PecceiQuinnU1}
The $\theta_{\rm QCD}$ term in the Lagrangian takes the form
${\cal L} \sim \theta_{\rm QCD} G\tilde G$ where $G$ and $\tilde G$
are the gluon
field strength and its dual. Such a term is a total
derivative, yet must be included in the path integral to account for
gluon field configurations of different windings.
Such topological considerations tell us
$\theta_{\rm QCD}$ is an angle. With non-vanishing quark masses, 
a non-zero angle
signals the breaking of CP which is severely constrained by experiments.
The idea of the QCD axion is to promote this angle to a dynamical field $\theta_{\rm QCD}
\rightarrow \phi/f$, thereby allowing a physical mechanism that relaxes
it to zero, as suggested by \citet{Peccei:1977hh}. 
The axion $\phi$ is the Goldstone boson 
associated with the breaking of a certain global symmetry,
Peccei-Quinn U(1), as
pointed out by 
\citet{Weinberg:1977ma,Wilczek:1977pj}.
See \citet{Dine:2000cj,Hook:2018dlk} for reviews on axions and
alternative solutions to the
strong CP problem.
} 
It has certain generic couplings to the standard model, 
allowing the possibility of experimental detection (see below).
More general examples---namely, axion-like-particles 
which have similar couplings to the standard model
but do not contribute to the resolution
of the strong CP problem---arise naturally in string theory
as the Kaluza-Klein zero modes of higher form fields when
the extra dimensions are compactified
\citep{Green:1987mn,Svrcek:2006yi,Arvanitaki:2009fg,Dine:2007zp,Halverson:2017deq,Bachlechner:2018gew}. 

\begin{marginnote}
\entry{Peccei-Quinn U(1)}{the symmetry associated with shifting $\phi$
by a constant. Its spontaneous breaking is what makes the axion
$\phi$ possible. Its small explicit breaking by non-perturbative effects gives $\phi$
a potential.}
\end{marginnote}

For illustration, consider a potential $V(\phi)$ of the following form:
\begin{equation}
\label{Vphi}
V(\phi) = \Lambda^4 (1 - {\,\rm cos\,}[\phi/f]) \, .
\end{equation}
(The QCD axion potential does not have this precise form, but shares similar qualitative
features.)
The cosine is consistent with the idea of $\phi/f$ being an angle. The
additive constant is not important for our considerations, and is chosen
merely to make $V$ vanish at the minimum $\phi=0$.
The mass of $\phi$ can be read off from expanding the cosine around
$\phi = 0$: $m = \Lambda^2/f$. Typically, $f$ is some high energy
scale up to Planck scale, while $\Lambda$ is exponentially suppressed
compared to that (see footnote \ref{instanton}), giving a small $m$. 
For instance, $f \sim 10^{17}$ GeV and $\Lambda \sim 100$ eV gives
$m \sim 10^{-22}$ eV. 
The QCD axion potential does not have the exact form
above \citep[for a recent computation, see][]{diCortona:2015ldu}, but $m \sim \Lambda^2/f$ remains true with $\Lambda$
being the QCD scale $\sim 100$ MeV.
For instance, $f \sim 10^{13}$ GeV gives $m \sim 10^{-6}$ eV for
the QCD axion. 

What determines the contribution of $\phi$ to the
energy content of the universe today? 
Here we outline the misalignment
mechanism \citep[reviewed in][]{Kolb:1990vq}.
Consider the equation of motion for a
{\it homogeneous} $\phi$ (following from Equation \ref{phiLag}) in an
expanding background): 
\begin{equation}
\label{phibareom}
\ddot\phi + 3 H \dot\phi + \partial_\phi V = 0 \, ,
\end{equation}
where $H$ is the Hubble expansion rate. 
In the early universe, when $H$ is large, Hubble friction is sufficient to
keep $\phi$ slowly rolling i.e. balancing the last two terms on the left.
Thus $V(\phi)$ plays the role of dark energy. The value of $\phi$ is
essentially stuck at its primordial value---we assume 
$\phi_{\rm primordial}/f$, the so called misalignment angle, is order
unity. 
\footnote{
An interesting variant of the idea, where the primordial $\phi$ has a
significant velocity, was proposed by \cite{Co:2019jts}.
}
The expansion rate drops as time goes on, 
until $H$ reaches $\sim m$.
After that $\phi$ rolls towards the minimum of the
potential and commences oscillations around it. 
The expansion of the universe takes energy out of such oscillations,
diminishing the oscillation amplitude. Subsequently, $\phi$ oscillates
close to zero, implying it is a good approximation to treat the
potential as:
\begin{equation}
\label{m2phi2}
V(\phi) \sim {1\over 2} m^2 \phi^2 \, .
\end{equation}
The energy density
contained in the $\phi$ oscillations is
\begin{equation}
\label{rho}
\rho = {1\over 2} \dot\phi {}^2 + {1\over 2} m^2 \phi^2 \, .
\end{equation}
It follows from Equation \ref{phibareom}
that $\rho$ redshifts like $a^{-3}$ where $a$ is the scale factor.
The $\phi$ oscillations, which can be interpreted as a set of
particles, therefore have the redshifting behavior of
(non-relativistic) matter, making this a suitable dark matter
candidate. Following this cosmological history, it can be shown
that the relic density today is \citep[e.g.,][]{Arvanitaki:2009fg,Marsh:2015xka,Hui:2016ltb}:
\begin{equation}
\label{relic}
\Omega_{\rm axion} \sim 0.1 \left( {f \over 10^{17} {\,\rm GeV}}
\right)^2 \left({m \over 10^{-22} {\,\rm eV}}\right)^{1/2} \,
\end{equation}
where $\Omega_{\rm axion}$ is the axion density today as a fraction of
the critical density. It is worth emphasizing the relic density is
more sensitive to the choice of $f$ than to $m$. The value of
$10^{17}$ GeV, close to but below the Planck scale, is motivated by
string theory constructions \citep{Svrcek:2006yi}. 
\footnote{See \cite{KM,Davoudiasl:2017jke,Alonso-Alvarez:2018tus} 
for recent explorations of model building.
}
But a slightly different $f$ would have to be paired with a quite different $m$, if
one were to insist on matching the observed dark matter abundance. 
Nonetheless, this relic abundance computation motivates the
consideration of light, even ultra-light, axions.

The reasoning above essentially follows the classic computation of the
QCD axion relic density
\citep{Preskill:1982cy,Abbott:1982af,Dine:1982ah}---the difference is
that while $V(\phi)$ is constant here, it is temperature dependent for
the QCD axion.
Besides the misalignment mechanism, it is also possible
axions arise from the decay of topological defects, if 
the Peccei-Quinn U(1) symmetry is broken after
inflation \citep[for recent lattice computations, see][]{Gorghetto:2020qws,Buschmann:2019icd}.

Aside from having the requisite relic abundance, a good dark matter
candidate should be cold and weakly interacting. The coldness
is implicit in the misalignment mechanism: the axion starts off as a
homogeneous scalar field in the early universe, with the homogeneity
guaranteed for instance by inflation. 
(There are inevitable small fluctuations as
well, which is discussed in Section \ref{implications}.) 
The weakly interacting nature is
implied by the large axion decay constant $f$. Possible interactions
include:\footnote{We list here only interactions for a
  pseudo-scalar like the axion. For a scalar, there are other possibilities; see e.g. 
\cite{Graham:2015ouw}.}
\begin{equation}
\label{phiInteractions}
{\cal L}^{\rm self}_{\rm int.} \sim {m^2 \over f^2} \phi^4 \quad , \quad 
{\cal L}^{\rm \gamma}_{\rm int.} \sim {\phi \over f} F^{\mu\nu} \tilde F_{\mu\nu}
\quad , \quad
{\cal L}^{\rm \Psi}_{\rm int.} \sim {\partial_\mu \phi \over f}
\bar\Psi \gamma^\mu \gamma_5 \Psi \, .
\end{equation}
The first interaction, a self-interaction of $\phi$, follows from
expanding out the potential $V(\phi)$ to quartic order; it is an
attractive interaction for the axion. The second interaction is with
the photon, $F$ and $\tilde F$ being the photon field strength and its
dual (there is an analogous interaction with gluon field strength and
its dual for the QCD axion). The third interaction is with a fermion $\Psi$, which could
represent quarks or leptons. The last two interactions are both
symmetric under a shift of $\phi$ by a constant, as befitting a
(pseudo) Goldstone boson. The generic expectation is that all three
coupling strengths are of the order shown, but models can be constructed that
deviate from it \citep{KM,Kaplan:2015fuy,Choi:2015fiu}.
The important point is that $f$ is expected
to be large, keeping these interactions weak, for both the QCD
axion and axion-like-particles.
For structure formation purpose, these interactions can be largely
ignored, though their presence is important for direct
detection and in certain extreme astrophysical environments, as we will
discuss below.


\section{Wave dynamics and phenomenology}
\label{dynamics}

The discussion above motivates us to consider a scalar field $\phi$
satisfying the Klein Gordon equation: 
\begin{equation}
\label{phiKG}
- \Box 
\phi + m^2 \phi = 0 \, ,
\end{equation}
which follows from Equation \ref{phiLag} with the potential approximated
by Equation \ref{m2phi2}. Much of the following discussion is not specific to
  axions---it applies to any scalar (or pseudo-scalar) 
  particle whose dominant
  interaction is gravitational. Occasionally, we will comment on
  features that are specific to axions, for instance in cases where their
  self-interaction is important.

Unlike in Equation \ref{phibareom}, here we are
interested in the possibility of $\phi$ having spatial fluctuations.
In the non-relativistic regime relevant for structure formation, it is
useful to introduce a complex scalar $\psi$ ($\phi$ is a real scalar):
\begin{equation}
\label{phipsi}
\phi = {1\over \sqrt{2m}} \left( \psi e^{-imt} + \psi^* e^{imt}
\right) \, .
\end{equation}
The idea is to factor out the fast time dependence of
$\phi$---oscillation with frequency $m$---and assume $\psi$
is slowly varying i.e. $|\ddot\psi | \ll m |\dot\psi|$. The Klein-Gordon
equation reduces to the Schr\"odinger equation:
\begin{equation}
\label{psiSchr}
i \, \partial_t \psi = - {\nabla^2 \over 2 m} \psi + m \Phi \psi \, .
\end{equation}
Several comments are in order. (1) In what sense is the assumption
of $\partial_t \ll m$ non-relativistic? From the Schr\"odinger
equation, we see $\partial_t \sim \nabla^2/m \sim k^2/m$.
Thus $\partial_t \ll m$ is equivalent to $k^2 / m \ll m$
i.e. momentum is small compared to rest mass.
(2) We introduce the gravitational potential $\Phi$.
Recall that $\Box = g^{\mu\nu}\nabla_\mu \nabla_\nu$ 
contains the
metric $g^{\mu\nu}$, thus gravitational interaction of $\phi$ is
implicit. For many applications,
this is the only interaction we need to include. 
\footnote{Wave dark matter described as such can be
  thought of as a minimalist version: the primary interaction
  is gravitational (though as we will see, 
  other interactions expected for an axion could be relevant in some cases).
  In the literature, there are studies of models where additional
  interactions play a crucial role
  e.g. \cite{RindlerDaller:2011kx,bk15,Fan:2016rda,Alexander:2016glq,Alexander:2019qsh}. 
Some of the phenomenology described here, such as wave interference,
applies to these models as well.
}
In principle, the
metric should account for the cosmic expansion, which we have
ignored to simplify the discussion. Cosmic counterparts of the equations
presented here can be found in \citep[e.g.,][]{Hu:2000ke,Hui:2016ltb}. 
(3) Despite the appearance of the Schr\"odinger equation, $\psi$
should be thought of as a (complex) classical field. The situation is
analogous to the case of electromagnetism: a state with
high occupancy is adequately described by the
classical electric and magnetic fields. 
We will on occasion refer to $\psi$ as the wavefunction, purely out of habit.

The non-relativistic 
dynamics of wave dark matter is completely described by
Equation \ref{psiSchr}, supplemented by the Poisson equation:
\begin{equation}
\label{poisson}
\nabla^2 \Phi = 4\pi G \rho \quad , \quad \rho = m |\psi|^2 \, .
\end{equation}
The expression for mass density $\rho$ can be justified by plugging 
Equation \ref{phipsi} into Equation \ref{rho}, taking the non-relativistic
limit and averaging over oscillations i.e. $|\psi|^2$ has the
meaning of particle number density.
Strictly speaking, the energy density should include
  gradient energy which is not contained in Equation \ref{rho}. 
The gradient energy contribution to $\rho$ is of order $|\nabla
\psi |^2/m$ which is negligible compared to
the rest mass contribution $m
|\psi|^2$ in the non-relativistic regime.

An alternative, fluid description of this wave system is instructive.
This is called the \citet{Madelung1927} formulation
\citep[see also][]{Feynman}. The mass density of the fluid is $\rho = m
|\psi|^2$ as discussed. The complex $\psi$ can be written
as $\psi = \sqrt{\rho/m} \, e^{i\theta}$. The fluid velocity $\vec v$
is related to the phase $\theta$ by:
\begin{equation}
\label{vdef}
\vec v = {1\over m} \vec \nabla \theta = {i \over 2m |\psi|^2} (\psi \vec\nabla
\psi^* - \psi^* \vec \nabla \psi) \, .
\end{equation}
Notice the fluid velocity is a gradient flow, resembling that of
a superfluid. (A superfluid can have vortices as topological defects,
see Section \ref{vortices}.)
With this identification of the fluid velocity,
what is normally understood as probability conservation
in quantum mechanics is now recast as mass conservation:
\begin{equation}
\label{masscons}
\partial_t \rho + \vec \nabla \cdot (\rho \vec v) = 0 \, .
\end{equation}
The Schr\"odinger equation possesses a U(1) symmetry,
the rotation of $\psi$ by a phase. In our context, 
conservation of the associated Noether current
expresses particle number conservation, or mass conservation, as
appropriate for the $\phi$ particles in the non-relativistic regime.

The Schr\"odinger equation is complex. Thus, besides
mass conservation, it implies an additional real equation, the Euler
equation: 
\begin{equation}
\label{euler}
\partial_t \vec v + (\vec v\cdot\vec\nabla)\,\vec v = - \vec \nabla \Phi + \frac{1}{
2 m^2} \vec \nabla \left( \frac{\nabla^2 \sqrt{\rho}}{ \sqrt{\rho}} \right).
\end{equation}
Equations \ref{masscons} and \ref{euler} serve as an alternative, fluid
description to the Schrodinger or wave formulation. The last term in
Equation \ref{euler} is often referred to as the quantum pressure term. It
is a bit of a misnomer (which we will perpetuate!),
for what we have is a classical system. Also,
the term arises from a stress tensor rather than
mere pressure:
\begin{equation}
\Sigma_{ij} = {1\over 4m^2} (\rho^{-1} \partial_i \rho \partial_j \rho
- \partial_i \partial_j \rho) = - {\rho \over 4
  m^2} \partial_i \partial_j {\,\rm ln\,}\rho \, ,
\end{equation}
i.e. $\partial_i (\nabla^2\sqrt{\rho} / \sqrt{\rho}) /(2m^2)
= - \rho^{-1} \partial_j \Sigma_{ij}$.
\footnote{
The Euler equation (combined with mass conservation)
can be re-expressed as $\partial_t (\rho v_i) + \partial_j (\rho v_i
v_j + \Sigma_{ij}) = - \rho \partial_i \Phi$. 
In other words, the standard energy-momentum tensor components
are:  $T^0 {}_0 = -\rho$, $T^0 {}_i = \rho v_i$, and 
$T^j {}_i = \rho v_i v_j +
\Sigma_{ij}$. It can be shown that
$T^j {}_i = T_{ji} = (4m)^{-1} (\partial_i \psi \partial_j \psi^* + \partial_i
\psi^* \partial_j \psi - \psi^* \partial_i \partial_j \psi -
\psi \partial_i \partial_j \psi^*)$. This $T^j {}_i$ can be rewritten
in a more familiar looking way by adding a tensor that is identically
conserved: $T^j {}_i \rightarrow (2m)^{-1} (\partial_i \psi \partial_j
\psi^* + \partial_i \psi^* \partial_j \psi - \delta_{ij} [\psi \nabla^2
\psi^*/2 + \psi^* \nabla^2 \psi /2 + \vec \nabla \psi \cdot \vec
\nabla \psi^*])$. 
Note the Euler equation 
in \citet{Hui:2016ltb} has
a factor of $\rho^{-1}$ missing in front of the divergence of the
stress tensor ($\sigma_{ij}$ there differs from $\Sigma_{ij}$ here by
an overall sign).
}
The stress tensor represents how the fluid description accounts for
the underlying wave dynamics. 
It shows in a clear way how the particle limit 
is obtained: for large $m$, the
Euler equation reduces to that for a pressureless fluid, as is
appropriate for particle dark matter.
We are interested in the opposite regime, where this stress tensor, or
the wave effects it encodes, plays an important role.

Incidentally, the insight that the {\it wave} formulation in the large
$m$ limit can be used to model {\it particle} cold dark matter
was exploited to good effect by
\citet{Widrow:1993qq}.
The wave description effectively reshuffles information in a phase-space Boltzmann
  distribution into a position-space wavefunction. It offers a number
  of insights that might otherwise be obscure \citep{CU2014,Uhlemann:2018gzz,Garny:2019noq}.

In the rest of this section, we deduce
a number of intuitive consequences 
from this system of equations---Equations \ref{psiSchr}
and \ref{poisson} in the wave description, or Equations \ref{masscons}
, \ref{euler} and \ref{poisson} in the fluid description.
Implications for observations and experiments are discussed
in Section \ref{implications}. 

\subsection{Perturbation theory}
\label{PT}

Suppose the density is approximately homogeneous with small
fluctuations: $\rho = \bar\rho (1 + \delta)$ where $|\delta| \ll 1$. 
We are interested in comparing the two terms---gravity and quantum pressure---on the right hand side of
the Euler equation (\ref{euler}). Taking the divergence of both, we
find:
\begin{equation}
- \nabla^2 \Phi + {1\over 4 m^2} \nabla^4 \delta \, ,
\end{equation}
where we have expanded out the quantum pressure term in small
$\delta$. Employing the Poisson equation 
$\nabla^2 \Phi = 4\pi G \bar\rho
\delta$,\footnote{The removal of $\bar\rho$ as a source for the
  Poisson equation (the so called Jeans swindle) can be
  justified in the cosmological context by considering perturbation
  theory around the Friedmann-Robertson-Walker background. Our
  expression is correct with $\nabla$ interpreted as derivative with
  respect to proper distance. Likewise, $k_J^{-1}$ given below is
  proper distance.} we see that the relative importance of
gravity versus quantum pressure is delineated by the Jeans scale:
\begin{equation}
\label{kJeans}
k_J = (16 \pi G\bar\rho)^{1/4} m^{1\over 2} \, ,
\end{equation}
where we have gone to Fourier space and 
replaced $\vec \nabla \rightarrow
i \vec k$. This gives $k_J \sim 70$/Mpc today for $m \sim 10^{-22}$ eV. 
On large length scales $k < k_J$, gravity dominates; on
small length scales $k > k_J$, quantum pressure wins. The sign
difference between the two terms makes clear quantum pressure 
suppresses fluctuations on small scales. This is the
prediction of linear perturbation theory---we will see in Section \ref{vortices} that the
opposite happens in the nonlinear regime. 

\begin{figure}[tb]
\centering
\includegraphics[width=1.0\textwidth,trim={2cm 13cm 1.5cm
  6.5cm},clip]{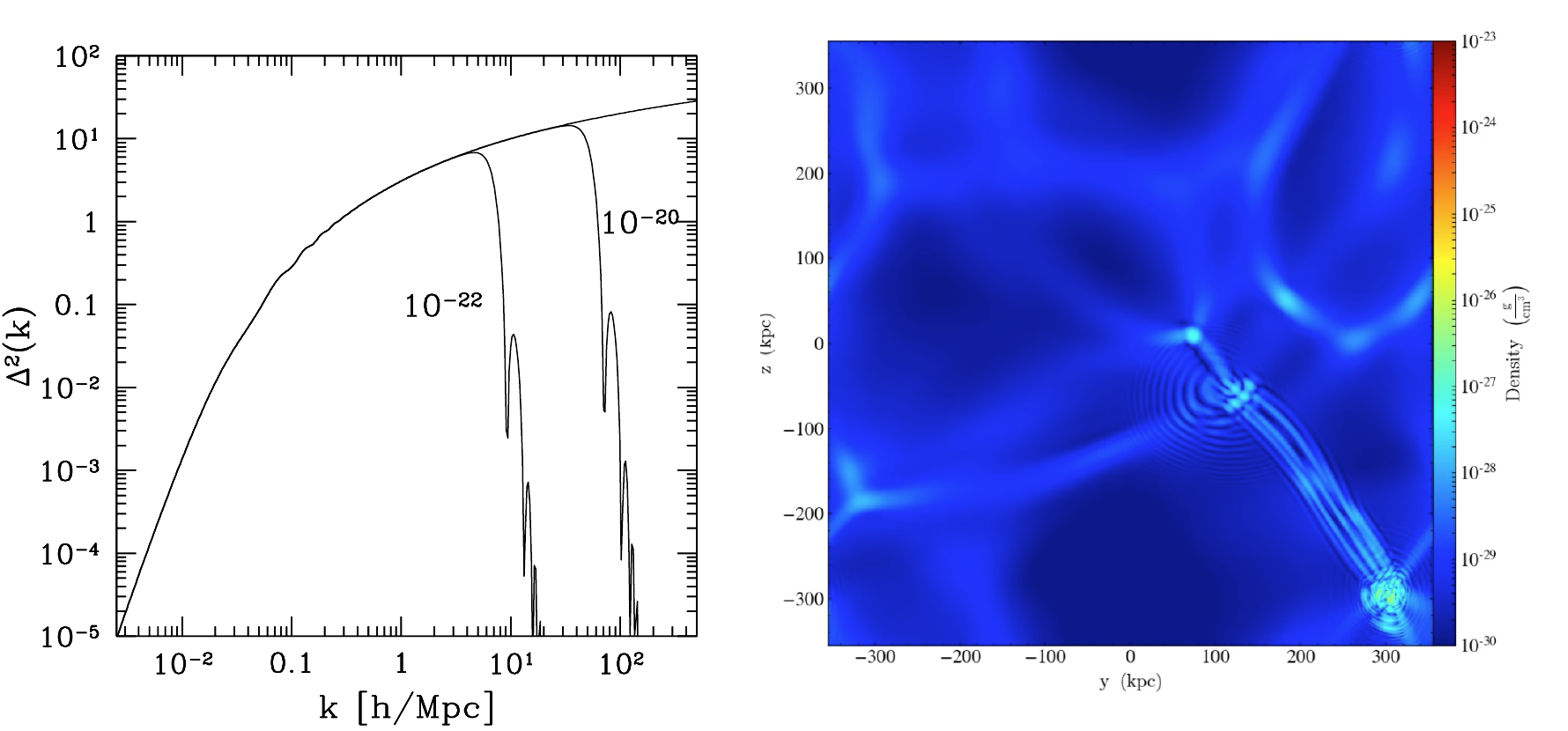}
\vspace{0.0cm}
\caption
{\small {\it Left panel:} the dimensionless linear mass power spectrum
  $\Delta^2(k) \equiv 4\pi k^3 P(k)/(2\pi)^3$, where $P(k$) is the
  dimensionful version, as a function of
  comoving momentum $k$. This is the linear power spectrum at redshift
  $z=0$. The top curve corresponds to that of conventional cold dark
  matter. The other two are for wave dark matter with $m =
  10^{-20}$ eV and $10^{-22}$ eV respectively, exhibiting the
  suppression of power on small scales (high $k$'s). The transfer
  function is taken from \citet{Hu:2000ke}. 
{\it Right panel:} a $z=5$ snapshot of
the dark matter density in a cosmological simulation of ultra-light
dark matter with $m=10^{-22}$ eV. The snapshot is $700$ kpc comoving
on a side. The color scale reflects the density
(in ${\rm \, g/cm^3}$).
Wave interference fringes can be seen along filaments and
in/around halos. Such interference patterns were first seen in
simulations by \citet*{Schive:2014dra}.
Snapshot produced by Xinyu Li \citep[][]{Li:2018kyk}.
}
\label{wavepicture}
\end{figure}

This reasoning tells us the linear power spectrum of wave dark
matter should match that of particle dark matter 
(or conventional cold dark
matter) at low $k$'s but be suppressed at sufficiently high
$k$'s. The precise transition scale differs from
$k_J$ given above---a proper computation must include the effect of
radiation in the early universe, and account for the full history, from
slow-roll to oscillations, outlined in Section \ref{motivations}. 
This was carried out by \cite{Hu:2000ke}, who gave
\begin{equation}
\label{khalf}
k_{1/2} = 4.5 \left({m \over 10^{-22} {\,\rm eV}}\right)^{4/9} {\,\rm
  Mpc}^{-1}  \, 
\end{equation}
as the (comoving) scale at which the linear power spectrum
is suppressed by a factor of two,
and beyond which the power drops 
precipitously ($\sim k^{-16}$). This is illustrated in the left panel
of Figure \ref{wavepicture}.
For more recent computations, see
\cite{Cookmeyer:2019rna,Hlozek:2016lzm,Hlozek:2014lca}.
If the scalar potential $V(\phi)$ is indeed of the form given
in Equation \ref{Vphi},
the computation should in principle account for the full shape of 
$V(\phi)$ rather than approximating it as quadratic,
especially if the primordial $\phi$ value is comparable to $f$.
This was investigated by \cite{Zhang:2017dpp,Arvanitaki:2019rax},
who found
that the predicted linear power spectrum is largely consistent with earlier
work, unless the primordial $\phi$ is extremely close to $\pi f$ i.e. the top of the
potential.
\footnote{Computations of the linear power spectrum discussed above
assume the fluctuations are adiabatic i.e. $\phi$ fluctuations,
like fluctuations in photons, baryons and
neutrinos, are all inherited from the curvature, or
inflaton, fluctuation. The scalar $\phi$ can in addition have its own 
isocurvature fluctuations (see Section \ref{implications}).}

The linear perturbative computation described above is phrased in the
fluid picture. 
A fluid perturbation theory computation up to third order in $\delta$
and $v$ was carried out in \cite{Li:2018kyk} to obtain the one-loop power
spectrum.
One could also consider perturbation theory in the wave formulation,
expanding in small $\delta\psi \equiv \psi - \bar\psi$, 
where $\bar\psi$ is the homogeneous contribution.
Wave perturbation theory turns out to break down at higher redshifts
compared to fluid perturbation theory \citep{Li:2018kyk}. 
\footnote{
Wave perturbation theory requires not only the smallness of 
$(\delta \psi + \delta
\psi^*)/\bar\psi$ (which equals $\delta$), but also
the smallness of $(\delta \psi - \delta \psi^*)/\bar\psi$ 
(it is related to the fluid velocity by $\vec v = \vec
\nabla (\delta\psi - \delta\psi^*) / (2im\bar\psi)$). 
In other words, wave perturbation theory assumes
small $\delta$ and $mv / k$, while fluid perturbation theory assumes 
small $\delta$ and $v$. In large scale structure, one is typically
interested in situations where $m / k \gg 1$. Thus perturbation
theory breaks down sooner in the wave formulation.
}

\subsection{Soliton/boson star}
\label{solitons}

The Euler equation is useful for intuiting
properties of certain nonlinear, bound objects, known as solitons or
boson stars
\citep{Kaup:1968zz,Ruffini:1969qy,Friedberg:1986tp,Friedberg:1986tq,Seidel:1993zk,Guzman:2006yc}. 
We are interested in objects in which quantum pressure
balances gravitational attraction i.e. the two terms on the right hand
side of Equation \ref{euler} cancel each other:
\begin{equation}
{GM \over R} \sim {1\over m^2 R^2} \, ,
\end{equation}
where $M$ is the total mass of the object and
$R$ is its radius, and we have replaced $\nabla \sim 1/R$ and dropped
factor of $2$.
This implies the size of the soliton/boson star is inversely
proportional to its mass:
\begin{eqnarray}
\label{Rsoliton}
&& R \sim {1\over GM m^2} \sim 100 {\,\rm pc} \, {10^9 {\,\rm M_\odot} \over M} 
\left( {10^{-22} {\,\rm eV} \over m} \right)^2 \nonumber \\
&& \quad \sim 300 {\,\rm km} \, {10^{-10} {\,\rm M_\odot} \over M} 
\left( {10^{-6} {\,\rm eV} \over m} \right)^2 
\sim 50 {\,\rm km} \, {5 {\, \rm M_\odot} \over M} 
\left( {10^{-11} {\,\rm eV} \over m} \right)^2 \, ,
\end{eqnarray}
where we give a few representative values of $M$ and
$m$.\footnote{This rough estimate is about a
factor of 4 smaller than the exact relation
\citep{Chavanis:2011zi}.
We focus on spherical solitons. Filamentary and pancake analogs
are explored in
\citet{Desjacques:2017fmf,Alexander:2019qsh,Mocz:2019pyf},
and 
rotating solitons are discussed in 
\citet{Hertzberg:2018lmt}.
}
The example of $m \sim 10^{-22}$ eV
corresponds to that of fuzzy dark matter---such a
soliton can form in the centers of galaxies
\citep[][see Section \ref{dynamics2} below]{Schive:2014dra,Schive:2014hza}. 
The example of $m \sim 10^{-6}$ eV
corresponds to that of the QCD axion---such an axion star (often 
called an axion
minicluster) could form
in the aftermath of 
Peccei-Quinn symmetry breaking after inflation \citep{Kolb:1993zz,Kolb:1995bu,Fairbairn:2017sil,Eggemeier:2019jsu,Buschmann:2019icd}.
The example of $m \sim 10^{-11}$ eV could be
an axion-like-particle---an object like this has been
studied as a possible gravitational wave event progenitor 
\citep{Helfer:2016ljl,Widdicombe:2018oeo}.

There is an upper limit to the mass of the soliton: $GM/R \, \lsim \, 1$ to
avoid collapse to a black hole. Plugging in the expression for $R$, we
deduce the maximum soliton mass (a Chandrasekhar mass of sort):
\begin{equation}
\label{Mmax1}
M_{\rm max} \sim {1\over G m} \sim 10^{12} {\,\rm M_\odot} \left( {10^{-22}
    {\,\rm eV} \over m} \right) \sim 10^{-4} {\,\rm M_\odot} \left( {10^{-6}
    {\,\rm eV} \over m} \right) \sim 10 {\,\rm M_\odot} \left( {10^{-11}
    {\,\rm eV} \over m} \right) \, .
\end{equation}
Strictly speaking, as one approaches the maximum mass, one should use
the relativistic Klein Gordon description rather than the Schr\"odinger
equation, but the above provides a reasonable estimate
\citep{Kaup:1968zz,Ruffini:1969qy, Friedberg:1986tq}.

Not all gravitationally bound objects are
solitons, of course. The argument above accounts for the two terms on the
right of the Euler equation (\ref{euler}). 
The velocity terms on the left could also play a role.
In other words, a bound object could exist by balancing
gravity against virialized motion instead i.e. 
$v^2 \sim GM/R > 1/(m^2 R^2)$. 
Most galaxies are expected to fall into this category, supported by
virialized motion except possibly at the core where a soliton could
condense (see Section \ref{dynamics2}). 

The discussion so far ignores the possibility of self-interaction. For
an axion, we expect a $m^2 \phi^4 / f^2$ contribution to the
Lagrangian (Equation \ref{phiInteractions}). It can be shown the relevant
quantities to compare are: $v^2$ (virialized motion), $1/(m^2 R^2)$
(quantum pressure) balancing against $GM/R$ (gravity) and $M/(m^2 f^2
R^3)$ (attractive self-interaction of the axion). 
This can be deduced by comparing the gravitational
  contribution to energy density $\rho \Phi$ with the self-interaction
  contribution $m^2 \phi^4/f^2 \sim \rho^2 / (m^2 f^2)$, and using
  $\Phi \sim GM/R$ and $\rho \sim M/R^3$.
The attractive self-interaction is destabilizing, going as $1/R^3$: 
if it dominates over gravity, there is nothing that would
stop $R$ from getting smaller and making the self-interaction 
even stronger. Demanding that the $M$-$R$ relation in
Equation \ref{Rsoliton} satisfies $GM/R > M/(m^2 f^2 R^3)$ modifies the
maximum soliton mass to
\citep{Eby:2015hsq,Eby:2015hyx,Helfer:2016ljl}:
\begin{eqnarray}
\label{Mmax2}
&& M_{\rm max} \sim {f \over G^{1/2} m} \sim 10^{10} {\,\rm M_\odot}
\left( {f \over 10^{17} {\,\rm GeV}} \right) 
\left( {10^{-22} {\,\rm eV} \over m} \right) \nonumber \\
&& \quad \quad \quad \sim 10^{-10} {\,\rm M_\odot} 
\left( {f \over 10^{13} {\,\rm GeV}} \right) 
\left( {10^{-6} {\,\rm eV} \over m} \right)  
\sim {\,\rm M_\odot} 
\left( {f \over 10^{18} {\,\rm GeV}} \right) 
\left( {10^{-11} {\,\rm eV} \over m} \right) \, .
\end{eqnarray}

\subsection{Numerical simulations}
\label{simulations}

Great strides have been made in 
numerical simulations of structure formation with wave dark
matter (the Schr\"odinger-Poisson system), 
starting with the work of \citet*{Schive:2014dra}. 
There are by now a number of different algorithms,
including spectral method and finite difference
\citep{Schive:2014dra,Schwabe:2016rze,Mocz:2017wlg,Du:2018qor,Li:2018kyk,Edwards:2018ccc,Mocz:2019pyf,Schwabe:2020eac}, 
often with adaptive mesh refinement. 
One key challenge to solving the Schr\"odinger-Poisson system
(Equations \ref{psiSchr} and \ref{poisson})
is the high demand for resolution. In cosmological applications, 
one is often interested in predictions on large scales, say length
scale $\lambda$.
To accurately describe bulk motion on such large
scales, say velocity $v$, one must include waves with the
corresponding wavelength $2\pi/(mv)$. The trouble is that 
one is often in situations where $2\pi/(mv) \ll \lambda$. 
For instance, with $m \sim 10^{-22}$ eV and a velocity of $100$ km/s,
the de Broglie wavelength $2\pi/(mv) \sim 1.2$ kpc is a lot smaller than
typical length scales of interest in large scale structure $\lambda >
1$ Mpc. A wave simulation, unlike an N-body simulation, 
thus must have high resolution even if
one is only interested in large scales.
This is why existing wave simulations are typically limited to small box
sizes. A related challenge is the requisite time-step: dimensional
analysis applied to the Schr\"odinger equation tells us the time-step scales as $m \times {\,\rm
  resolution}^2$, i.e. the time-step has to be less than the de
Broglie wavelength divided by the typical velocity. Contrast this with
the requirement for an N-body simulation---a time step of $\, \lsim
\,\lambda/v$ suffices. A recent $\sim 10$ Mpc box,
de-Broglie-scale-resolved, wave simulation was described by
\cite{May:2021wwp}.

An alternative is to simulate the fluid formulation,
expressed in Equations \ref{poisson}, \ref{masscons} and \ref{euler} 
\citep{Mocz:2015sda,Veltmaat:2016rxo,Nori:2018hud,Nori:2018pka}. 
With $\rho$ and $\vec v$ as variables (related to the amplitude and
phase of $\psi$), there is no need to have high spatial resolution
just to correctly capture the large scale flows. 
The downside is that the fluid formulation is ill-defined at places
where $\rho = 0$. This can be seen by looking at the form of the
quantum pressure term in the Euler equation (\ref{euler}), or more
simply, by noting that the phase of the wavefunction $\psi$
(which determines $\vec v$) becomes ill-defined at
locations where $\rho = m |\psi|^2$ vanishes. One might think
occurrences of vanishing $\rho$ must be rare and have a negligible
impact; this turns out to be false \citep{Li:2018kyk,Hui:2020hbq}---we
will have more to say about this in Section \ref{vortices}.
A promising approach to overcome this and the resolution challenge is a
hybrid scheme, where the large scale evolution proceeds according to
the fluid formulation or an N-body code (the vanishing-$\rho$ issue
does not arise on large scales), and the small scale evolution
follows the wave formulation \citep{Veltmaat:2018dfz}.

Recall that the Schr\"odinger equation originates as 
a non-relativistic approximation to the Klein-Gordon equation.
If one is interested in applications where relativity plays a role,
such as a soliton close to its maximum possible mass
(Section \ref{solitons}), or the scalar field close to black holes or in the
early universe, a Klein-Gordon code (or more generally, a code to
evolve a scalar with arbitrary potential) should be used. There are
many examples in the literature:
\citet{Felder:2000hq,Easther:2009ft,Giblin:2010bd,Amin:2011hj,Helfer:2016ljl,Widdicombe:2018oeo,Buschmann:2019icd,Eggemeier:2019jsu}.

Much of the recent progress in understanding halo substructure for
wave dark matter comes from numerical simulations, often in the
ultra-light regime of $m \sim 10^{-22}$ eV. 
Many of the
qualitative features carry over to higher masses; 
the quantitative implications for
observations/experiments are mass specific of course, as we will discuss.


\subsection{Wave interference---granules and vortices}
\label{vortices}

The right panel of Figure \ref{wavepicture} shows the dark matter density in a snapshot of a
cosmological wave simulation \citep{Li:2018kyk}. A striking feature is
the presence of interference fringes, a characteristic prediction of
wave dark matter, first demonstrated in cosmological simulations by 
\citet*{Schive:2014dra}, and subsequently
confirmed by many groups
\citep{Schive:2014dra,Schwabe:2016rze,Veltmaat:2016rxo,Mocz:2017wlg,Du:2018qor,Li:2018kyk,Edwards:2018ccc,Nori:2018hud,Veltmaat:2018dfz,Mocz:2019pyf,Schwabe:2020eac}.
The interference patterns are particularly obvious in the nonlinear
regime, along filaments and in/around collapsed halos. 
In these nonlinear objects, wave interference causes order one
fluctuations in density: blobs of constructive interference of de
Broglie size (sometimes called granules) interspersed between patches
of destructive interference. 

As a simple model of a galactic halo, consider a
superposition of plane waves:
\begin{equation}
\label{psiRandom}
\psi (t, \vec x) = \sum_{\vec k} A_{\vec k} e^{iB_{\vec k}} e^{i{\vec k} \cdot {\vec x} - i \omega_k t} \, ,
\end{equation}
where $A_{\vec k}$ and $B_{\vec k}$ are the amplitude and phase of
each plane wave of momentum $\vec k$. 
\footnote{\label{eigenstates} 
Here, $\omega_k = |\vec k|^2 /(2m)$. 
A more realistic model would superimpose eigenstates of a
  desired gravitational potential \citep{Lin:2018whl,Li:2020ryg}, in which case
  $\omega_k$ would be the energy of each eigenmode (labeled abstractly
  by $k$), with $e^{i\vec k \cdot \vec x}$ replaced by the
  corresponding eigenfunction.}
In a virialized halo, it is reasonable to expect, as a zero order
approximation, that the phases $B_{\vec k}$'s are randomly
distributed. This is the analog of assuming random orbital
phases for stars in a halo.
We refer to this as the random phase halo model.
The amplitudes $A_{\vec k}$'s should reflect
the velocity (or momentum) dispersion within the halo. For instance
we can adopt $A_{\vec k} \propto e^{-k^2/k_0^2}$ (where $k =
|\vec k|$), resembling an isothermal distribution, with a de Broglie
wavelength $\propto 1/k_0$. The density is:
\begin{equation}
\rho = m |\psi|^2 = m \sum_{\vec k} A_{\vec k}^2 + m \sum_{\vec k \ne \vec
  k'} A_{\vec k} A_{\vec k'} e^{i (B_{\vec k} - B_{\vec k'})}
e^{i({\vec k} - {\vec k'}) \cdot \vec x - i (\omega_k - \omega_{k'})t}
\, .
\end{equation}
The first term comes from squaring each Fourier mode and summing
them. The second represents the contribution from interference between 
different Fourier modes.\footnote{If we had built a more realistic model where
the plane waves are replaced by energy eigenstates (see footnote
\ref{eigenstates}), the first term would be ${\vec x}$ dependent, but
would remain time independent.} It is the second term that is
responsible for the appearance of interference fringes in numerical
simulations such as shown in Figure \ref{wavepicture}. 
The typical difference in momenta between different Fourier modes is
of the order of $k_0$, which fixes the characteristic size of the
interference fringes or granules i.e. the de Broglie wavelength $\sim
2\pi/k_0$. The typical difference in energy between the modes is of
the order of $\sim k_0^2/(2m) \sim k_0 v/2$, where $v$ is the velocity
dispersion. This determines the characteristic time scale over which
the interference pattern changes i.e. the de Broglie time:
\begin{eqnarray}
\label{tcohere}
&& t_{\rm dB} \equiv {2\pi \over mv^2} = 1.9 \times 10^6 {\,\rm yr.} 
\left( {10^{-22} {\,\rm eV} \over m} \right) \left( {250 {\,\rm km/s} \over
  v} \right)^2 \nonumber \\
&& \quad = 5.9 \times 10^{-3} {\,\rm s} \left(
  {10^{-6} {\,\rm eV} \over m} \right) \left( {250 {\,\rm km/s} \over
  v} \right)^2 \, .
\end{eqnarray}
There is some arbitrariness in the choice of the prefactor
  $2\pi$. Reasonable choices range within factor of a few.

In other words, wave interference produces de-Broglie-scale, order unity density
fluctuations which vary on time scale of $t_{\rm dB}$. Such fluctuations can
in principle take the density all the way to zero i.e. complete destructive interference. 
What is interesting is that (1) such occurrences are not
rare, and (2) the locations of complete destructive interference are
vortices. This was explored in \cite{2011JPhB...44k5101C,Hui:2020hbq}.
\footnote{More generally, vortices in dark matter were studied in
  \cite{Silverman:2002qx,Brook:2009ku,Kain:2010rb,RindlerDaller:2011kx,Zinner:2011if,Banik:2013rxa,Alexander:2016glq,Alexander:2019puy}. Most
  of the studies focused on a regime where
  self-interaction dominates over quantum pressure. Here, we describe
  the opposite regime, relevant for weakly-coupled dark matter with a
  long de Broglie wavelength, where gravity and quantum pressure completely
  describe the physics. Vortices have long been studied
in other contexts, such as high energy and condensed matter
physics
\citep{Nielsen:1973cs,Luscher:1980ac,1949NCim....6S.279O,LUND1991245,
2008LaPhy..18....1F}.
}
Below we summarize the findings, following the line of reasoning  in
\cite{Hui:2020hbq}. 

\begin{figure}[tb]
\centering
\includegraphics[width=1.0\textwidth,trim={1cm 13cm 1cm
  9cm},clip]{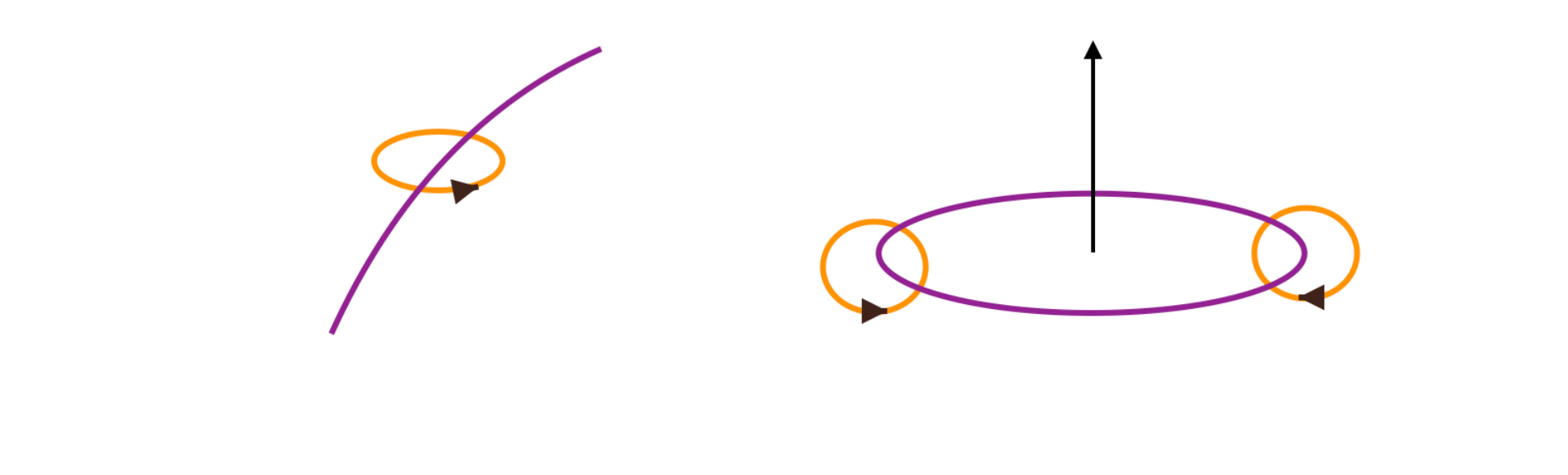}
\vspace{-0.5cm}
\caption
{\small Schematic illustration of vortices. {\it Left panel:} a vortex
  line, or segment thereof (purple line). 
  The loop with arrow indicates velocity
  circulation (or phase winding) around the vortex. 
  {\it Right panel:} a vortex ring (purple
  line). The loops with arrows indicate velocity circulation. The
  arrow in the middle indicates the bulk motion of the ring.
}
\label{winding}
\end{figure}


In three spatial dimensions, the set of points where the real part of
the wavefunction vanishes generically forms a surface. Likewise
for the imaginary part. Demanding both parts of the wavefunction
vanish thus gives a line, where the two surfaces cross.
The purple line in the left panel of Figure \ref{winding}
depicts such a line of vanishing $\psi$ (i.e. the amplitude of $\psi$
is zero and the phase is ill-defined on the line).
Consider a loop going around this line: 
for the wavefunction to be single-valued, the phase of
the wavefunction must wind by integers of $2\pi$. 
Recall the fluid velocity is given by the gradient of the
phase (Equation \ref{vdef}); integrating the velocity around a loop
encircling the line of vanishing $\psi$ gives:
\begin{equation}
\label{phasewinding}
{\rm circulation\,} \equiv \oint d\vec x \cdot \vec v = {2\pi n \over m} \, ,
\end{equation}
where $n$ is an integer. The line of vanishing $\psi$ is therefore a
vortex. 
\footnote{Note that the vortex is distinct from the axion string.
The relevant $U(1)$ for an axion string is the Peccei-Quinn $U(1)$,
while that for a vortex is the $U(1)$ associated with particle number conservation
in the non-relativistic limit. This raises the interesting question of
how to view the vortex from the perspective of the full $\phi$
theory. See discussions in \cite{Hui:2020hbq}.
}
It is helpful to consider a Taylor expansion around a point on the
vortex (let's take it to be the origin):
\begin{equation}
\psi (\vec x) \sim \vec x \, \cdot \vec\nabla \psi |_0 \, ,
\label{psiTaylor}
\end{equation}
assuming $\vec \nabla \psi |_0$, the
derivative evaluated at $x=0$, does not
vanish.
It can be shown the winding number $n = \pm 1$ as long as
$\vec \nabla \psi |_0$ does not vanish. 
If it vanishes, one would have to consider the next higher order
term in the Taylor expansion, yielding higher winding.
A vortex line, much like a magnetic field line, cannot end, and so one
expects generically a vortex ring, depicted in the right panel of
Figure \ref{winding}. 
It can be further shown that, in addition to velocity circulation around the
ring, the ring itself moves with a bulk velocity that scales
inversely with its size. Analytic solutions illustrating this behavior
(and more) can be found in \citet{2000PhRvA..61c2110B,Hui:2020hbq}. 

A number of features of vortices in wave dark
matter are worth stressing. 
(1) One might think these locations of
chance, complete destructive interference must be rare, but they are
actually ubiquitous: on average there is about one vortex ring per de
Broglie volume in a virialized halo. 
This has been verified analytically in the random
phase halo model, and in numerical wave simulations of halos that form from
gravitational collapse.\footnote{
In a numerical simulation, checking that the density is low is not
enough to ascertain that one has a vortex (keep in mind the density
almost never exactly vanishes numerically). A better diagnostic is to look for
non-vanishing velocity circulation, or phase winding---this is also 
more robust against varying resolution.
} 
Note that gravity plays an important role in the formation of vortices
in the cosmology setting. In the early universe, the density (and the 
wavefunction) is roughly homogeneous with very small fluctuations;
this means nowhere does the wavefunction
vanish. It is only after gravity amplifies the density fluctuations,
to order unity or larger, is complete destructive interference possible.
(2) Vortex rings in a realistic halo
are not nice round circles, but rather deformed loops. Nonetheless,
certain features are robust. Close to a vortex,
the velocity scales as $1/r$ where $r$ is distance from vortex
(following from Equation \ref{phasewinding}), and the 
density scales as $r^2$ (following from Equation \ref{psiTaylor}).
\footnote{More generally, the density scales as $r^{2|n|}$ where $n$ is
  the winding number. However, simulations suggest $|n|=1$ is the generic
  expectation: it is rare to have $\psi$ {\it and} $\vec\nabla \psi$
  vanish at the same time.
}
Moreover, a segment of a ring moves with a
velocity that scales with the curvature i.e. curvier means faster.
(3) Vortex rings come in a whole range of sizes: the distribution is
roughly flat below the de Broglie wavelength, but is exponentially
suppressed beyond that.
(4) Vortex rings are transient, in the same sense that wave
interference patterns are. The coherence time is roughly the de
Broglie time (Equation \ref{tcohere}).
Vortex rings cannot appear or disappear in an arbitrary way, though.
A vortex ring can appear by first nucleating as a point, and then
growing to some finite size. It can disappear only by shrinking back
to a point (or merge with another ring). This behavior can be
understood as a result of Kelvin's theorem: recall that the fluid
description is valid away from vortices; conservation of
circulation tells us that vortices cannot be arbitrarily removed or
created.

To summarize, wave interference substructures, of which vortices are
a dramatic manifestation, are a unique signature of wave dark
matter. It is worth stressing that while the wave nature of dark
matter leads to a suppression of small scale power in the linear regime
(Section \ref{PT}), it leads to the opposite effect in the nonlinear
regime, by virtue of interference.
We discuss the implications for observations and
experiments in Section \ref{implications}.

\subsection{Dynamical processes---relaxation, oscillation, evaporation, friction and heating}
\label{dynamics2}

An interesting phenomenon in a wave dark matter halo is
soliton condensation, first pointed out by
\cite{Schive:2014dra,Schive:2014hza}. It is observed that virialized
halos in a cosmological simulation 
tend to have a core that resembles the soliton
discussed in Section \ref{solitons}, with a soliton mass that scales with
the halo mass as:
\begin{equation}
\label{solitonhalorelation}
M_{\rm soliton} \sim 6.7 \times 10^7 {\,\rm M_\odot} {10^{-22} {\,\rm eV} \over m}
\left( {M_{\rm halo} \over 10^{10} {\,\rm M_\odot}} \right)^{1/3} \, .
\end{equation}
\footnote{\label{solitonhalo}
It is worth emphasizing that this relation is well-tested only over a
limited range of halo mass: $\sim 10^9 - 10^{11} {\,\rm M_\odot}$, because
of the difficulty in simulating large boxes (Section \ref{simulations}). 
The relation can be roughly
understood as follows \citep{Schive:2014hza}.
Recall that $R_{\rm
  soliton} \propto 1/M_{\rm soliton}$ (Equation \ref{Rsoliton}). Thus,
the gravitational potential of the soliton $
\sim GM_{\rm soliton}/R_{\rm soliton} \propto M_{\rm soliton}^2$. 
Equating this with the gravitational potential of the halo $\sim GM_{\rm
  halo}/R_{\rm halo}$, and assuming $M_{\rm halo} / R_{\rm halo}^3$ 
is constant i.e. $R_{\rm halo} \propto M_{\rm halo}^{1/3}$, the relation
$M_{\rm soliton} \propto M_{\rm halo}^{1/3}$ follows.
That the gravitational potential of the soliton and of the halo roughly
match can be interpreted as some sort of isothermal condition.
It would be useful to check if the kinetic approach of
\cite{Levkov:2018kau} can reproduce this.
See \cite{Bar:2018acw} for further discussions.
}
The condensation process was studied by solving the Landau kinetic
equation in \cite{Levkov:2018kau} \citep[see
also][]{Seidel:1993zk,hmt03,gul06,Schwabe:2016rze}. 
Here, we describe a heuristic
derivation of the condensation, or relaxation, time scale \citep{Hui:2016ltb}. 
Consider the part of a halo interior to radius $R$, with velocity
dispersion $v$. Suppose there is no soliton yet. Wave interference as
described in Section \ref{vortices} inevitably produces granules of de
Broglie size $\lambda_{\rm dB}$. In this region, we have $\sim
(2R/\lambda_{\rm dB})^3$ 
such granules or quasi-particles. The relaxation time for such a
gravitational system is roughly a tenth of the crossing time 
$2R/v$ times the number of granules i.e.
\begin{eqnarray}
\label{trelax}
t_{\rm relax} \sim  0.1 {2R\over v} \left( {2R \over \lambda_{\rm dB}} \right)^3 \sim
10^{8} {\,\rm yr} 
\left( {R \over 2 {\,\rm kpc}} \right)^4
\left( {v \over 100 {\,\rm km/s}} \right)^2 
\left( {m \over 10^{-22} {\,\rm eV}} \right)^3 \nonumber \\
\sim 10^{8} {\,\rm yr} 
\left( {0.14 {\,\rm M_\odot} /{\,\rm pc}^3 \over \rho} \right)^2
\left( {v \over 100 {\,\rm km/s}} \right)^6
\left( {m \over 10^{-22} {\,\rm eV}} \right)^3 
\, .
\end{eqnarray}
In essence, we have adapted the standard relaxation time for a
gravitational system \citep{BT} 
by replacing the number of particles/stars by the
number of de Broglie granules.
The above estimate suggests the condensation of solitons
quickly becomes inefficient for larger values of $m$. 
It remains to be verified, though, whether this 
is indeed the relevant time scale for soliton formation 
in a cosmological setting where halos undergo repeated mergers. 
For instance, in a numerical study of six halos by \cite{Veltmaat:2018dfz}, 
all halos have substantial cores from the moment of halo formation,
though two of them exhibit some core growth over time.

\begin{figure}[tb]
\centering
\includegraphics[width=1.0\textwidth,trim={2.4cm 12.5cm 2.3cm
  7.5cm},clip]{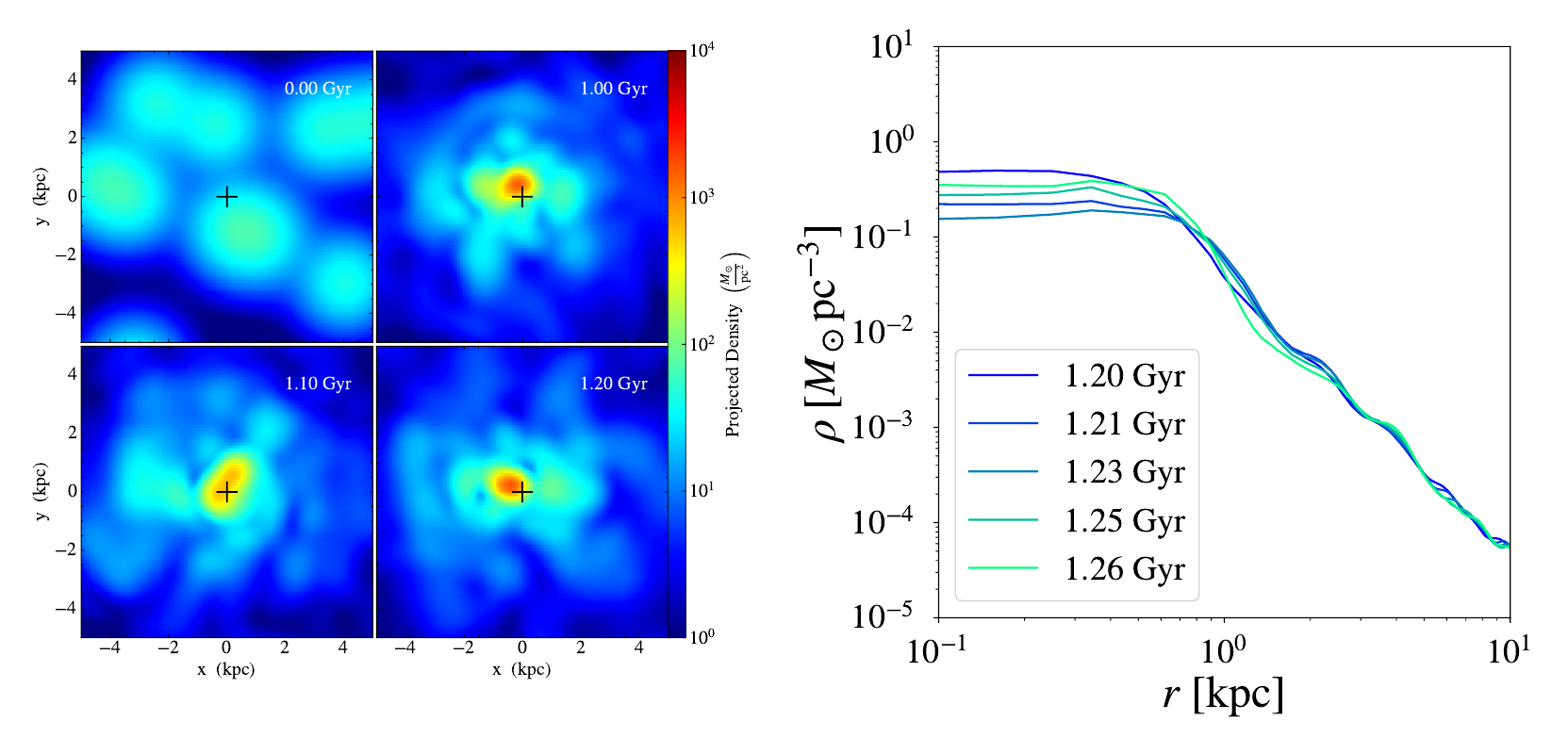}
\vspace{0.0cm}
\caption
{\small {\it Left panel:} Snapshots of the formation of a halo.
  Clockwise from top-left: initial moment, $1$ Gyr, $1.2$ Gyr and
  $1.1$ Gyr. Each snapshot is $10$ kpc on a side. Color coding denotes the projected density in
  ${\rm M_\odot/\,\rm pc^2}$. The cross in the middle denotes the center
  of mass. Note how the soliton core wanders.
  {\it Right panel:} Spherically averaged density profile (density in
  ${\rm M_\odot/\,\rm pc}^3$ as a function of radius in kpc) at several
  different moments, from $1.2$ Gyr to $1.26$ Gyr.
  The soliton core exhibits persistent oscillations. 
  Soliton oscillations and random walk were first observed in
  simulations by
  \citet{Veltmaat:2018dfz,Schive:2019rrw}.
  Figure adapted from \citet{Li:2020ryg}.
}
\label{oscill}
\end{figure}

Detailed studies of simulations suggest the core of a fuzzy dark
matter halo is not an exact soliton. \citet*{Veltmaat:2018dfz} pointed
out that the core object has persistent oscillations, and
\citet*{Schive:2019rrw} demonstrated that it random walks
(see Figure \ref{oscill}).
This is another manifestation of wave interference. Think of the halo
gravitational potential as approximately constant (in time); the halo
can be decomposed into a superposition of energy eigenstates
\citep{Lin:2018whl}. The ground state (i.e. the solitonic state)
contributes substantially to the density around the halo center, but
it is not the only state that does. Interference between the ground
state and excited states approximately matches the core oscillations
and random walk observed in simulations
\citep{Li:2020ryg,Padmanabhan:2020}.

It is well known that a subhalo embedded inside a larger parent halo 
can be tidally disrupted. The tidal radius is roughly where the
average interior density of the subhalo matches that of the parent
halo. Quantum pressure adds a new twist to this story: even mass
within the tidal radius of the subhalo is unstable to disruption.
The evaporation time scale of a soliton inside a host halo was
computed in \cite{Hui:2016ltb}: a soliton would evaporate in 
$\, \lsim \, 10$ orbits if its density is $\,\lsim \, 60$ times the
host density. This was verified in wave simulations by
\cite{Du:2018qor}.

The wave nature of dark matter also has an impact on dynamical
friction. Recall how dynamical friction works: a heavy object ploughs
through a sea of dark matter particles; gravitational scattering
creates an overdense tail of particles in its wake; the overdense tail
gravitationally pulls on the heavy object, effecting friction.
For wave dark matter, one expects a smoothing of the overdense tail
on the de Broglie scale. The dynamical
friction is thus suppressed. A computation, neglecting self-gravity of
the dark matter and assuming the unperturbed background is homogenous, 
is described in \cite{Hui:2016ltb} \citep[see also][]{lora12}:
while the frictional force is
$4\pi \rho (GM/v)^2 \left({\,\rm ln} [2r/(GM/v^2)] - 1\right)$ in
the particle limit, it is
$4\pi \rho (GM/v)^2 \left({\,\rm ln} [2r mv] - 1 + \gamma\right)$
in the wave limit.
\footnote{The result is derived by integrating momentum flux over a
  sphere surrounding $M$, as opposed to a cylinder like in
  Chandrasekhar's classic computation, hence a small difference in the
  Coulomb logarithm in the particle limit. Also, $r mv \gg 1$ is
  assumed. See \cite{Hui:2016ltb} for details.
}
Here, $\rho$ is the background mass density, $M$
is the mass of the heavy object (such as a globular cluster), $v$ is
the velocity of the heavy object, 
$r$ is the size of the galactic halo or the orbital radius of $M$ in
the halo,
and $\gamma = 0.577...$ is the
Euler-Mascheroni constant. The distinction between the particle limit
(i.e. Chandrasekhar) and the wave limit comes
down to comparing two length scales: 
$GM/v^2$ (the impact parameter at which significant deflection occurs) 
versus the de Broglie
scale $\sim 1/(mv)$. The wave limit applies when the former is less
than the latter i.e. if the following ratio is small:
\begin{equation}
{GM/v^2 \over (1/mv)} = 0.002 \left({M \over 10^6 {\,\rm M_\odot}}\right)
\left( {100 {\rm \, km/s} \over v} \right) 
\left( {m \over 10^{-22} {\,\rm eV}} \right) \, .
\end{equation}
Depending on the parameters of interest, dynamical
friction can be suppressed significantly, if $m$ is in the
ultra-light range.
A computation of dynamical friction in more general fluid dark
matter is carried out in \cite{Berezhiani:2019pzd}. Investigations of
dynamical friction in fuzzy dark matter in more realistic settings---
inhomogeneous background, with de Broglie granules---can be found in
\cite{Du:2016zcv,Bar-Or:2018pxz,Lancaster:2019mde}.

We close this section with a discussion of one more dynamical effect
from the wave nature of dark matter. Recall from Section \ref{vortices}
that the wave interference pattern of granules and vortices is
transient, on time scale of $t_{\rm dB}$ (Equation \ref{tcohere}). The
fluctuating gravitational potential leads to the heating and scattering of stars
\citep{Hui:2016ltb,Amorisco:2018dcn,Bar-Or:2018pxz,Church:2018sro,Marsh:2018zyw,Schive:2019rrw}. A
rough estimate can be obtained as follows.
Consider a star undergoing deflection by a de Broglie blob:
the angle of (weak) deflection is $\sim 2 G M / (b v^2)$ where $M$ is the
mass of the blob and $b$ is the impact parameter. The deflection 
imparts a kick to the velocity of the star, perpendicular to the
original direction of motion: $\Delta v \sim 2 G M / (b v)$. 
Using $M \sim 4\pi \rho (\lambda_{\rm dB}/2)^3 / 3$ and $b \sim \lambda_{\rm
  dB}/2$, one finds\footnote{Note that an underdensity, such as around a
  vortex ring, would effectively cause a deflection of the opposite
  sign compared to an overdensity. We are not keeping track of this sign.
  Note also if we were more careful, we should have integrated over a range
  of impact parameters instead of setting $b\sim \lambda_{\rm dB}/2$,
  yielding some Coulomb logarithm.}
\begin{equation}
\Delta v \sim 0.08 {\,\rm km/s} \left( {\rho \over 0.01 {\,\rm M_\odot
      \, pc^{-3}}} \right) \left( {250 {\,\rm km/s} \over v} \right)^3 \left(
    {10^{-22} {\,\rm eV} \over m} \right)^2 \, .
\end{equation}
This is a stochastic kick, and its rms value accumulates in a root $N$
fashion, where $N$ is the number of de Broglie blobs the star
encounters, which is roughly $T v / \lambda_{\rm dB}$ where $T$ is the
time over which such encounters take place. Thus,
\begin{equation}
{\rm rms\,} \Delta v \sim 4 {\,\rm km/s} \left({T \over 5 {\,\rm
      Gyr}}\right)^{1/2} \left( {\rho \over 0.01 {\,\rm M_\odot
      \, pc^{-3}}} \right) \left( {250 {\,\rm km/s} \over v} \right)^2 \left(
    {10^{-22} {\,\rm eV} \over m} \right)^{3/2} \, .
\end{equation}
See \cite{Bar-Or:2018pxz,Church:2018sro} for more careful analyses of
such heating. We discuss the implications for tidal streams,
galactic disks and stellar clusters in Section \ref{implications}.

\subsection{Compact objects and relativistic effects---black hole
  accretion, superradiance and potential oscillation}
\label{compactobj}

What happens to wave dark matter around compact objects, such as black
holes? First of all, accretion onto black holes should occur. This
includes accretion of both mass and angular momentum. Second, for a
spinning black hole, the reverse can happen: mass and angular momentum
can be extracted out of a Kerr black hole, an effect known as
superradiance.

To study these phenomena properly, because relativistic effects become
relevant close to the horizon, one needs to
revert to the Klein-Gordon description i.e. $\phi$ obeying
Equation \ref{phiKG}. 
There is a long history of studying solutions to the Klein-Gordon
equation in a Schwarzschild or Kerr background
\citep{1973JETP3728S,unruh1976,Detweiler:1980uk,Bezerra:2013iha,
Vieira:2014waa,Konoplya:2006br,Dolan:2007mj,Arvanitaki:2009fg,Arvanitaki:2010sy,Barranco:2012qs,Arvanitaki:2016qwi}. 
The treatments generally differ in the boundary
conditions assumed: while the boundary condition at the horizon is
always ingoing, that far away can be outgoing (for studying
quasi-normal modes), asymptotically vanishing (for studying
superradiance clouds), or infalling (for studying 
accretion), or combination of infalling and outgoing (for studying
scattering). 

For a black hole immersed in a wave dark matter halo, the infalling
boundary condition is the most relevant. In particular, the stationary
accretion flow around a black hole
was investigated in \cite{Clough:2019jpm,Hui:2019aqm,Bamber:2020bpu} i.e. 
the time-dependence of $\phi$ is a linear combination of 
$e^{\pm imt}$ at all radii.
The Klein-Gordon equation in a Schwarzschild background takes the form:
\begin{equation}
\left[\partial_t^2 -\partial_{r_*}^2 + U(r) \right] (r\phi) = 0 \quad , \quad U(r)
\equiv \left(1 - {r_s \over r}\right) \left( m^2 + 
  {\ell (\ell + 1) \over r^2} + {r_s \over r^3}  \right) \, ,
\end{equation}
where $t$ and $r$ are the time and radial coordinates of the
Schwarzschild metric, $r_s$ is the Schwarzschild radius, and 
$r_*$ is the tortoise coordinate: $r_* =  r + r_s {\,\rm log\,} (r/r_s
- 1)$. We have
assumed the angular dependence of $\phi$ is given by a spherical
harmonic of some $\ell$. 
For $\phi \propto e^{\pm imt}$, this resembles the Schr\"odinger
equation with some potential.
For $\ell=0$, the radial profile of $\phi$ goes roughly as follows:
(1) for $r_s^{-1} \,\lsim\,m$, we have $\phi \sim r^{-3/4}$ i.e. there is a pile-up of the scalar
towards the horizon;\footnote{
This is the particle limit, in that the Compton wavelength is smaller
than the horizon size. Note that here the
relevant wavelength is Compton, not de Broglie.
The $r^{-3/4}$ behavior can be
understood as follows. A stationary accretion flow should have $r^2
\rho v =$ constant, where $v$ is the radial velocity, and $\rho$ is
the dark matter density. Energy
conservation for the dark matter particle means $v^2 \sim 1/r$. 
Thus, $\rho \sim r^{-3/2}$. Noting that $\rho \sim \phi^2$ tells us
$\phi \sim r^{-3/4}$. Such a dark matter spike around a black hole was
discussed in \cite{Gondolo:1999gy,Ullio:2001fb}.
}
(2) for $m \, \lsim \, v_{\rm halo} \, r_s^{-1}$, where $v_{\rm halo}$
is the velocity dispersion of the ambient halo, the
scalar profile is more or less flat;
(3) for $m$ in between these two limits, $\phi$ exhibits both
particle behavior (the $r^{-3/4}$ pile-up) and wave behavior in the
form of standing waves.
\footnote{The stationary accretion flow of $\phi$ onto the black hole
  can be thought of as some sort of hair. The classic no-scalar-hair
  theorem of \citet{Bekenstein:1971hc,Bekenstein:1972ky}
  assumes $\phi$ vanishes far away from the
  black hole, which is violated in this case. The boundary
  condition of $e^{\pm i m t}$ can be thought of as a generalization
  of the $\phi \sim t$ boundary condition considered by 
  \cite{Jacobson:1999vr} \citep[see also][]{Horbatsch:2011ye,Wong:2019yoc}.
}
The computation described above assumes the black hole dominates
gravitationally: one can check that, for astrophysically relevant
parameters, the pile-up of the scalar towards the horizon does not lead to significant
gravitational backreaction. 
There is, however, the possibility that
self-interaction (the quartic interaction for the axion) might be non-negligible close
to the horizon due to the pile-up. 
As one goes to
larger distances from the black hole, the dark matter (and baryons) 
eventually dominates gravitationally. An interesting setting is the
wave dark matter soliton at the center of a galaxy which also hosts a
supermassive black hole \citep{Brax:2019npi}. Investigations of how the black hole
modifies the soliton can be found in
\cite{Chavanis:2019bnu,Bar:2019pnz,Davies:2019wgi}. 

Even though the instantaneous gravitational backreaction of the scalar
is small close to the black hole, the cumulative accreted mass
could be significant. The accretion rate in the low $m$ regime (for
$\ell = 0$) is:
\begin{equation}
\dot M_{\rm BH} = 4\pi r_s^2 \rho_{\rm halo} \sim 4 \times 10^{-9} {\,\rm M_\odot}
{\,\rm yr}^{-1} \left( {M_{\rm BH} \over 10^9 {\,\rm M_\odot}}
\right)^2 \left( {\rho_{\rm halo} \over 0.1 {\,\rm M_\odot \,
      pc^{-3}}} \right)
\end{equation}
where $M_{\rm BH}$ is the mass of the black hole, and $\rho_{\rm
  halo}$ is the ambient dark matter halo density.\footnote{
This is simple to understand: in the low mass regime, there is
essentially no pile-up towards the horizon. Thus, the dark matter
density at horizon is roughly the same as $\rho_{\rm halo}$, the
density far away. At the horizon, dark matter flows into the black
hole at the speed of light, which is unity in our convention. Hence
the expression for $\dot M$.
} In the high $m$ regime, the pile-up enhances this by a factor of
$\sim 1/v_{\rm halo}^3$. For $v_{\rm halo} \sim 10^{-3}$, we
see that $\dot M_{\rm BH}$ goes up to $4 {\,\rm M_\odot /yr}$ in the high $m$
limit, though it should be kept in mind this estimate assumes $\ell=0$. 
(Note that $r_s^{-1} = 6.7 \times 10^{-20} {\,\rm eV}
(10^9 {\,\rm M_\odot\,}/M_{\rm BH})$.)

Suppose one solves the Klein-Gordon equation with a different boundary
condition far away from the black hole: that $\phi$ vanishes.  In that case,
assuming the time dependence is given by $e^{-i\omega t}$, the
allowed frequency $\omega$ forms a discrete spectrum, much like the
energy spectrum of a hydrogen atom. For a spinning black hole, some
of these $\omega$'s are complex with a positive imaginary part,
signaling an instability, known as superradiance
\citep{1972JETP351085Z,Bardeen:1972fi,Press:1972zz,1973JETP3728S,Damour:1976kh,
  Dolan:2007mj,Arvanitaki:2009fg,Arvanitaki:2010sy,
  Arvanitaki:2016qwi,Endlich:2016jgc}. The superradiance condition is:
\begin{equation}
{\,\rm Re\,} \omega < {a m_J \over {r_s r_+}} 
\end{equation}
where $r_s = 2 GM$, $r_+ = (r_s/2) + \sqrt{(r_s/2)^2 - a^2}$ is 
the horizon, $a$ is the black hole angular momentum per unit mass
(the dimensionless spin is $2a/r_s$, between $0$ and $1$), and
$m_J$ is the angular momentum quantum number
of the mode in question.
\footnote{Re $\omega$ is
always of the order of the mass of the particle $m$, 
and Im $\omega$ is maximized for the $\ell = m_J = 1$ mode
and $m r_s/2 \sim 0.1 - 0.5$ depending on the value of $a$. It is a weak
instability in the sense that Im $\omega$ is at best about $10^{-6}
m$. See \cite{Dolan:2007mj}.}
A superradiant mode
extracts energy and angular momentum from the black hole. That this
mode grows with time means the scalar need not be dark
matter at all---
even quantum fluctuations could provide the initial seed to
grow a whole superradiance cloud around the black hole.
In the process, the black hole loses mass and angular
momentum (much of which occurs when the cloud is big). 
At some point, the black hole's mass and spin are such that the mode
in question is no longer unstable, and in fact some of the lost energy
and angular momentum flow back into the black hole, until another
superradiant mode---one that grows more slowly, typically higher
$\ell$---takes over \citep[see e.g.][]{Ficarra:2018rfu}.
The implied net black hole spin-down is used to
put constraints on the existence of light scalars, using black holes
with spin measurements \citep[for recent discussions, see
e.g.][]{Stott:2018opm,Davoudiasl:2019nlo}.
Other phenomena associated with the black hole superradiance cloud
includes gravitational wave emission, and run-away explosion when 
self-interaction becomes important
\citep{Arvanitaki:2010sy,Yoshino:2013ofa,Hannuksela:2018izj}.

It is worth stressing that these constraints do not assume the scalar
in question is the dark matter. An interesting question is how the
constraints might be modified if the scalar is the dark matter.
For instance there can be accretion of angular momentum from the ambient
dark matter, much like the accretion of mass discussed earlier.
\footnote{There can also be accretion of baryons, discussed in e.g. \citet{Barausse:2014tra}.}
The cloud surrounding the black hole is thus a combination of
superradiant unstable and stable modes. This was explored in
\citet{Ficarra:2018rfu}: if the initial seed cloud (of both
unstable and stable modes) is
large enough, the long term evolution of the black hole mass and spin
can be quite different from the case of a small initial seed.
\footnote{\label{nonlinearKG} It is worth stressing that, while the Klein-Gordon
  equation is linear in $\phi$, the evolution of the combined
  black-hole-scalar-cloud system is nonlinear. As the black hole mass
  and spin evolve due to accretion/extraction, the background geometry
  for the Klein-Gordon equation is modified, which affects the scalar evolution. This
  feedback loop has non-negligible effects, even though
  at any given moment in time, the geometry is dominated by the black hole rather than the cloud.
}
This is particularly relevant if the
scalar in question is the dark matter, and therefore present
around the black hole from the beginning. It would be worth
quantifying how existing superradiance constraints might be modified
in this case. There are also interesting investigations on how such a cloud
interacts with a binary system
\citep{Baumann:2018vus,Zhang:2019eid,Annulli:2020lyc}. 

We close this section with the discussion of one more relativistic
effect, pointed out by 
\citet{Khmelnitsky:2013lxt}. The energy density associated with the
oscillations of $\phi$ (which can be interpreted as a collection of
$\phi$ particles) is $\rho = (\dot \phi^2 + m^2 \phi^2)/2$
(Equation \ref{rho}). It can be shown the corresponding pressure is
$P = (\dot \phi^2 - m^2 \phi^2)/2$. For $\phi \sim {\,\rm sin}(mt)$
  or ${\,\rm cos}(mt)$, we see that $\rho$ is constant while $P$
  oscillates with frequency $2m$. Einstein equations tell us this sources an oscillating
  gravitational potential. In Newtonian gauge, with the spatial part
  of the metric as $g_{ij} = (1 - 2\Psi) \delta_{ij} $, the
  gravitational potential $\Psi$ has a constant piece that obeys the
  usual Poisson equation $\nabla^2 \Psi = 4 \pi G \rho$, and an
  oscillating part obeying $-\ddot \Psi \sim  4\pi G P$. Thus
  $\Psi$ oscillates with frequency $2m$ and amplitude $\pi G \rho
  /m^2$. In other words, the
  oscillating part of $\Psi$ is suppressed compared to the constant
  part by $k^2 / m^2$. The typical (constant part of) gravitational
  potential is of the order $10^{-6}$ in the Milky Way; the oscillating
  part is then about $10^{-12}$. For $m$ in the
  ultra-light range, recalling $m^{-1} \sim 0.2 {\,\rm yr} \, (10^{-22}
  {\,\rm eV} / m)$, pulsar timing arrays are well suited to search for
  this effect, as proposed by \cite{Khmelnitsky:2013lxt}. See further
  discussions in Section \ref{probesCombojb}. 
 
\section{Observational/experimental implications and constraints}
\label{implications}

In this section, we discuss the observational and experimental
implications of the wave dynamics and phenomenology explained above. 
The discussion serves a dual function. One is to summarize current
constraints---because of the wide scope,
the treatment is more schematic than in previous sections, but 
provides entry into the literature.
The other is to point out the limitations of current constraints, how
they might be improved, and to highlight promising new directions.
Astrophysical observations are relevant mostly, though not
exclusively, for the ultra-light end
of the spectrum. 
Axion detection experiments, on the other hand,
largely probe the heavier masses, though new experiments are rapidly
expanding the mass range.
Much of the discussion applies to any wave dark matter candidate
whose dominant interaction is gravitational. Some of it---on axion
detection experiments for instance---
applies specifically to axions with their expected
non-gravitational interactions (Equation \ref{phiInteractions}). 

Sections \ref{linearpowerspectrum} and \ref{galacticConstraints} 
focus on ultra-light wave dark matter i.e. fuzzy dark matter.
Table \ref{tab1} summarizes some of the corresponding
astrophysical constraints.
Sections \ref{earlyuniverse}, \ref{probesCombojb}, 
\ref{photonpropagation} and \ref{axionExp} cover more general wave
dark matter, with Section \ref{axionExp} on axion detection experiments.

\subsection{Early universe considerations}
\label{earlyuniverse}

Within the inflation paradigm, the light scalar $\phi$ associated with wave dark
matter has inevitable quantum fluctuations which are stretched to
large scales by an early period of accelerated expansion
\citep{Axenides:1983hj,Linde:1985yf,Seckel:1985tj,Turner:1990uz}.
These are isocurvature fluctuations, distinct from the usual adiabatic
fluctuations associated with the inflaton $\varphi$, which is another light
scalar. 
The relevant power
spectra are \citep[e.g.,][]{Baumann:2009ds,Marsh:2013taa}:
\begin{equation}
\Delta_\zeta^2 = {1 \over 8\pi^2 \epsilon} {H_{\rm infl}^2 \over
  m_{\rm pl}^2} \quad , \quad \Delta_{\phi}^2 = {1 \over
  \pi^2} {H_{\rm infl}^2 \over \phi_{i}^2} \, ,
\end{equation}
where $\Delta_\zeta^2$ is the (adiabatic) curvature power spectrum, 
$\Delta_{\phi}^2$ is the (isocurvature) density power spectrum for
$\phi$, $H_{\rm
  infl}$ is the Hubble scale during inflation, $m_{\rm pl} \equiv
1/\sqrt{8\pi G} \sim 2.4 \times 10^{18} {\,\rm GeV}$ is the reduced
Planck mass, $\phi_{i}$ is the (axion)
scalar field value during inflation, and $\epsilon$ is the first
slow-roll parameter.
\footnote{\label{tilt} The dimensionless power spectrum $\Delta^2 (k)$
  is related to the dimensionful power spectrum $P(k)$ by
$\Delta^2 \equiv 4\pi k^3 P(k)/(2\pi)^2$. 
We have suppressed a $k$ dependent factor that depends on the spectral index
$n$ i.e. $\Delta^2 \propto k^{n-1}$. 
For single field slow roll inflation, $n-1 = 2\eta - 6\epsilon$, where
$\epsilon \equiv ({\cal V}_{,\varphi} m_{\rm pl} / {\cal V})^2/2 = -\dot H_{\rm
  infl}/H_{\rm infl}^2$ and $\eta \equiv m_{\rm
  pl}^2 {\cal V}_{,\varphi\varphi} / {\cal V}$ are the first and second slow roll
parameters, with ${\cal V}$ being the inflaton potential.
The spectral tilt for $\zeta$ is
observed to be $n \sim 0.97$ \citep{Hinshaw:2012aka,Aghanim:2018eyx}.
}
Microwave background anisotropies bound $\Delta_\phi^2 /
\Delta_\zeta^2 \,\lsim\, 0.05$ \citep{Hinshaw:2012aka,Aghanim:2018eyx}, 
implying $8\epsilon (m_{\rm pl}/\phi_{i})^2 \, \lsim \, 0.05$. 
Consider for instance $\phi_i \sim 10^{17}$ GeV (see
Equation \ref{relic}, where $\phi_i \sim f$). In that case,
observations require $\epsilon \,\lsim\,
10^{-5}$.\footnote{Given that the scalar spectral index is
observed to be $n-1=2\eta - 6 \epsilon \, \sim \, 0.97$. The
smallness of $\epsilon$ means the requisite inflation model is one
where $\eta \gg \epsilon$. For recent model building in this
direction, see \cite{Schmitz:2018nhb}.}
Since $\Delta^2_\zeta$ is observed to be about $10^{-9}$, this implies
$H_{\rm infl}/m_{\rm pl} \,\lsim\, 10^{-6}$. This is a low inflation
scale, suggesting a low level of gravitational waves, or 
tensor modes \citep{Lyth:1989pb}. One can see this more
directly by recalling that tensor modes suffer the same level of
fluctuations as a spectator scalar like $\phi$:
\begin{equation}
\Delta^2_{\rm tensor} = {2\over \pi^2} {H_{\rm infl}^2 \over m_{\rm
    pl^2}} \quad , \quad r \equiv {\Delta^2_{\rm tensor} \over \Delta^2_\zeta}
= 16\epsilon
\end{equation}
where $\Delta_{\rm tensor}^2$ resembles $\Delta_{\phi}^2$, with $\phi_{i}$ replaced by
$m_{\rm pl}$, and a factor of $2$ for the $2$ polarizations. 
The tensor-to-scalar ratio $r$ is thus constrained by the isocurvature bound
to be: $r \,\lsim\, 0.1 (\phi_{i} / m_{\rm pl})^2$. 
For $\phi_{i} \sim 10^{17}$ GeV, this means $r \,\lsim\,
2\times 10^{-4}$, making tensor modes challenging to observe with future
microwave background experiments. Most axion models have lower
$\phi_i$'s which would strengthen the bound.
This is thus a general requirement: to satisfy the existing isocurvature
bound, the inflation scale $H_{\rm infl}$ must be sufficiently low,
implying a low primordial gravitational wave background. This holds as long as
the scalar dark matter derives its abundance from 
the misalignment mechanism, with the misalignment angle in place
during inflation. A way to get around this is to consider models where
  the scalar $\phi$ becomes heavy during inflation \citep{Higaki:2014ooa}.

The requirement does not apply in cases where the relic abundance is
determined by other means. For instance, for the QCD axion, it could
happen that the Peccei-Quinn symmetry is broken only after inflation
(recall the axion as a Goldstone mode exists only after spontaneous
breaking of the symmetry), in
which case the relic abundance is determined by the decay of axion
strings and domain walls
\citep{Kolb:1990vq,Buschmann:2019icd,Gorghetto:2020qws}.
There are also proposals for vector, as opposed to scalar, wave dark
matter: isocurvature vector perturbations are relatively harmless
because they decay \citep{Graham:2015rva,Kolb:2020fwh}.

The above discussion includes only the gravitational interaction
of scalar dark matter. Other early universe effects are
possible with non-gravitational interactions.
For instance, \cite{Sibiryakov:2020eir} pointed out if the scalar
has a dilaton-like coupling to the standard model,
Helium-4 abundance
from big bang nucleosynthesis can be significantly
altered. 
\footnote{Such a scalar coupling to the standard model must be close to being
  universal to satisfy stringent equivalence principle violation
  constraints \citep{Wagner:2012ui,Graham:2015ifn}. 
  The pseudo-scalar coupling to fermions (Equation
  \ref{phiInteractions}) gives rise to a spin-dependent force that
  can also be probed experimentally \citep{Terrano:2015sna}.
}

\subsection{Linear power spectrum and early structure formation}
\label{linearpowerspectrum}

As discussed in Section \ref{PT}, light scalar dark matter---produced out of a
transition process from slow-roll to oscillations---has a primordial
power spectrum suppressed on small scales (high $k$'s). 
For fuzzy dark matter, the suppression scale is around $k \sim 5$/Mpc
(Equation \ref{khalf}). Observations of the Lyman-alpha forest
are sensitive to power on such scales. The Lyman-alpha forest is the part of the
spectrum of a distant object (usually a quasar) between Lyman-alpha
and Lyman-beta in its rest frame. Intergalactic neutral hydrogen
causes absorption, with measurable spatial fluctuations.
With suitable modeling, the spatial fluctuations
can be turned into statements about the dark matter power spectrum \citep{Croft:1997jf,Hui:1998hq,McDonald:2004xn,Palanque-Delabrouille:2013gaa}.
With this technique,
a limit of $m \,\gsim \, 3 \times 10^{-21}$ eV was obtained by
\citet{Irsic:2017yje,Kobayashi:2017jcf,Armengaud:2017nkf}.  
\citet{Rogers:2020ltq} found a stronger bound of $2 \times
10^{-20}$ eV---among the differences in analysis are
assumptions on the reionization history.

In this type of investigation, often the only effect of fuzzy dark
matter accounted for is its impact
on the primordial power spectrum.
One might worry about the effect of quantum
pressure on the subsequent dynamics, but this was shown to be
a small effect at the scales and redshifts for the
Lyman-alpha forest
\citep{Nori:2018pka,Li:2018kyk}. Another assumption is
that the observed fluctuations in neutral hydrogen reflect
fluctuations in the dark matter. This need not be true, since
astrophysical fluctuations modulate the neutral hydrogen
distribution, such as
fluctuations in the ionizing background
\citep{Croft:2003qn,McDonald:2004xp,DAloisio:2016dcv}, the 
temperature-density relation
\citep{Hui:1997dp,Cen2009,
Keating:2017lgk,
Wu:2019sgk,Onorbe:2018zoi} and from galactic winds
\citep{McDonald:2004xp,Viel:2012sd}.
Measurements of the power spectrum growth from
the forest suggest the astrophysical fluctuations are
sub-dominant, that gravity is sufficient to account for the observed
growth \citep{McDonald:2004xn}. Nonetheless, it is worth stressing
for the bound on $m$, one has to worry about
systematic effects at the few percent level.
\footnote{For instance, 
the Lyman-alpha absorption power spectrum for $m = 10^{-21}$ eV
fuzzy dark matter differs from that for
conventional cold dark matter at the few percent level (at $z \sim 5$;
smaller as one goes to lower redshifts),
{\it if} one allows the intergalactic medium parameters (especially
the temperature) to float to fit the data. If the latter parameters
were held fixed,
the two model predictions differ significantly, up to factor of a
few. But that is not the relevant comparison. Since the intergalactic
medium parameters are unknown and need to be fit from the data,
the relevant comparison is between fuzzy dark matter at its best fit
and conventional dark matter at its best fit---they differ at the few
percent level. Thanks are due to Rennan Barkana, Vid Ir\v{s}i\v{c} and
Matteo Viel for discussions on this point.
}
The astrophysical fluctuations were accounted for in the following way
in deriving constraints
\citep{Irsic:2017yje,Kobayashi:2017jcf,Armengaud:2017nkf}. 
Simulations with these astrophysical fluctuations are
compared against those without; the {\it scale and redshift dependence} of the fractional difference in the
predicted Lyman-alpha power spectrum is then fixed, while
the {\it amplitude} of the difference is treated as a free parameter to be
determined from the data. The question is to what extent
simulations of the astrophysical fluctuations have enough variety to
account for the range of possible {\it scale and redshift dependence}. The variety in
question derives from the distribution of ionizing sources, the reionization
history and the strength and form of galactic feedback.
\footnote{The Lyman-alpha forest can also be used to constrain scenarios where
  Peccei-Quinn symmetry breaking occurs after inflation. See
  \cite{Irsic:2019iff}.}

Formation of the first nonlinear objects in the universe is also
sensitive to the small scale power spectrum. Recall in hierarchical
structure formation, it is the small, less massive objects that form first.
A suppression of small scale power implies fewer nonlinear objects at
high redshifts, delaying reionization \citep{Barkana:2001gr}.
The EDGES experiment \citep{Bowman:2018yin} announced the
detection of an absorption feature around $78$ MHz that may result
from the hyperfine transition (21cm) of hydrogen at redshift around $15 -
20$. This suggests the spin temperature of the 21cm line is coupled to
the gas temperature at such high redshifts, and points to early star
formation which produces the requisite radiation to do so.
This was used to place bounds on fuzzy dark matter
$m \,\gsim \, 5 \times 10^{-21}$ eV
\citep{Safarzadeh:2018hhg,Schneider:2018xba,Lidz:2018fqo}.
A few considerations should be kept in mind.
The EDGES detection remains to be confirmed \citep{Hills:2018vyr}. 
These bounds assume (1) star formation tracking halo formation,
and (2) an upper limit on the fraction of halo baryons that turn into stars
\citep[$0.05$ in][]{Lidz:2018fqo}. Another important assumption is that
the halo mass function can be reliably predicted from the linear power
spectrum by the standard Press-Schechter or Sheth-Tormen
relations \citep{Press:1973iz,Sheth:1999mn,Marsh:2013ywa,Kulkarni:2020pnb}.
\footnote{
The idea is to map the mass of a halo to a comoving length scale.
The number density of halos at that mass (i.e. the mass function)
is then related to the {\it linear} power spectrum at the corresponding length scale.
}
These relations have been checked for fuzzy dark matter
models using only N-body, as opposed to wave, simulations, i.e. the
``fuzziness'' enters only through the primordial power
spectrum \citep{Schive:2015kza}. 
Typical wave simulations use too small a box size to give
a reliable halo mass function. 
It is conceivable that wave
interference phenomena might help make more smaller objects than expected
from Press-Schechter type arguments.

Looking towards the future, spectral distortion measurements of the
microwave background hold the promise of measuring the linear power
spectrum down to very small scales, comoving $k$ as high as
$10^4$/Mpc \citep{Kogut:2019vqh,Chluba:2019nxa}.
\footnote{An experiment like PIXIE can probe excess power over the conventional
cold dark matter prediction. To check if there is a power deficit,
from wave dark matter for instance, would
require something more ambitious, Super-PIXIE \citep{Chluba:2019nxa}.}
From Equation \ref{khalf}, this kind of experiment can thus probe 
a wave dark matter mass as high as $\sim 10^{-15}$ eV.

\begin{table}[tb]
\caption{Some constraints in the literature on fuzzy dark matter}
\label{tab1}
\begin{center}
\begin{tabular}{@{}l|c|c|c@{}}
\hline
Method & Constraint & Sources of systematic uncertainties & Refs.\\
\hline
Lyman-alpha forest & m $> 3 \times 10^{-21}$ eV & Ionizing
                                                background/temp.
                                                fluctuations & 1 \\ \hline
Density profile & m $> 10^{-21}$ eV & Baryonic feedback/black hole & 2
  \\ \hline
Satellite mass & m $> 6 \times 10^{-22}$ eV & Tidal stripping & 3 \\
  \hline
Satellite abundance & m $> 2.9 \times 10^{-21}$ eV & Subhalo mass
                                                     function
                                                     prediction & 4 \\ \hline
\end{tabular}
\end{center}
\begin{tabnote}
References:
1=\citet{Irsic:2017yje,Kobayashi:2017jcf,Armengaud:2017nkf},
2=\citet{Bar:2018acw},
3=\cite{Safarzadeh:2019sre},
4=\cite{Nadler:2020prv}.
See text on the methodology and systematic uncertainties of each constraint.
\end{tabnote}
\end{table}

\subsection{Galactic dynamics and structure---density profile, stellar
scattering, dynamical friction, subhalo mass function and  interference substructures}
\label{galacticConstraints}

There is a wide variety of methods to constrain wave dark matter
from galactic structure or dynamics,
especially at the ultra-light end of the spectrum.

{\it Density profile.} Wave simulations demonstrate that fuzzy dark
matter halos generically have a solitonic core, and an 
NFW-like outer density profile \citep{Schive:2014hza}. 
There is a substantial literature on comparing this prediction
against observations. Investigations focusing on the inner density
profile (i.e. within the purported soliton) of Milky Way dwarf
satellites found reasonable agreement with $m \sim 10^{-22} -
10^{-21}$ eV \citep{Chen:2016unw,Calabrese:2016hmp}. 
A $10^9 \, {\,\rm M_\odot}$ soliton at the center of the Milky Way was
reported by \cite{DeMartino:2018zkx}, though there is substantial
uncertainty because of the dominance of baryons \citep{Li:2020qva}.
Investigations bearing on how the soliton connects with the outer halo
generally found tension with data, for $m \, \lsim 
10^{-21}$ eV.
Taking the soliton-halo relation
(Equation \ref{solitonhalorelation}) seriously, one expects an inner
circular velocity that matches the outer asymptotic value (a
reflection of the rough equality of the soliton potential and halo
potential; see footnote \ref{solitonhalo}), something not seen in
observations of disk galaxies \citep{Bar:2018acw}. 
Moreover, dynamical measurements of Milky Way dwarf satellites, 
when used to fit for solitonic cores, predict halo masses that are too
large, incompatible with their survival under dynamical friction, 
giving a bound of $m \, > \, 6 \times
10^{-22}$ eV \citep{Safarzadeh:2019sre}. 
It was also pointed out by \cite{Burkert:2020laq} that
low mass galaxies have a universal core surface density $\sim 75 \,
{\,\rm M_\odot / pc^2}$ while
spanning a large range in core radius; this conflicts with the soliton
scaling of $M \propto 1/R$ (Equation \ref{Rsoliton}) implying a surface
density $\propto 1/R^3$. 
On the other hand, \cite{Pozo:2020ukk} pointed out 
that the stellar density profile of dwarfs matches well the mass density
profile in fuzzy dark matter simulations.

Overall, it appears the fuzzy dark matter soliton does
not in a straightforward way match galaxy 
cores seen in dynamical data, 
when viewed in the larger context of the host halo.
A number of possible mitigating factors should be kept in mind.
The relaxation time for forming a soliton scales as $m^3$
(Equation \ref{trelax}), which can get quite long for the higher masses.
Some of the galaxies investigated are in dense environments; tidal
interactions could perturb them in significant ways that should be
taken into account (see Section \ref{dynamics2}). Inference of galaxy
density profiles from dynamical data is subject to uncertainty from the velocity
anisotropy profile \citep[see e.g.,][]{Walker:2009zp,Amorisco:2011hb},
or possible non-circular motions \citep{Oman:2017vkl}.
Baryons and central supermassive black holes 
could affect galaxy density profiles in non-negligible ways.
There has been a lot of work in this direction for
conventional cold dark matter, with some success and some remaining
puzzles e.g. \cite{Oman:2015xda}. 
\footnote{See also \cite{Kaplinghat:2019dhn} on the self-interacting
  dark matter model.
}
These considerations are
likely relevant for testing fuzzy dark matter from density profiles
\citep{Bar:2019bqz,Bar:2019pnz}.

{\it Heating/scattering of stars.} Transient, de Broglie size
substructures due to wave interference heat up stars in a galaxy
(Section \ref{dynamics2}). Such heating of the Milky Way disc was
investigated by \cite{Church:2018sro} who put a bound
$m \, > \, 0.6 \times 10^{-22}$ eV to avoid overheating.
Stellar streams from tidally disrupted globular clusters can be heated
up in a similar way, leading to thickening. A bound of $m \, > \,
1.5 \times 10^{-22}$ eV was placed by \cite{Amorisco:2018dcn} based on
this argument. The stellar cluster at the center of the ultra-faint dwarf
Eridanus II was used to place constraints on $m$ by
\cite{Marsh:2018zyw}. Solitons in wave simulations are observed to
have oscillations \citep{Veltmaat:2018dfz}. The oscillation time scale
would be shorter than the dynamical time scale of the stellar cluster
for $m \, \gsim \, 10^{-21}$ eV, leading to heating and disruption of
the stellar cluster for $m$ up to $10^{-20}$ eV.
\footnote{For $m \,\lsim\, 10^{-21}$ eV, the long soliton oscillation
  time ($\sim 1/(mv^2)$) means the impact on the stellar cluster is
  adiabatic i.e. no heating. For $m \,\gsim\, 10^{-20}$,
  \cite{Marsh:2018zyw} derived constraints not from heating by soliton
  oscillation, but from heating by de Broglie granules.
}
The observation of soliton oscillations was based on simulations of
isolated halos, while Eridanus II is a Milky Way satellite subject to
tidal forces. Recently, a simulation including an external tidal field was
described in \cite{Schive:2019rrw}. They showed that tidal disruption
of the outer halo surrounding the soliton leads to suppressed heating of
a stellar cluster in the soliton.\footnote{
It was pointed out by \cite{Schive:2019rrw} that the soliton in
general undergoes random walks as well as oscillates. Tidal stripping
of the outer halo appears to suppress excitations associated with such processes.
} 
Analytic arguments suggest the same
\citep{Li:2020ryg}. 

{\it Dynamical friction.} The wave nature of dark matter can lead to a
suppression of dynamical friction, as explained in
Section \ref{dynamics2}. It was argued by \cite{Hui:2016ltb} that 
a fuzzy dark matter mass of $m \sim 10^{-22}$ eV helps explain the
survival of globular clusters against orbital decay in the halo of
Fornax \citep{tre76,oh2000}.
See \cite{Lancaster:2019mde}  for a numerical exploration
of this phenomenon, and \cite{Bar-Or:2018pxz} on how the
suppression of dynamical friction is tempered by diffusion.
It is worth noting that within the conventional cold dark
matter model, a possible solution to this dynamical friction problem is to
invoke core-stalling
\citep{Goerdt:2006rw,Read:2006fq,Inoue:2009wd,Cole:2012ns}.
Dynamical data with higher precision, and
on more systems, would be very helpful.

{\it Subhalo mass function.} Fuzzy dark matter, with its suppressed
power on small scales, predicts fewer low mass halos compared with conventional
cold dark matter. The same is expected to be true for subhalos of a
parent galaxy, such as the Milky Way. Several different ways to probe
the subhalo mass function have been discussed in the literature.
One way is to infer the subhalo mass function from
the observed luminosity function of Milky Way satellites, using
abundance matching. This was carried out by \cite{Nadler:2020prv}
who obtained the bound $m \, > \, 2.9 \times 10^{-21}$ eV.
Another method is to use stellar streams from tidally disrupted
globular clusters or satellites in our galaxy
\citep{Johnston:2001wh,Ibata:2001iv}. 
Observed perturbations of streams
were used to place
constraints on the subhalo mass function, which were then turned into
constraints on warm dark matter \citep{Banik:2019smi} and fuzzy dark
matter \citep{Schutz:2020jox}, obtaining $m \, > \, 2.1 \times
10^{-21}$ eV. Yet another method is to use flux anomaly in strongly
lensed systems to probe subhalos in the lensing galaxies
\citep{Dalal:2001fq}. This was used by \cite{Gilman:2019nap}  to
constrain warm dark matter and \cite{Schutz:2020jox} to limit fuzzy
dark matter, obtaining $m \, > \, 2.1 \times
10^{-21}$ eV. A natural question for these investigations is to what
extent the subhalo mass function for fuzzy dark matter is accurately known.
It is typically computed using Press-Schechter type formalism, meaning
the effect of fuzzy dark matter enters only through the initial power
spectrum (i.e. its suppression on small scales). Dynamical effects due
to wave interference could influence the subsequent evolution, and
thus the subhalo mass function. It
would be useful to quantify it with wave simulations (see discussion
at the end of Section \ref{linearpowerspectrum}).
Moreover, wave interference granules---not
virialized subhalos---could by themselves give rise to
these signals, such as the scattering of stellar streams
\citep{Dalal:2020mjw}. Their effects should be taken into account.

{\it Probing interference substructures.} 
One generic prediction of wave dark matter is the existence of
interference substructures in halos. These are
de Broglie scale, order unity density fluctuations. The fluctuation can
take the density all the way to zero (complete destructive interference
i.e. vortices; see Section \ref{vortices}). There are different ways
to probe these interference substructures.
One is through the heating and scattering of stars, already discussed
above. The other is through gravitational lensing by the
substructures. For instance, a de Broglie size blob in our own galaxy
passing over the line of sight to some distant object would cause the
apparent position of that object to shift 
\citep{Weiner:2019zrg,Mondino:2020rkn,Mishra-Sharma:2020ynk,Hui:2020hbq}.
The effect is small---\cite{Mishra-Sharma:2020ynk} proposed the
correlated shifts of many distant objects could be used to look for small
signals. Another context where a gravitational lensing signal can be searched
for is cases of strong lensing. The lensing flux anomaly refers to the
phenomenon that strongly magnified images of a distant source have
flux ratios that are discordant with expectations from a smooth
lensing halo \citep{Mao:1997ek,Chiba:2001wk,Metcalf:2001ap,Dalal:2001fq,Hezaveh:2014aoa,Alexander:2019puy,Dai:2020rio}. For instance, two images close to a critical line
(corresponding to a fold caustic) are
expected to have the same magnification, barring substructures
on scales smaller than the image separation. It has been shown that interference
substructures can cause a $\sim 10 \%$ difference
in cases of high magnification $\sim 100$
\citep{Chan:2020exg,Hui:2020hbq}. Since subhalos also give rise to
such flux anomaly, to distinguish between fuzzy dark matter and
conventional cold dark matter, a measurement of the anomaly as a
function of image separation would be helpful. The anomaly power
spectrum of fuzzy dark matter would have a feature around the de
Broglie scale.

\subsection{Probes using compact objects---superradiance, solitons,
  potential oscillation and stellar cooling}
\label{probesCombojb}

{\it Superradiance.} Superradiance constraints on the existence
of light scalars, or light bosons more generally---
not necessarily dark matter---were summarized in
\cite{Stott:2018opm}. The idea is to use the measured spin of black
holes to put limits on scalars which could drain away their angular momentum,
if their Compton wavelength roughly matches the horizon size (see
Section \ref{compactobj}). The boson mass probed this way covers a wide
range, from $\sim 10^{-13} - 10^{-12} {\,\rm eV}$ for black holes at
tens of solar mass, to $\sim 10^{-18} - 10^{-21} {\,\rm eV}$ for
supermassive black holes. It was pointed out by \cite{Davoudiasl:2019nlo}
that the spin constraint on the M87 supermassive black hole, reported
by the Event Horizon Telescope (EHT) collaboration
\citep{Akiyama:2019fyp}, disfavors ultra-light bosons around
$10^{-21}$ eV. It is worth noting that the EHT constraint comes not
from measurement of the famous shadow, but from modeling of the
jet coming out of the galactic nucleus. 

The existing superradiance constraints were obtained by assuming the
superradiance cloud grows from a small initial seed of
superradiance-unstable modes (produced by quantum fluctuations for instance).
As pointed out by \cite{Ficarra:2018rfu}, the existence of additional
superradiance-stable modes could significantly modify the long term
evolution of the cloud, and therefore the mass and spin of the black
hole (see footnote \ref{nonlinearKG}). Such stable modes are naturally present if the light boson in
question were the dark matter. Dark matter mass and angular momentum accretion
onto the black hole inevitably occurs 
\citep{Clough:2019jpm,Hui:2019aqm,Bamber:2020bpu}.
It would be useful to revisit the superradiance constraints for cases
where the light boson is the dark matter. 
It is also worth noting that enhanced interactions of the axion
could lead to relaxation of the superradiance constraints
\citep{Mathur:2020aqv}.

{\it Boson stars.} Light boson dark matter can be probed astrophysically in a different
way, by the boson stars or solitons that could form in the early
universe. Using the Chandrasekhar-like maximum mass as a guide
(Equations \ref{Mmax1} or \ref{Mmax2}), the interesting boson star mass
could range from $10^{-10} {\,\rm M_\odot}$ to $10^{10} {\,\rm
  M_\odot}$, for dark matter mass from $10^{-6}$ eV to $10^{-22}$ eV.
Gravitational lensing could be used to detect or constrain a
population of such objects \citep{Kolb:1995bu,Fairbairn:2017sil}. They could also contribute to
merger events seen by gravitational wave experiments if they are sufficiently compact
\citep{Macedo:2013jja,Palenzuela:2017kcg,Clough:2018exo,Helfer:2018vtq}.
The computation of the early universe production of boson stars, specifically axion
stars, was pioneered by \citet{Kolb:1993zz}. 
Termed axion miniclusters, they form due to large fluctuations from
the breaking of the Peccei-Quinn symmetry after inflation. 
The mass function of boson stars subsequently evolves, due to mergers and 
condensation processes \citep{Fairbairn:2017sil,Eggemeier:2019jsu}.
Further computations to firm up the prediction of the
eventual mass distribution of boson stars would be helpful.

{\it Gravitational potential oscillations.} An oscillating scalar
produces an oscillating gravitational potential at
frequency $2m$, as pointed out by
\citet{Khmelnitsky:2013lxt}. This effect can be searched for in pulsar
timing array data, which has a frequency coverage that probes
$m \sim 10^{-24} - 10^{-22}$ eV. The oscillating potential scales as 
$\rho/m^2$ (see Section \ref{compactobj}) so the constraints are
stronger at smaller $m$'s.
A bound of $\rho \, < \, 6 {\,\rm GeV/cm^3}$ for
$m \le 10^{-23}$ eV was obtained by \cite{Porayko:2018sfa} from the
Parkes Pulser Timing Array data.
A bound of $\rho \, < \, 2 {\,\rm GeV/cm^3}$ for
$m \sim 10^{-23}$ eV was obtained by
\cite{Kato:2019bqz} from the NANOGrav data.
These are proofs of concept, since the local dark matter
density is already known to be $\rho \sim 0.4 {\,\rm GeV / cm^3}$ 
\citep{Bovy:2012tw,Sivertsson:2017rkp,McKee2015}.
As a probe of wave dark matter, this method is interesting because it
directly probes the scalar field oscillations at frequency $m$, and has very different
systematics from other astrophysical probes.
The solar system ephemeris turns out to be an important source of
systematic error.
Forecasts of future improvements,
with the planned Square Kilometre Array, can be found in
\cite{Porayko:2018sfa}. To place meaningful limits on $m \sim
10^{-22}$ eV, it is important to have high cadence in addition to long
integration time. 

{\it Stellar axion emission.} To close this sub-section on compact
objects, we mention one
classic probe: axion bounds from the cooling of stars.
Axion couples to photons, gluons and fermions in the standard
model (Equation \ref{phiInteractions}). The interaction strength is weak,
but deep in the interior of stars, there can be enough axion production to
affect stellar structure and evolution. (The weak interaction strength also makes it
relatively easy for the axion to escape from the star.)
This has been applied to the Sun \citep{Schlattl:1998fz}, red giants
\citep{Raffelt:1987yu}, supernova 1987A 
\citep{Raffelt:1987yt,Ellis:1987pk,Turner:1987by,Mayle:1987as}
and neutron star mergers
\citep{Dietrich:2019shr}.
\footnote{
For 1987A, the axion constraint comes from its effect on the neutrino burst
duration. For ways to evade such supernova or stellar cooling bounds,
see \cite{Bar:2019ifz,DeRocco:2020xdt}.
}
There are also experiments built specifically to detect solar axions
such as CAST \citep{Anastassopoulos:2017ftl}.
Phrased in terms of the axion decay constant $f$ (larger $f$ means
weaker coupling; see Equation \ref{phiInteractions}), 
the strongest constraint from these
considerations is about $f \, \gsim \, 10^{9}$ GeV. Note that these
constraints on the axion assume only its existence, {\it not} its viability
as a dark matter candidate.
A comprehensive recent review can be found in \cite{Raffelt:2006cw}.
There are also proposals to detect axion dark
matter from the production of photons in strong magnetic fields around
neutron stars \citep{Bai:2017feq,Hook:2018iia,Foster:2020pgt}.

\subsection{Photon propagation in axion background}
\label{photonpropagation}

The axion coupling to $\vec E \cdot \vec B$
(Equation \ref{phiInteractions}) affects the propagation of photons in
the universe
if dark matter is indeed made up of axions.
To be concrete, suppose the Lagrangian for the photon consists of
\begin{equation}
\label{Lphoton}
{\cal L} = -{1\over 4} F_{\mu\nu} F^{\mu\nu} + {1\over 4} g_\gamma
\phi F_{\mu\nu} \tilde F^{\mu\nu} \,
\end{equation}
where $F_{\mu\nu}$ is the photon field strength and $\tilde F^{\mu\nu}
= \epsilon^{\mu\nu\alpha\beta} F_{\alpha\beta}/2$. The coupling
constant $g_\gamma$ plays the role of $\sim 1/f$ in
Equation \ref{phiInteractions}.
%
The modified Maxwell equations, setting
$\vec E$ and $\vec B$ proportional to $e^{-i\omega t + i
  {\vec k} \cdot {\vec x}}$, imply a 
dispersion relation of the form \citep{Harari:1992ea}:
\begin{equation}
\omega = |\vec k| \pm {1\over 2} g_\gamma (\partial_t \phi + \hat k \cdot
\vec \nabla \phi ) \, ,
\end{equation}
for the two circular polarizations ($\pm$).
This is obtained assuming the WKB limit (i.e. $\partial^2 \phi \, \ll
\, \omega \partial \phi$), and small $g_\gamma$.
The fact that the two circular polarizations have different dispersion
relations means a linearly polarized photon rotates in
polarization as it propagates. One can phrase this in terms of the
phase difference between the two circular polarizations:
\begin{equation}
\label{DeltaS}
\Delta S = g_\gamma \int dt {D\phi \over Dt} \, ,
\end{equation}
where $D/Dt$ is a total time derivative: $\partial_t + \hat k \cdot
\vec \nabla$ i.e. the phase for the respective polarization is
$S = -|\vec k|t + \vec k \cdot \vec x \pm \Delta S/2$. 
There have been several attempts or proposals to search for this
birefringence effect in astronomical data, for instance
the polarization of radio galaxies
\citep{Carroll:1989vb,Harari:1992ea,Nodland:1997cc,Carroll:1997tc}
and the microwave background
\citep{Harari:1992ea,Lue:1998mq,Liu:2016dcg,Fedderke:2019ajk}.
\footnote{
See also \cite{Agrawal:2019lkr} for a proposal to look for axion
strings in the microwave background polarization data.
}
Recently, \cite{Ivanov:2018byi} proposed and searched for a
polarization signal that oscillates in time in 
observations of jets in active galaxies \citep[see
also][]{Caputo:2019tms, Fedderke:2019ajk}. 
The frequency $m$ oscillations in $\phi$ cause the
linear polarization angle to oscillate, 
which can be searched for in data.
A limit of $g_\gamma \,\lsim\, 10^{-12} {\,\rm GeV}^{-1}$  was obtained for $m \sim
5 \times 10^{-23} - 1.2 \times 10^{-21}$ eV. 
Note that the birefringence signal does not depend on the distance over
which the photon travels; it depends only on the values of $\phi$ at
the source and at the observer. A source in a high dark matter
density environment (therefore large $\phi$), such as at the center of a galaxy, is
therefore a promising target.

The fact that rotation of the linear polarization angle is independent
of propagation distance means one could also search for this effect in
the laboratory where high precision measurements are possible
e.g. \cite{Liu:2018icu,DeRocco:2018jwe,Martynov:2019azm,Blas:2019qqp}.
This brings us naturally to the subject of the next section.
We close by mentioning that the same coupling of
the axion to photons (Equation \ref{Lphoton}) gives rise to a different
effect that can be searched for: the conversion of photons into axions
in an environment with magnetic fields
\citep{Raffelt:1987im,Mirizzi:2006zy}. This effect does not require the axions to be
dark matter.

\subsection{Experimental detection of axions}
\label{axionExp}

The experimental detection of axions is a large subject we cannot hope
to do justice here. For recent comprehensive reviews, see e.g.
\cite{Graham:2015ouw,Irastorza:2018dyq,Sikivie:2020zpn}.
We instead focus on aspects of
the detection that have to do with the wave nature of axion dark
matter. 
This sub-section is less about summarizing current
constraints, and more about discussing ways to probe or take advantage
of the wave dynamics
and interference substructures. 
\footnote{In this sub-section, we pick a few experiments to illustrate
  how the wave nature of axions is relevant to detection. There is a
  tremendous diversity in the variety of axion experiments. Some
  aim to detect dark matter; some probe the existence of an axion regardless
  of whether it is dark matter. See \cite{Graham:2015ouw,Irastorza:2018dyq,Sikivie:2020zpn}.
}
There are a number of papers on this
subject. Novel observables for the detection of the axion as a field (or wave) rather
than as a particle were discussed by \citet{CASPEr}.
Stochastic properties of the axion field were computed by \citet{Derevianko:2016vpm}
and \citet*{Foster:2017hbq}. Implications for
the design and interpretation of experiments were discussed by them,
and by \citet{Roberts:2017hla,
Savalle:2019jsb,Centers:2019dyn,Hui:2020hbq,Foster:2020fln}. 
The discussion here follows that in \cite{Hui:2020hbq}.

A good place to start is to remind ourselves of the relation between the
axion $\phi$ and the wavefunction $\psi$:
\begin{equation}
\label{phipsiB}
\phi (t, \vec x)= {1\over \sqrt{2m}} \left( \psi (t, \vec x) e^{-imt}
  + \psi^*(t, \vec x) e^{imt}
\right) \, .
\end{equation}
Axion detection experiments measure $\phi$ or its derivatives via its
coupling to photons
(${\cal L} \sim g_\gamma \phi F \tilde F$) and fermions such as 
quarks or leptons
(${\cal L} \sim g_\Psi \partial_\mu\phi \bar\Psi \gamma^\mu \gamma_5
\Psi$).\footnote{
The coupling constants $g_\gamma$ and $g_\Psi$ play the role of $1/f$
in Equation (\ref{phiInteractions}). There is also the coupling to
gluons, related to an oscillating electric dipole moment for nucleons
\citep{CASPEr}.
}
Writing $\phi$ in terms of $\psi$ reminds us there are two
time scales of interest: one is the fast Compton time
scale $\sim m^{-1}$ of $\phi$ oscillations; the other is the slow de
Broglie time scale
$\sim (mv^2)^{-1}$ of $\psi$ fluctuations due to wave
interference ($v$ is the velocity
dispersion of dark matter; see discussion around Equation
\ref{tcohere}):
\begin{eqnarray}
 && t_{\rm osc.} \equiv {2\pi \over m} = 1.3 {\,\rm yr.} \left(
  {10^{-22} {\,\rm eV} \over m} \right) = 4.1 \times 10^{-9} {\,\rm s} \left(
  {10^{-6} {\,\rm eV} \over m} \right) \, , \nonumber \\
&& t_{\rm dB} \equiv {2\pi \over mv^2} = 1.9 \times 10^6 {\,\rm yr.} 
\left( {10^{-22} {\,\rm eV} \over m} \right) \left( {250 {\,\rm km/s} \over
  v} \right)^2 \, \nonumber \\
&& \quad \quad = 5.9 \times 10^{-3} {\,\rm s} \left(
  {10^{-6} {\,\rm eV} \over m} \right) \left( {250 {\,\rm km/s} \over
  v} \right)^2 \, .
\end{eqnarray}
%

\begin{figure}[tb]
\centering
\includegraphics[width=0.9\textwidth, trim={2.4cm 12cm 1.5cm
  6cm},clip]{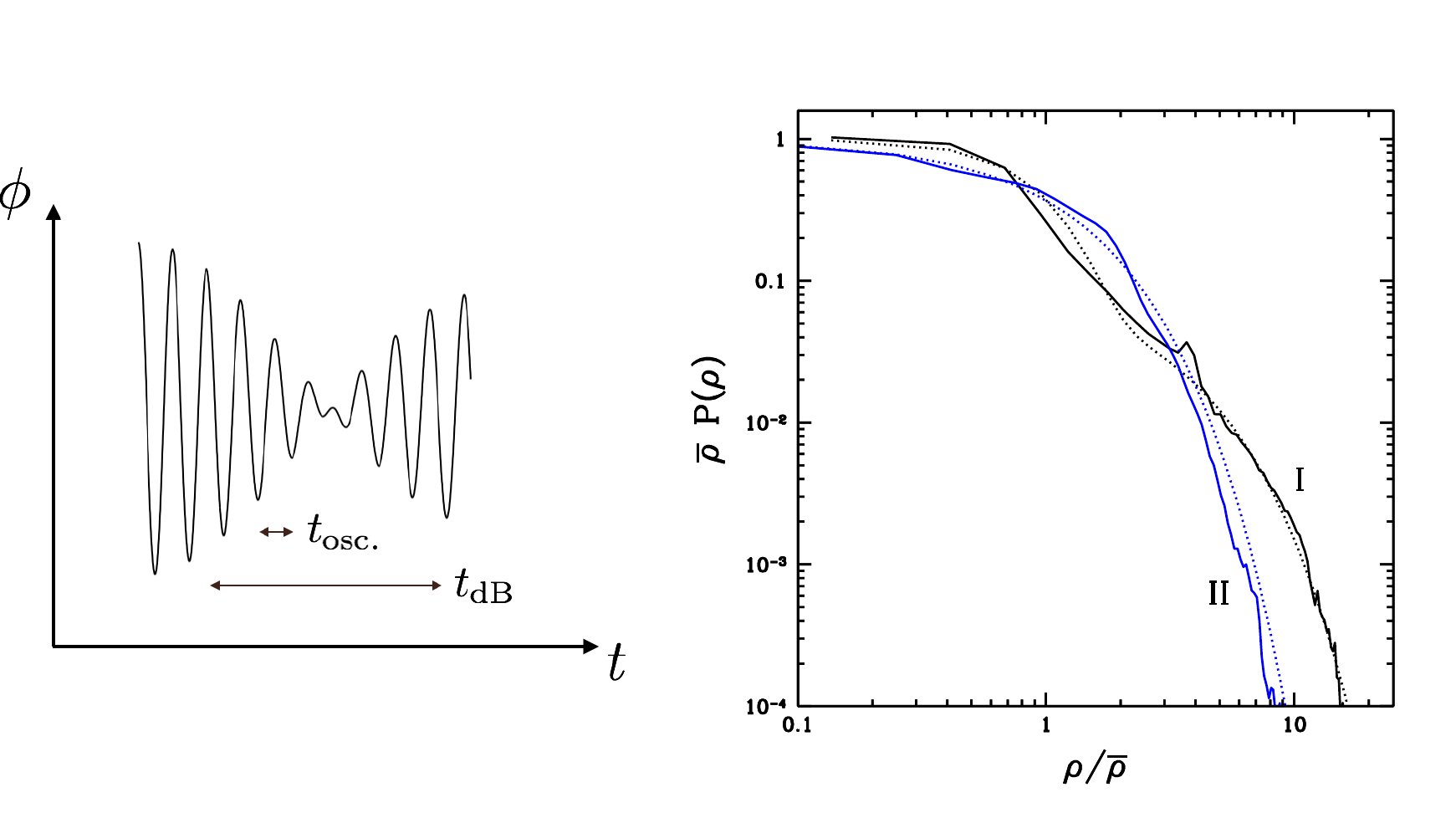}
\vspace{-0.1cm}
\caption
{\small {\it Left panel:} a schematic illustration of the time dependence of the scalar
  $\phi$ at some fixed location. It has short time scale 
  $t_{\rm osc.} = 2\pi/m$ oscillations (around $\phi = 0$),
  and long time scale $t_{\rm dB} = 2\pi/(mv^2)$ modulations.
  In practice, $t_{\rm dB} \gg t_{\rm osc.}$. 
{\it Right panel:} the one-point probability distribution of density
in two wave dark matter halos. Here, $P(\rho) d\rho$ gives the
probability that the density $\rho$ takes the
  values within the interval $d\rho$ and $\bar\rho$ is the (local) mean
  density. The solid lines are measured from
  numerical wave simulations of two halos that form from mergers of
  smaller seed halos and gravitational collapse. The blue line (II) is
  for a case where the halo is well-mixed, and the black line (I) is
  for a case where the halo retains some memory of the initial
  conditions. The blue dotted line shows the analytic prediction from
  the random phase halo model, $\bar\rho P(\rho) 
  = e^{-\rho/\bar\rho}$, which describes case II well. The black
  dotted line is an approximate fit to case I: 
  $\bar\rho P(\rho) =
  0.9 \, e^{-1.06(\rho/\bar\rho)^2} + 0.1 \, e^{-0.42
    (\rho/\bar\rho)}$. 
  Figure adapted from \cite{Hui:2020hbq}.
}
\label{tosctdBfig}
\end{figure}
The time variation of $\phi$ at a fixed location is depicted in the
left panel of Figure \ref{tosctdBfig}.
In addition, $\phi$ fluctuates spatially because $\psi$ does, on 
the de Broglie length scale $\lambda_{\rm dB}$
(Equation \ref{lambdadB} and Figure \ref{wavepicture}).
In other words, because the halo is composed of a superposition of
waves of largely random phases, the wavefunction $\psi$ is essentially a stochastic
field, which imprints $\sim t_{\rm dB}$ temporal modulations 
and $\sim \lambda_{\rm dB}$ spatial fluctuations on the axion $\phi$.
Existing experiments are sensitive to a wide range of axion masses,
from $m \sim 10^{-22}$ to $10^{-3}$ eV, though with significant gaps
\citep{Graham:2015ouw,Irastorza:2018dyq,Sikivie:2020zpn}.
In many cases, time scales from $t_{\rm osc.}$ to $t_{\rm dB}$ and beyond
are accessible to experiments.

A simple starting point for thinking about the stochastic
fluctuations is the random phase halo model, spelled out
in Equation \ref{psiRandom}: $\psi$ consists of a set of plane waves each
with an amplitude $A_{\vec k}$ that depends on momentum $\vec k$, and a random
phase. A simple distribution of momentum would be $A_{\vec k} \propto
e^{-k^2/k_0^2}$, essentially an isothermal one, though other
distributions are possible. 
In the random phase model, $\psi$ is a Gaussian random field obeying:
\footnote{Note how the random phase for each plane wave is sufficient to
  guarantee the {\it complex} $\psi$ is Gaussian random, even if
  $A_{\vec k}$ is non-stochastic.}
\begin{equation}
\langle \psi (t_1, \vec x_1) \psi^* (t_2, \vec x_2) \rangle
= \sum_{\vec k} A_{\vec k}^2 \, e^{i \vec k \cdot (\vec x_1 - \vec x_2) -
  i \omega_k (t_1 - t_2)} \quad , \quad
\langle \psi (t_1, \vec x_1) \psi (t_2, \vec x_2) \rangle = 0 \, .
\end{equation}
The higher point correlation
functions obey Wick's theorem, expressible as products of the two-point
function. From this, all statistical properties of the axion $\phi$
follow, such as:
\begin{equation}
\label{phi2pt}
\langle \phi (t_1, \vec x_1) \phi (t_2, \vec x_2) \rangle
= {1\over 2m} \left( \langle \psi (t_1 , \vec x_1) \psi^* (t_2 , \vec
  x_2) \rangle e^{-im(t_1 - t_2)} + {\,\rm c.c.} \right) \, ,
\end{equation}
where ${\,\rm c.c.}$ represents complex conjugate.
The Gaussian random nature of $\psi$ tells us 
the one-point probability distribution is Gaussian,
specifically a two-dimensional one since $\psi$ has real and imaginary
parts i.e. the Gaussian probability density $ {\,\rm
  exp}[-|\psi|^2/(2\Gamma^2)]$, where $\Gamma^2 \equiv \sum_{\vec k}
A_{\vec k}^2/2$,
should come with the measure $d{\rm Re}\psi \, d{\rm Im}\psi = 2\pi
|\psi| d|\psi|$. In other words,
\begin{equation}
d|\psi|  {|\psi| \over \Gamma^2} {\,\rm
  exp}\left[ -{|\psi|^2 \over 2\Gamma^2} \right] \, ,
\end{equation}
gives the probability that $|\psi|$ takes the values within the
interval $d|\psi|$ \citep{Centers:2019dyn}.
It can be checked that this is properly normalized.
Recalling the density is $\rho = m |\psi|^2$, so
average density is $\bar\rho = m \langle |\psi|^2 \rangle = m^2 \langle
\phi^2 \rangle = 2 m \Gamma^2$, the one-point distribution of density is
thus: 
\footnote{This distribution can be derived directly from $\phi$
  without going through $\psi$, but it is important to remember
$\rho = (\dot\phi {}^2 + m^2 \phi^2)/2$ is determined not by $\phi$
alone, but also by its time derivative. Spatial gradient energy also
contributes to $\rho$ but is sub-dominant in the non-relativistic limit.}
\begin{equation}
{d\rho\over \bar\rho} e^{-\rho/\bar\rho} \, .
\end{equation}
There is a non-negligible
probability for the density to fluctuate to low values, indeed all the
way to zero (i.e. at sites of complete destructive interference or
vortices). The right panel of Figure \ref{tosctdBfig} shows a comparison of this analytic prediction
with results from 
numerical simulations of two halos that form from mergers and 
gravitational collapse, taken from
\cite{Hui:2020hbq}.
The analytic prediction works
reasonably well, especially in the case (II) where the halo is well 
mixed. It works less well in the case (I) where 
some memory of the initial conditions persists---the halo
has coherent substructures in the form of subhalos.
See also \cite{Veltmaat:2018dfz} for correlation function measurements
from numerical simulations.

The stochastic nature of the axion field $\phi$ and its derivatives
has rich implications for axion detection. 
For instance, given the average local density $\bar\rho$ ($\sim 0.4
{\,\rm GeV/cm^3}$), an axion experiment would
sample from the whole distribution of $\rho$'s depicted in
Figure \ref{tosctdBfig}, {\it if} time scales longer than 
the de Broglie time $t_{\rm dB}$ were accessible.
In particular, there would be a non-negligible probability of sampling $\rho <
\bar\rho$. As pointed out by \cite{Centers:2019dyn}, experimental
constraints on the axion couplings, such as $g_\gamma$ or $g_\Psi$,
should take this into account. The full implications remain to be
explored---depending on the experiment of
interest, the relevant correlation function can be
obtained by taking suitable derivatives of Equation \ref{phi2pt}.

Moreover, the stochastic nature of $\phi$ suggests it would be useful
to measure correlation functions.
For instance, the signal for ADMX 
\citep{Du:2018uak}
is often expressed in terms of the
power output in a microwave cavity, which is proportional to $\phi^2$, 
or $\phi^2$ averaged over the rapid, frequency $m$
oscillations.\footnote{
The idea was proposed by \cite{Sikivie:1983ip}. It involves
looking for photons produced by axions in the presence of a magnetic field.
} 
One can consider the following correlation function in
time (coincident location):
\begin{equation}
\langle \phi(t_1)^2 \phi(t_2)^2 \rangle - \langle \phi^2 \rangle^2 
= {1\over m^2} |\langle \psi(t_1) \psi^*(t_2) \rangle |^2
=  {\bar\rho^2 \over m^4} \left(1 + {k_0^4 (t_1 - t_2)^2 \over 16 m^2} \right)^{-3/2}\, ,
\end{equation}
where we have implicitly averaged $\phi^2(t)$ over the rapid
oscillations, and assumed the random phase model.
Here, $k_0$ is the rms (3D) momentum times $2/\sqrt{3}$, following from
the distribution $A_{\vec k}^2 \propto e^{-2k^2/k_0^2}$. 
This correlation function 
can be measured in a microwave cavity experiment.
The characteristic power-law decay at large time separation might be
helpful in pulling signal out of noisy data.
Some experiments measure $\dot\phi$ by searching for a time varying
magnetic flux produced by the oscillating axion in the presence of an
external magnetic field, such as
ABRACADABRA \citep{Kahn:2016aff,Ouellet:2018beu}.
Others are sensitive to $\vec
\nabla \phi$, such as CASPEr \citep{CASPEr,Budker:2013hfa} or spin
pendulum experiments \citep{Terrano:2019clh}.
The idea is to measure the spin precession around the direction picked
out by $\vec \nabla \phi$, using the axion-fermion coupling (Equation
\ref{phiInteractions}).
Correlation functions thereof can be obtained by
differentiating Equation \ref{phi2pt}. 

More generally, with a network of detectors, one can measure
the correlation function in space-time:
\be
\begin{aligned}
\langle \phi(t_1, \vec x_1)^2\phi(t_2, \vec x_2)^2 \rangle - \langle \phi^2 \rangle^2  = {\bar\rho {}^2\over m^4}\left(1+\frac{k_0^4
    (t_1-t_2)^2}{16m^2}\right)^{-3/2} \exp\left(-\frac{4k_0^2m^2\lvert
    \vec x_1-\vec x_2\rvert^2}{16m^2+k_0^4(t_1-t_2)^2}\right) \, ,
\end{aligned}
\ee
where again we have implicitly averaged over the rapid oscillations.
The difference in dependence on time-separation versus
space-separation originates from the fact $\omega_k$, the frequency
for a Fourier mode, goes as $k^2$ rather than $k$. 
The idea of using a network of detectors, much like an interferometry
array in radio astronomy,
has been discussed in \cite{Pustelny:2013rza} for GNOME, and
in \cite{Derevianko:2016vpm,Foster:2017hbq, Roberts:2017hla,
Savalle:2019jsb,Centers:2019dyn,Hui:2020hbq,Foster:2020fln}. 
Experiments that measure the rotation of photon polarization in an
axion background naturally measures $\phi$ at points separated in time
and/or space \citep{Liu:2018icu,DeRocco:2018jwe,Martynov:2019azm}.

It is worth pointing out that different experiments respond
differently to the passing of a vortex. As discussed in
Section \ref{vortices}, at the location of a vortex, $\psi$ vanishes but
its gradient generically does not. This implies experiments that
probe $\phi$ or $\dot\phi$ have a vanishing signal while those that probe
$\vec\nabla \phi$ have a non-vanishing one.
\footnote{In the non-relativistic limit, $\dot\phi$ and $\phi$ are
practically equivalent i.e. $\phi \sim \psi e^{-imt} + {\,\rm c. c.}$
while $\dot\phi \sim -im \psi e^{-imt} + {\,\rm c.c.}$.}
Perhaps more interesting is how the generic existence of vortices
(one vortex ring per de Broglie volume) points to interesting
structures in the {\it phase} of the axion oscillations. Plugging
$\psi = \sqrt{\rho/m} \, e^{i\theta}$ into Equation \ref{phipsiB}, the
axion field $\phi$ can be expressed as:
\begin{equation}
\label{phitheta}
\phi (t, \vec x) = m^{-1} \sqrt{2\rho(t, \vec x)} {\,\rm cos\,}
\left[ mt
- \theta(t,\vec x) \right] \, .
\end{equation}
Dark matter detection, for good reasons, generally focuses on
measuring the amplitude of the axion oscillations, which tells us
about the density of dark matter $\rho$.
The arguments in Section \ref{vortices} tell us 
wave interference generically produces non-trivial structures
in the oscillation phase $\theta(t, \vec x)$ i.e. winding around vortices.
It would be useful to explore how such winding could be measured, 
how it might be exploited to enhance detection sensitivity.
Doing so likely requires a network of detectors, possibly combining
different detection techniques that get at different derivatives of
$\phi$ \citep{Hui:2020hbq}.

\section{Discussion---theory exploration, numerical simulations,
  astrophysical probes and experimental detection}
\label{conclude}

We have reviewed the particle physics motivations for considering wave
dark matter, and the observational and experimental implications, with
the axion as the prime example. We close with a list of open questions
and directions for further research.

{\it Theory exploration.} The dark matter sector could well be as rich
as the visible sector, with different kinds of particles. This has a
certain plausibility in string theory, which generically predicts a
variety of axions. Most of them would be too massive to be
a suitable dark matter candidate. But if one of them is light enough
to be dark matter, perhaps there maybe more 
\citep{Arvanitaki:2009fg,Bachlechner:2018gew,Broadhurst:2018fei}? 
And if these light axions are coupled, how is the relic abundance
computation modified? 
What is the impact on galactic substructures if there is a
mixture of wave and particle dark matter, or a mixture of wave dark
matter of different masses \citep{Schwabe:2020eac}?
If the axion as a field exists during inflation, it has inevitable
isocurvature fluctuations---if the energy
scale of inflation is high enough to saturate the existing
isocurvature bound, what are the implications for structure formation
(Section \ref{earlyuniverse})? 


{\it Numerical simulations.} There is a great need for more and better 
simulations of wave dark matter structure formation. 
Some of the existing constraints at the ultra-light end of the
spectrum ($10^{-22} - 10^{-20}$ eV, fuzzy dark matter) rely on the halo
or subhalo mass function that has not been checked
with wave simulations (Section \ref{galacticConstraints}). Current estimates of the halo/subhalo mass
function account for the wave nature of dark matter primarily through its
impact on the initial condition i.e. the primordial power spectrum
(Section \ref{linearpowerspectrum}).
It is important to quantify how the wave dynamics affects the
subsequent evolution. Further simulations would also be useful 
for interpreting constraints from galaxy density profiles (by including the
effects of baryons and tidal forces), and constraints from the
Lyman-alpha forest (by exploring the variety of fluctuations from the
ionizing background, reionization history and galactic winds). 
There is also room for improvement in numerical algorithm: 
it is challenging to carry out wave simulations in large boxes with
the requisite de-Broglie-scale resolution (Section \ref{simulations}).
The hybrid scheme of \cite{Veltmaat:2018dfz} is one promising approach.
In addition, there is a need for more simulations of the early universe. If
the Peccei-Quinn symmetry is broken after inflation, large
fluctuations are expected to lead to axion star
formation \citep{Kolb:1993zz}. An accurate mass function of such objects,
accounting for the effect of subsequent mergers
\citep{Eggemeier:2019jsu}, would be very useful. The axion in question
can span a large range in mass and need not be ultra-light
(Sections \ref{solitons}
and \ref{probesCombojb}).

{\it Astrophysical probes.} A striking prediction of wave dark matter
is the interference substructures inside a halo. These are order unity
density fluctuations on the scale of the de Broglie wavelength. 
The density can even vanish, where
complete destructive interference occurs. These are locations of
vortices---a unique wave phenomenon (Section \ref{vortices}). 
Such interference patterns are distinct from subhalos as a form of halo substructure.
Some observational signatures, for ultra-light masses, have been worked out, such as the
scattering of stars and gravitational lensing
(Section \ref{galacticConstraints}). 
Recent measurements of the density power spectrum
along globular cluster tidal streams GD-1 and Palomar 5, from Gaia and
Pan-STARRS data, suggest consistency with
scattering by subhalos in conventional cold dark matter
\citep{Bovy:2016irg,Banik:2019cza,Banik:2019smi}.
\footnote{For more background on the streams and the data, see
\citet{Grillmair:2006bd,Ibata_2016,GAIA,chambers2019panstarrs1}.}
Are the same measurements consistent with
fuzzy dark mater? To answer this question, one
must account for scattering by both the subhalo
contents \citep{Schutz:2020jox} {\it and} the
interference substructures \citep{Dalal:2020mjw}. In addition,
it is important to clarify to what extent the tidal stream
density fluctuations can be attributed to the tidal disruption process
itself \citep{Kuepper:2009sg,Ibata_2020}. More measurements spanning different
orbital radii would be
helpful in differentiating between models: scattering by interference
substructures is expected to be more important at small radii relative
to scattering by subhalos \citep{Dalal:2020mjw}. It is also 
worth noting there are
other statistics that might have different sensitivity to the
mass and compactness of subhalos \citep[e.g.][]{Bonaca:2018fek}.
Improvement in stellar stream data is expected from further Gaia data
release and the
upcoming Vera Rubin Observatory \citep{Ivezic:2008fe}.

Anomalous flux ratios between gravitationally lensed images
have been used to constrain
substructures in galaxy lenses
\citep{Hezaveh:2016ltk,Hsueh:2019ynk,Gilman:2019vca,Dai:2020rio}.
See Section \ref{galacticConstraints}.
Typically these constraints are obtained by fitting 
the data with a parametrized model of subhalos, which is then checked
against the prediction of conventional cold dark matter. For fuzzy
dark matter, two issues should be addressed. One is a proper {\it wave} computation of
the subhalo mass function, discussed earlier. The other is the
inclusion of wave interference substructures as an additional source
of flux anomaly \citep{Chan:2020exg,Hui:2020hbq}. 
This is a promising technique given the expected improvement
in lensing data, e.g. from ALMA \citep{ALMA,Hezaveh:2016ltk}. 

Observations of the high redshift ($z > 5$) universe have the potential to probe
the linear power spectrum on small scales, and therefore constrain
fuzzy dark matter, as discussed in Section
\ref{linearpowerspectrum}. Promising future data include those from
the James Webb Space Telescope \citep{Gardner:2006ky,Hirano:2017bnu} and 21cm
experiments \citep{DeBoer:2016tnn,Bull:2018lat,Bowman:2018yin}.
To take full advantage of these data, the fuzzy dark matter predictions for
early structure formation should be refined using wave
simulations in larger boxes \citep{Mocz:2019pyf,May:2021wwp}.

Another area where more data are needed is the study of dynamical
friction.
The Fornax dwarf galaxy is the main example where there is possibly a
dynamical friction problem---that its globular clusters
survive in its halo despite efficient dynamical friction \citep{tre76,oh2000}.
One resolution is to invoke fuzzy dark matter to weaken dynamical
friction, though it appears core stalling might also do the job
(see Sections \ref{dynamics2} and
\ref{galacticConstraints}).
Data on more such systems would be instructive. 

{\it Detection experiments.} The interference substructures are a
robust prediction of wave dark matter, regardless of the dark matter
mass. Away from the ultra-light end of the spectrum, the corresponding
de Broglie wavelength is small, making the interference
substructures challenging to observe astrophysically. 
But the substructures remain relevant for
axion detection experiments which are sensitive to much smaller
scales. The axion field is effectively stochastic, in
a halo made out of a superposition of waves with random phases.
At a minimum, this stochastic nature should be accounted for in
deriving constraints. Moreover, the stochastic
nature motivates the measurement of correlation functions of the axion
field. The correlation can involve both time and space separations,
further motivating the idea of a network of
detectors, like in radio interferometry.
An under-explored area is the information contained in the
phase of the axion oscillations (Equation \ref{phitheta}). That vortices
generically exist tells us there are non-trivial structures in the
phase, such as winding. An interesting question is whether searching for
such structures might help extract signal out of noisy data
(Section \ref{axionExp}).

\section*{DISCLOSURE STATEMENT}
The author is not aware of any affiliations, memberships, funding, or financial holdings that
might be perceived as affecting the objectivity of this review. 

\section*{ACKNOWLEDGMENTS}
Thanks are due to my collaborators for
teaching me much of the subject: Jamie Bamber, Jo Bovy,
Greg Bryan, Katy Clough, Neal Dalal, Pedro Ferreira, Austin Joyce, Dan Kabat,
Michael Landry, Albert Law, Macarena Lagos, Xinyu Li,  Adam Lidz, Jerry
Ostriker, Klaas Parmentier, Luca Santoni, Guanhao Sun, Gianmaria Tomaselli,
Scott Tremaine, Enrico Trincherini, Edward Witten, Sam Wong and
Tomer Yavetz. Thanks to Eric Adelberger, Emanuele Berti, 
Tom Broadhurst, Vitor Cardoso, Gary Centers, Andy Cohen, Vincent
Desjacques, Sergei
Dubovsky, Mark Hertzberg, 
Vid Ir\u si\u c, Dima Levkov, Eugene Lim, Doddy Marsh, Philip Mocz,
Alberto Nicolis, Jens
Niemeyer, Adi Nusser, Marco Peloso, Massimo Pietroni,
Alessandro Podo, Riccardo Rattazzi, Leslie Rosenberg, Hsi-Yu
Schive, Sergei Sibiryakov, Pierre Sikivie, Will Terrano, Cora
Uhlemann, Tanmay Vachaspati, Jacqueline van Gorkom, Matteo Viel 
and Dennis Zaritsky for useful discussions.
Special thanks to Xinyu Li for providing some of
the figures, and to
Kfir Blum, Jo Bovy, Tom Broadhurst, Katy Clough, Neal Dalal,
Anson Hook, Vid Ir\u si\u c, Eliot Quataert, Jerry Ostriker, Surjeet Rajendran, Leslie Rosenberg,
David Spergel, Will Terrano, 
Scott Tremaine, Matteo Viel and Dennis Zaritsky 
for comments and suggestions on the manuscript. 
Support by a Simons Fellowship in Theoretical Physics and the 
Department of Energy
DE-SC0011941 is gratefully acknowledged.

\bibliographystyle{ar-style2}

\bibliography{references}

\begin{thebibliography}{}
\expandafter\ifx\csname natexlab\endcsname\relax\def\natexlab#1{#1}\fi

\bibitem[{Abbott \& Sikivie(1983)}]{Abbott:1982af}
Abbott L, Sikivie P. 1983.
\textit{Phys. Lett. B} 120:133--136

\bibitem[{Aghanim et~al.(2020)}]{Aghanim:2018eyx}
Aghanim N, et~al. 2020.
\textit{Astron. Astrophys.} 641:A6

\bibitem[{Agrawal et~al.(2020)Agrawal, Hook \& Huang}]{Agrawal:2019lkr}
Agrawal P, Hook A, Huang J. 2020.
\textit{JHEP} 07:138

\bibitem[{Akiyama et~al.(2019)}]{Akiyama:2019fyp}
Akiyama K, et~al. 2019.
\textit{Astrophys. J. Lett.} 875:L5

\bibitem[{Alexander et~al.(2019)Alexander, Bramburger \&
  McDonough}]{Alexander:2019qsh}
Alexander S, Bramburger JJ, McDonough E. 2019.
\textit{Phys. Lett. B} 797:134871

\bibitem[{Alexander \& Cormack(2017)}]{Alexander:2016glq}
Alexander S, Cormack S. 2017.
\textit{JCAP} 1704:005

\bibitem[{Alexander et~al.(2020)Alexander, Gleyzer, McDonough, Toomey \&
  Usai}]{Alexander:2019puy}
Alexander S, Gleyzer S, McDonough E, Toomey MW, Usai E. 2020.
\textit{Astrophys. J.} 893:15

\bibitem[{Allali \& Hertzberg(2020)}]{Allali:2020ttz}
Allali I, Hertzberg MP. 2020.
\textit{JCAP} 07:056

\bibitem[{Alonso-\'Alvarez \& Jaeckel(2018)}]{Alonso-Alvarez:2018tus}
Alonso-\'Alvarez G, Jaeckel J. 2018.
\textit{JCAP} 10:022

\bibitem[{Amendola \& Barbieri(2006)}]{Amendola:2005ad}
Amendola L, Barbieri R. 2006.
\textit{Phys. Lett.} B642:192--196

\bibitem[{Amin et~al.(2012)Amin, Easther, Finkel, Flauger \&
  Hertzberg}]{Amin:2011hj}
Amin MA, Easther R, Finkel H, Flauger R, Hertzberg MP. 2012.
\textit{Phys. Rev. Lett.} 108:241302

\bibitem[{Amorisco \& Evans(2012)}]{Amorisco:2011hb}
Amorisco N, Evans N. 2012.
\textit{Mon. Not. Roy. Astron. Soc.} 419:184--196

\bibitem[{Amorisco \& Loeb(2018)}]{Amorisco:2018dcn}
Amorisco NC, Loeb A. 2018.
\textit{arXiv:1808.00464}

\bibitem[{Anastassopoulos et~al.(2017)}]{Anastassopoulos:2017ftl}
Anastassopoulos V, et~al. 2017.
\textit{Nature Phys.} 13:584--590

\bibitem[{Annulli et~al.(2020)Annulli, Cardoso \& Vicente}]{Annulli:2020lyc}
Annulli L, Cardoso V, Vicente R. 2020.
\textit{Phys. Rev. D} 102:063022

\bibitem[{Aoki \& Mukohyama(2016)}]{Aoki:2016zgp}
Aoki K, Mukohyama S. 2016.
\textit{Phys. Rev. D} 94:024001

\bibitem[{Armengaud et~al.(2017)Armengaud, Palanque-Delabrouille, Y\`eche,
  Marsh \& Baur}]{Armengaud:2017nkf}
Armengaud E, Palanque-Delabrouille N, Y\`eche C, Marsh DJ, Baur J. 2017.
\textit{Mon. Not. Roy. Astron. Soc.} 471:4606--4614

\bibitem[{Arvanitaki et~al.(2017)Arvanitaki, Baryakhtar, Dimopoulos, Dubovsky
  \& Lasenby}]{Arvanitaki:2016qwi}
Arvanitaki A, Baryakhtar M, Dimopoulos S, Dubovsky S, Lasenby R. 2017.
\textit{Phys. Rev. D} 95:043001

\bibitem[{Arvanitaki et~al.(2010)Arvanitaki, Dimopoulos, Dubovsky, Kaloper \&
  March-Russell}]{Arvanitaki:2009fg}
Arvanitaki A, Dimopoulos S, Dubovsky S, Kaloper N, March-Russell J. 2010.
\textit{Phys. Rev.} D81:123530

\bibitem[{Arvanitaki et~al.(2020)Arvanitaki, Dimopoulos, Galanis, Lehner,
  Thompson \& Van~Tilburg}]{Arvanitaki:2019rax}
Arvanitaki A, Dimopoulos S, Galanis M, Lehner L, Thompson JO, Van~Tilburg K.
  2020.
\textit{Phys. Rev. D} 101:083014

\bibitem[{Arvanitaki \& Dubovsky(2011)}]{Arvanitaki:2010sy}
Arvanitaki A, Dubovsky S. 2011.
\textit{Phys. Rev.} D83:044026

\bibitem[{Axenides et~al.(1983)Axenides, Brandenberger \&
  Turner}]{Axenides:1983hj}
Axenides M, Brandenberger RH, Turner MS. 1983.
\textit{Phys. Lett. B} 126:178--182

\bibitem[{Bachlechner et~al.(2019)Bachlechner, Eckerle, Janssen \&
  Kleban}]{Bachlechner:2018gew}
Bachlechner TC, Eckerle K, Janssen O, Kleban M. 2019.
\textit{JCAP} 1909:062

\bibitem[{Bai \& Hamada(2018)}]{Bai:2017feq}
Bai Y, Hamada Y. 2018.
\textit{Phys. Lett. B} 781:187--194

\bibitem[{Baldeschi et~al.(1983)Baldeschi, Ruffini \&
  Gelmini}]{Baldeschi:1983mq}
Baldeschi MR, Ruffini R, Gelmini GB. 1983.
\textit{Phys. Lett.} 122B:221--224

\bibitem[{Bamber et~al.(2020)Bamber, Clough, Ferreira, Hui \&
  Lagos}]{Bamber:2020bpu}
Bamber J, Clough K, Ferreira PG, Hui L, Lagos M. 2020.
\textit{arXiv:2011.07870}

\bibitem[{Banik et~al.(2019{\natexlab{a}})Banik, Bovy, Bertone, Erkal \&
  de~Boer}]{Banik:2019cza}
Banik N, Bovy J, Bertone G, Erkal D, de~Boer T. 2019{\natexlab{a}}.
\textit{arXiv:1911.02662}

\bibitem[{Banik et~al.(2019{\natexlab{b}})Banik, Bovy, Bertone, Erkal \&
  de~Boer}]{Banik:2019smi}
Banik N, Bovy J, Bertone G, Erkal D, de~Boer T. 2019{\natexlab{b}}.
\textit{arXiv:1911.02663}

\bibitem[{Banik \& Sikivie(2013)}]{Banik:2013rxa}
Banik N, Sikivie P. 2013.
\textit{Phys. Rev.} D88:123517

\bibitem[{Bar et~al.(2018)Bar, Blas, Blum \& Sibiryakov}]{Bar:2018acw}
Bar N, Blas D, Blum K, Sibiryakov S. 2018.
\textit{Phys. Rev.} D98:083027

\bibitem[{Bar et~al.(2020)Bar, Blum \& D'Amico}]{Bar:2019ifz}
Bar N, Blum K, D'Amico G. 2020.
\textit{Phys. Rev. D} 101:123025

\bibitem[{Bar et~al.(2019{\natexlab{a}})Bar, Blum, Eby \& Sato}]{Bar:2019bqz}
Bar N, Blum K, Eby J, Sato R. 2019{\natexlab{a}}.
\textit{Phys. Rev. D} 99:103020

\bibitem[{Bar et~al.(2019{\natexlab{b}})Bar, Blum, Lacroix \&
  Panci}]{Bar:2019pnz}
Bar N, Blum K, Lacroix T, Panci P. 2019{\natexlab{b}}.
\textit{JCAP} 07:045

\bibitem[{Bar-Or et~al.(2019)Bar-Or, Fouvry \& Tremaine}]{Bar-Or:2018pxz}
Bar-Or B, Fouvry JB, Tremaine S. 2019.
\textit{Astrophys. J.} 871:28

\bibitem[{Barausse et~al.(2014)Barausse, Cardoso \& Pani}]{Barausse:2014tra}
Barausse E, Cardoso V, Pani P. 2014.
\textit{Phys. Rev. D} 89:104059

\bibitem[{Bardeen et~al.(1972)Bardeen, Press \& Teukolsky}]{Bardeen:1972fi}
Bardeen JM, Press WH, Teukolsky SA. 1972.
\textit{Astrophys. J.} 178:347

\bibitem[{Barkana et~al.(2001)Barkana, Haiman \& Ostriker}]{Barkana:2001gr}
Barkana R, Haiman Z, Ostriker JP. 2001.
\textit{Astrophys. J.} 558:482

\bibitem[{Barranco et~al.(2012)Barranco, Bernal, Degollado, Diez-Tejedor,
  Megevand et~al.}]{Barranco:2012qs}
Barranco J, Bernal A, Degollado JC, Diez-Tejedor A, Megevand M, et~al. 2012.
\textit{Phys. Rev. Lett.} 109:081102

\bibitem[{Baumann(2011)}]{Baumann:2009ds}
Baumann D. 2011.
\textit{{Inflation}}. In \textit{{Theoretical Advanced Study Institute in
  Elementary Particle Physics}: {Physics of the Large and the Small}}

\bibitem[{Baumann et~al.(2019)Baumann, Chia \& Porto}]{Baumann:2018vus}
Baumann D, Chia HS, Porto RA. 2019.
\textit{Phys. Rev. D} 99:044001

\bibitem[{Bekenstein(1972{\natexlab{a}})}]{Bekenstein:1972ky}
Bekenstein J. 1972{\natexlab{a}}.
\textit{Phys. Rev. D} 5:2403--2412

\bibitem[{Bekenstein(1972{\natexlab{b}})}]{Bekenstein:1971hc}
Bekenstein JD. 1972{\natexlab{b}}.
\textit{Phys. Rev. D} 5:1239--1246

\bibitem[{{Bennett} et~al.(2013){Bennett}, {Larson}, {Weiland}, {Jarosik},
  {Hinshaw} et~al.}]{WMAP2013}
{Bennett} CL, {Larson} D, {Weiland} JL, {Jarosik} N, {Hinshaw} G, et~al. 2013.
\textit{\apjs} 208:20

\bibitem[{Berezhiani et~al.(2019)Berezhiani, Elder \&
  Khoury}]{Berezhiani:2019pzd}
Berezhiani L, Elder B, Khoury J. 2019.
\textit{JCAP} 10:074

\bibitem[{Berezhiani \& Khoury(2015{\natexlab{a}})}]{Berezhiani:2015bqa}
Berezhiani L, Khoury J. 2015{\natexlab{a}}.
\textit{Phys. Rev.} D92:103510

\bibitem[{Berezhiani \& Khoury(2015{\natexlab{b}})}]{bk15}
Berezhiani L, Khoury J. 2015{\natexlab{b}}.
\textit{Phys. Rev.} D92:103510

\bibitem[{Bezerra et~al.(2014)Bezerra, Vieira \& Costa}]{Bezerra:2013iha}
Bezerra VB, Vieira HS, Costa AA. 2014.
\textit{Class. Quant. Grav.} 31:045003

\bibitem[{{Bialynicki-Birula} et~al.(2000){Bialynicki-Birula},
  {Bialynicka-Birula} \& {{\'S}liwa}}]{2000PhRvA..61c2110B}
{Bialynicki-Birula} I, {Bialynicka-Birula} Z, {{\'S}liwa} C. 2000.
\textit{Phys. Rev. {\bf A}} 61:032110

\bibitem[{{Binney} \& {Tremaine}(2008)}]{BT}
{Binney} J, {Tremaine} S. 2008.
\textit{{Galactic Dynamics, 2nd ed.}}
Princeton, NJ, Princeton University Press

\bibitem[{Bird et~al.(2016)Bird, Cholis, Mu\~noz, Ali-Ha\"\i{}moud,
  Kamionkowski et~al.}]{Bird:2016dcv}
Bird S, Cholis I, Mu\~noz JB, Ali-Ha\"\i{}moud Y, Kamionkowski M, et~al. 2016.
\textit{Phys. Rev. Lett.} 116:201301

\bibitem[{Blas et~al.(2020)Blas, Caputo, Ivanov \& Sberna}]{Blas:2019qqp}
Blas D, Caputo A, Ivanov MM, Sberna L. 2020.
\textit{Phys. Dark Univ.} 27:100428

\bibitem[{Bonaca et~al.(2018)Bonaca, Hogg, Price-Whelan \&
  Conroy}]{Bonaca:2018fek}
Bonaca A, Hogg DW, Price-Whelan AM, Conroy C. 2018.
\textit{arXiv:1811.03631}

\bibitem[{Bovy et~al.(2017)Bovy, Erkal \& Sanders}]{Bovy:2016irg}
Bovy J, Erkal D, Sanders JL. 2017.
\textit{Mon. Not. Roy. Astron. Soc.} 466:628--668

\bibitem[{Bovy \& Tremaine(2012)}]{Bovy:2012tw}
Bovy J, Tremaine S. 2012.
\textit{Astrophys. J.} 756:89

\bibitem[{Bowman et~al.(2018)Bowman, Rogers, Monsalve, Mozdzen \&
  Mahesh}]{Bowman:2018yin}
Bowman JD, Rogers AEE, Monsalve RA, Mozdzen TJ, Mahesh N. 2018.
\textit{Nature} 555:67--70

\bibitem[{Brax et~al.(2020)Brax, Cembranos \& Valageas}]{Brax:2019npi}
Brax P, Cembranos JA, Valageas P. 2020.
\textit{Phys. Rev. D} 101:023521

\bibitem[{Brook \& Coles(2009)}]{Brook:2009ku}
Brook MN, Coles P. 2009.
\textit{arXiv:0902.0605}

\bibitem[{Budker et~al.(2014)Budker, Graham, Ledbetter, Rajendran \&
  Sushkov}]{Budker:2013hfa}
Budker D, Graham PW, Ledbetter M, Rajendran S, Sushkov A. 2014.
\textit{Phys. Rev.} X4:021030

\bibitem[{Burkert(2020)}]{Burkert:2020laq}
Burkert A. 2020.
\textit{arXiv:2006.11111}

\bibitem[{Buschmann et~al.(2020)Buschmann, Foster \& Safdi}]{Buschmann:2019icd}
Buschmann M, Foster JW, Safdi BR. 2020.
\textit{Phys. Rev. Lett.} 124:161103

\bibitem[{Calabrese \& Spergel(2016)}]{Calabrese:2016hmp}
Calabrese E, Spergel DN. 2016.
\textit{Mon. Not. Roy. Astron. Soc.} 460:4397--4402

\bibitem[{Caputo et~al.(2019)Caputo, Sberna, Frias, Blas, Pani
  et~al.}]{Caputo:2019tms}
Caputo A, Sberna L, Frias M, Blas D, Pani P, et~al. 2019.
\textit{Phys. Rev. D} 100:063515

\bibitem[{Carroll \& Field(1997)}]{Carroll:1997tc}
Carroll SM, Field GB. 1997.
\textit{Phys. Rev. Lett.} 79:2394--2397

\bibitem[{Carroll et~al.(1990)Carroll, Field \& Jackiw}]{Carroll:1989vb}
Carroll SM, Field GB, Jackiw R. 1990.
\textit{Phys. Rev. D} 41:1231

\bibitem[{Cen et~al.(2009)Cen, McDonald, Trac \& Loeb}]{Cen2009}
Cen R, McDonald P, Trac H, Loeb A. 2009.
\textit{Astrophys. J.} 706:L164--L167

\bibitem[{Centers et~al.(2019)}]{Centers:2019dyn}
Centers GP, et~al. 2019.
\textit{arXiv:1905.13650}

\bibitem[{Chambers et~al.(2019)Chambers, Magnier, Metcalfe, Flewelling, Huber
  et~al.}]{chambers2019panstarrs1}
Chambers KC, Magnier EA, Metcalfe N, Flewelling HA, Huber ME, et~al. 2019.
\textit{arXiv:1612.05560}

\bibitem[{Chan et~al.(2020)Chan, Schive, Wong, Chiueh \&
  Broadhurst}]{Chan:2020exg}
Chan JH, Schive HY, Wong SK, Chiueh T, Broadhurst T. 2020.
\textit{Phys. Rev. Lett.} 125:111102

\bibitem[{Chavanis(2011)}]{Chavanis:2011zi}
Chavanis PH. 2011.
\textit{Phys. Rev.} D84:043531

\bibitem[{Chavanis(2019)}]{Chavanis:2019bnu}
Chavanis PH. 2019.
\textit{Eur. Phys. J. Plus} 134:352

\bibitem[{Chen et~al.(2017)Chen, Schive \& Chiueh}]{Chen:2016unw}
Chen SR, Schive HY, Chiueh T. 2017.
\textit{Mon. Not. Roy. Astron. Soc.} 468:1338--1348

\bibitem[{Chiba(2002)}]{Chiba:2001wk}
Chiba M. 2002.
\textit{Astrophys. J.} 565:17

\bibitem[{{Chiueh} et~al.(2011){Chiueh}, {Woo}, {Jian} \&
  {Schive}}]{2011JPhB...44k5101C}
{Chiueh} T, {Woo} TP, {Jian} HY, {Schive} HY. 2011.
\textit{Journal of Physics B} 44:115101

\bibitem[{Chluba et~al.(2019)}]{Chluba:2019nxa}
Chluba J, et~al. 2019.
\textit{arXiv:1909.01593}

\bibitem[{Choi \& Im(2016)}]{Choi:2015fiu}
Choi K, Im SH. 2016.
\textit{JHEP} 01:149

\bibitem[{Church et~al.(2019)Church, Ostriker \& Mocz}]{Church:2018sro}
Church BV, Ostriker JP, Mocz P. 2019.
\textit{Mon. Not. Roy. Astron. Soc.} 485:2861--2876

\bibitem[{Clough et~al.(2018)Clough, Dietrich \& Niemeyer}]{Clough:2018exo}
Clough K, Dietrich T, Niemeyer JC. 2018.
\textit{Phys. Rev. D} 98:083020

\bibitem[{Clough et~al.(2019)Clough, Ferreira \& Lagos}]{Clough:2019jpm}
Clough K, Ferreira PG, Lagos M. 2019.
\textit{Phys. Rev.} D100:063014

\bibitem[{{Clowe} et~al.(2006){Clowe}, {Brada{\v{c}}}, {Gonzalez},
  {Markevitch}, {Randall} et~al.}]{Clowe2006}
{Clowe} D, {Brada{\v{c}}} M, {Gonzalez} AH, {Markevitch} M, {Randall} SW,
  et~al. 2006.
\textit{\apjl} 648:L109--L113

\bibitem[{Co et~al.(2020)Co, Hall \& Harigaya}]{Co:2019jts}
Co RT, Hall LJ, Harigaya K. 2020.
\textit{Phys. Rev. Lett.} 124:251802

\bibitem[{Cole et~al.(2012)Cole, Dehnen, Read \& Wilkinson}]{Cole:2012ns}
Cole DR, Dehnen W, Read JI, Wilkinson MI. 2012.
\textit{Mon. Not. Roy. Astron. Soc.} 426:601

\bibitem[{Cookmeyer et~al.(2020)Cookmeyer, Grin \& Smith}]{Cookmeyer:2019rna}
Cookmeyer J, Grin D, Smith TL. 2020.
\textit{Phys. Rev. D} 101:023501

\bibitem[{Croft et~al.(1998)Croft, Weinberg, Katz \& Hernquist}]{Croft:1997jf}
Croft R, Weinberg DH, Katz N, Hernquist L. 1998.
\textit{Astrophys. J.} 495:44--62

\bibitem[{Croft(2004)}]{Croft:2003qn}
Croft RA. 2004.
\textit{Astrophys. J.} 610:642--662

\bibitem[{Dai et~al.(2020)Dai, Kaurov, Sharon, Florian, Miralda-Escud\'e
  et~al.}]{Dai:2020rio}
Dai L, Kaurov AA, Sharon K, Florian MK, Miralda-Escud\'e J, et~al. 2020.
\textit{Mon. Not. Roy. Astron. Soc.} 495:3192--3208

\bibitem[{Dalal et~al.(2020)Dalal, Bovy, Hui \& Li}]{Dalal:2020mjw}
Dalal N, Bovy J, Hui L, Li X. 2020.
\textit{arXiv:2011.13141}

\bibitem[{Dalal \& Kochanek(2002)}]{Dalal:2001fq}
Dalal N, Kochanek CS. 2002.
\textit{Astrophys. J.} 572:25--33

\bibitem[{D'Aloisio et~al.(2018)D'Aloisio, McQuinn, Davies \&
  Furlanetto}]{DAloisio:2016dcv}
D'Aloisio A, McQuinn M, Davies FB, Furlanetto SR. 2018.
\textit{Mon. Not. Roy. Astron. Soc.} 473:560--575

\bibitem[{Damour et~al.(1976)Damour, Deruelle \& Ruffini}]{Damour:1976kh}
Damour T, Deruelle N, Ruffini R. 1976.
\textit{Lett. Nuovo Cim.} 15:257--262

\bibitem[{Davies \& Mocz(2020)}]{Davies:2019wgi}
Davies EY, Mocz P. 2020.
\textit{Mon. Not. Roy. Astron. Soc.} 492:5721--5729

\bibitem[{Davoudiasl \& Denton(2019)}]{Davoudiasl:2019nlo}
Davoudiasl H, Denton PB. 2019.
\textit{Phys. Rev. Lett.} 123:021102

\bibitem[{Davoudiasl \& Murphy(2017)}]{Davoudiasl:2017jke}
Davoudiasl H, Murphy CW. 2017.
\textit{Phys. Rev. Lett.} 118:141801

\bibitem[{De~Martino et~al.(2020)De~Martino, Broadhurst, Tye, Chiueh \&
  Schive}]{DeMartino:2018zkx}
De~Martino I, Broadhurst T, Tye SHH, Chiueh T, Schive HY. 2020.
\textit{Phys. Dark Univ.} 28:100503

\bibitem[{DeBoer et~al.(2017)}]{DeBoer:2016tnn}
DeBoer DR, et~al. 2017.
\textit{Publ. Astron. Soc. Pac.} 129:045001

\bibitem[{Derevianko(2018)}]{Derevianko:2016vpm}
Derevianko A. 2018.
\textit{Phys. Rev.} A97:042506

\bibitem[{DeRocco et~al.(2020)DeRocco, Graham \& Rajendran}]{DeRocco:2020xdt}
DeRocco W, Graham PW, Rajendran S. 2020.
\textit{Phys. Rev. D} 102:075015

\bibitem[{DeRocco \& Hook(2018)}]{DeRocco:2018jwe}
DeRocco W, Hook A. 2018.
\textit{Phys. Rev. D} 98:035021

\bibitem[{Desjacques et~al.(2018)Desjacques, Kehagias \&
  Riotto}]{Desjacques:2017fmf}
Desjacques V, Kehagias A, Riotto A. 2018.
\textit{Phys. Rev. D} 97:023529

\bibitem[{Detweiler(1980)}]{Detweiler:1980uk}
Detweiler SL. 1980.
\textit{Phys. Rev.} D22:2323--2326

\bibitem[{Dietrich \& Clough(2019)}]{Dietrich:2019shr}
Dietrich T, Clough K. 2019.
\textit{Phys. Rev. D} 100:083005

\bibitem[{Dine(2000)}]{Dine:2000cj}
Dine M. 2000.
\textit{{TASI lectures on the strong CP problem}}. In \textit{{Theoretical
  Advanced Study Institute in Elementary Particle Physics (TASI 2000): Flavor
  Physics for the Millennium}}

\bibitem[{Dine(2016)}]{Dine:2007zp}
Dine M. 2016.
\textit{{Supersymmetry and String Theory}: {Beyond the Standard Model}}.
Cambridge University Press

\bibitem[{Dine \& Fischler(1983)}]{Dine:1982ah}
Dine M, Fischler W. 1983.
\textit{Phys. Lett. B} 120:137--141

\bibitem[{Dine et~al.(1981)Dine, Fischler \& Srednicki}]{Dine:1981rt}
Dine M, Fischler W, Srednicki M. 1981.
\textit{Phys. Lett. B} 104:199--202

\bibitem[{Dolan(2007)}]{Dolan:2007mj}
Dolan SR. 2007.
\textit{Phys. Rev.} D76:084001

\bibitem[{Du et~al.(2018{\natexlab{a}})}]{Du:2018uak}
Du N, et~al. 2018{\natexlab{a}}.
\textit{Phys. Rev. Lett.} 120:151301

\bibitem[{Du et~al.(2017)Du, Behrens \& Niemeyer}]{Du:2016zcv}
Du X, Behrens C, Niemeyer JC. 2017.
\textit{Mon. Not. Roy. Astron. Soc.} 465:941--951

\bibitem[{Du et~al.(2018{\natexlab{b}})Du, Schwabe, Niemeyer \&
  B\"urger}]{Du:2018qor}
Du X, Schwabe B, Niemeyer JC, B\"urger D. 2018{\natexlab{b}}.
\textit{Phys. Rev. D} 97:063507

\bibitem[{Dvali \& Zell(2018)}]{Dvali:2017ruz}
Dvali G, Zell S. 2018.
\textit{JCAP} 07:064

\bibitem[{Easther et~al.(2009)Easther, Giblin, Hui \& Lim}]{Easther:2009ft}
Easther R, Giblin John~T. J, Hui L, Lim EA. 2009.
\textit{Phys. Rev. D} 80:123519

\bibitem[{Eby et~al.(2016{\natexlab{a}})Eby, Kouvaris, Nielsen \&
  Wijewardhana}]{Eby:2015hsq}
Eby J, Kouvaris C, Nielsen NG, Wijewardhana L. 2016{\natexlab{a}}.
\textit{JHEP} 02:028

\bibitem[{Eby et~al.(2016{\natexlab{b}})Eby, Suranyi \&
  Wijewardhana}]{Eby:2015hyx}
Eby J, Suranyi P, Wijewardhana L. 2016{\natexlab{b}}.
\textit{Mod. Phys. Lett. A} 31:1650090

\bibitem[{Edwards et~al.(2018)Edwards, Kendall, Hotchkiss \&
  Easther}]{Edwards:2018ccc}
Edwards F, Kendall E, Hotchkiss S, Easther R. 2018.
\textit{JCAP} 1810:027

\bibitem[{Eggemeier \& Niemeyer(2019)}]{Eggemeier:2019jsu}
Eggemeier B, Niemeyer JC. 2019.
\textit{Phys. Rev. D} 100:063528

\bibitem[{Ellis \& Olive(1987)}]{Ellis:1987pk}
Ellis JR, Olive KA. 1987.
\textit{Phys. Lett. B} 193:525

\bibitem[{Endlich \& Penco(2017)}]{Endlich:2016jgc}
Endlich S, Penco R. 2017.
\textit{JHEP} 05:052

\bibitem[{Fairbairn et~al.(2018)Fairbairn, Marsh, Quevillon \&
  Rozier}]{Fairbairn:2017sil}
Fairbairn M, Marsh DJE, Quevillon J, Rozier S. 2018.
\textit{Phys. Rev. D} 97:083502

\bibitem[{Fan(2016)}]{Fan:2016rda}
Fan J. 2016.
\textit{Phys. Dark Univ.} 14:84--94

\bibitem[{Fedderke et~al.(2019)Fedderke, Graham \&
  Rajendran}]{Fedderke:2019ajk}
Fedderke MA, Graham PW, Rajendran S. 2019.
\textit{Phys. Rev. D} 100:015040

\bibitem[{Felder \& Tkachev(2008)}]{Felder:2000hq}
Felder GN, Tkachev I. 2008.
\textit{Comput. Phys. Commun.} 178:929--932

\bibitem[{{Fetter}(2008)}]{2008LaPhy..18....1F}
{Fetter} AL. 2008.
\textit{Laser Physics} 18:1--11

\bibitem[{Feynman et~al.(1963)Feynman, Leighton \& Sands}]{Feynman}
Feynman RP, Leighton RB, Sands M. 1963.
\textit{{The Feynman Lectures on Physics}}.
Addison Wesley Longman

\bibitem[{Ficarra et~al.(2019)Ficarra, Pani \& Witek}]{Ficarra:2018rfu}
Ficarra G, Pani P, Witek H. 2019.
\textit{Phys. Rev. D} 99:104019

\bibitem[{Foster et~al.(2020{\natexlab{a}})Foster, Kahn, Macias, Sun, Eatough
  et~al.}]{Foster:2020pgt}
Foster JW, Kahn Y, Macias O, Sun Z, Eatough RP, et~al. 2020{\natexlab{a}}.
\textit{Phys. Rev. Lett.} 125:171301

\bibitem[{Foster et~al.(2020{\natexlab{b}})Foster, Kahn, Nguyen, Rodd \&
  Safdi}]{Foster:2020fln}
Foster JW, Kahn Y, Nguyen R, Rodd NL, Safdi BR. 2020{\natexlab{b}}.
\textit{arXiv:2009.14201}

\bibitem[{Foster et~al.(2018)Foster, Rodd \& Safdi}]{Foster:2017hbq}
Foster JW, Rodd NL, Safdi BR. 2018.
\textit{Phys. Rev.} D97:123006

\bibitem[{Freeman(1970)}]{Freeman:1970mx}
Freeman K. 1970.
\textit{Astrophys. J.} 160:811

\bibitem[{Friedberg et~al.(1987{\natexlab{a}})Friedberg, Lee \&
  Pang}]{Friedberg:1986tp}
Friedberg R, Lee T, Pang Y. 1987{\natexlab{a}}.
\textit{Phys. Rev. D} 35:3640

\bibitem[{Friedberg et~al.(1987{\natexlab{b}})Friedberg, Lee \&
  Pang}]{Friedberg:1986tq}
Friedberg R, Lee T, Pang Y. 1987{\natexlab{b}}.
\textit{Phys. Rev. D} 35:3658

\bibitem[{Garcia-Bellido \& Ruiz~Morales(2017)}]{Garcia-Bellido:2017mdw}
Garcia-Bellido J, Ruiz~Morales E. 2017.
\textit{Phys. Dark Univ.} 18:47--54

\bibitem[{Gardner et~al.(2006)}]{Gardner:2006ky}
Gardner JP, et~al. 2006.
\textit{Space Sci. Rev.} 123:485

\bibitem[{Garny et~al.(2020)Garny, Konstandin \& Rubira}]{Garny:2019noq}
Garny M, Konstandin T, Rubira H. 2020.
\textit{JCAP} 04:003

\bibitem[{Giblin et~al.(2010)Giblin, Hui, Lim \& Yang}]{Giblin:2010bd}
Giblin John~T. J, Hui L, Lim EA, Yang IS. 2010.
\textit{Phys. Rev. D} 82:045019

\bibitem[{Gilman et~al.(2020)Gilman, Birrer, Nierenberg, Treu, Du \&
  Benson}]{Gilman:2019nap}
Gilman D, Birrer S, Nierenberg A, Treu T, Du X, Benson A. 2020.
\textit{Mon. Not. Roy. Astron. Soc.} 491:6077--6101

\bibitem[{Gilman et~al.(2019)Gilman, Birrer, Treu, Nierenberg \&
  Benson}]{Gilman:2019vca}
Gilman D, Birrer S, Treu T, Nierenberg A, Benson A. 2019.
\textit{Mon. Not. Roy. Astron. Soc.} 487:5721--5738

\bibitem[{Glauber(1963)}]{Glauber:1963fi}
Glauber RJ. 1963.
\textit{Phys. Rev.} 130:2529--2539

\bibitem[{Goerdt et~al.(2006)Goerdt, Moore, Read, Stadel \&
  Zemp}]{Goerdt:2006rw}
Goerdt T, Moore B, Read J, Stadel J, Zemp M. 2006.
\textit{Mon. Not. Roy. Astron. Soc.} 368:1073--1077

\bibitem[{Gondolo \& Silk(2000)}]{Gondolo:1999gy}
Gondolo P, Silk J. 2000.
\textit{Nucl. Phys. B Proc. Suppl.} 87:87--89

\bibitem[{Goodman(2000)}]{Goodman:2000tg}
Goodman J. 2000.
\textit{New Astron.} 5:103

\bibitem[{Gorghetto et~al.(2020)Gorghetto, Hardy \&
  Villadoro}]{Gorghetto:2020qws}
Gorghetto M, Hardy E, Villadoro G. 2020.
\textit{arXiv:2007.04990}

\bibitem[{Graham et~al.(2015)Graham, Irastorza, Lamoreaux, Lindner \& van
  Bibber}]{Graham:2015ouw}
Graham PW, Irastorza IG, Lamoreaux SK, Lindner A, van Bibber KA. 2015.
\textit{Ann. Rev. Nucl. Part. Sci.} 65:485--514

\bibitem[{Graham et~al.(2016{\natexlab{a}})Graham, Kaplan, Mardon, Rajendran \&
  Terrano}]{Graham:2015ifn}
Graham PW, Kaplan DE, Mardon J, Rajendran S, Terrano WA. 2016{\natexlab{a}}.
\textit{Phys. Rev. D} 93:075029

\bibitem[{Graham et~al.(2016{\natexlab{b}})Graham, Mardon \&
  Rajendran}]{Graham:2015rva}
Graham PW, Mardon J, Rajendran S. 2016{\natexlab{b}}.
\textit{Phys. Rev. D} 93:103520

\bibitem[{Graham \& Rajendran(2013)}]{CASPEr}
Graham PW, Rajendran S. 2013.
\textit{Phys. Rev.} D88:035023

\bibitem[{Green et~al.(1988)Green, Schwarz \& Witten}]{Green:1987mn}
Green MB, Schwarz J, Witten E. 1988.
\textit{{SUPERSTRING THEORY. VOL. 2: LOOP AMPLITUDES, ANOMALIES AND
  PHENOMENOLOGY}}

\bibitem[{Grilli~di Cortona et~al.(2016)Grilli~di Cortona, Hardy, Pardo~Vega \&
  Villadoro}]{diCortona:2015ldu}
Grilli~di Cortona G, Hardy E, Pardo~Vega J, Villadoro G. 2016.
\textit{JHEP} 01:034

\bibitem[{Grillmair \& Dionatos(2006)}]{Grillmair:2006bd}
Grillmair CJ, Dionatos O. 2006.
\textit{Astrophys. J. Lett.} 643:L17--L20

\bibitem[{Guth et~al.(2015)Guth, Hertzberg \& Prescod-Weinstein}]{Guth:2014hsa}
Guth AH, Hertzberg MP, Prescod-Weinstein C. 2015.
\textit{Phys. Rev.} D92:103513

\bibitem[{Guzman \& Urena-Lopez(2006{\natexlab{a}})}]{Guzman:2006yc}
Guzman F, Urena-Lopez L. 2006{\natexlab{a}}.
\textit{Astrophys. J.} 645:814--819

\bibitem[{Guzman \& Urena-Lopez(2006{\natexlab{b}})}]{gul06}
Guzman FS, Urena-Lopez LA. 2006{\natexlab{b}}.
\textit{Astrophys. J.} 645:814--819

\bibitem[{Halverson et~al.(2017)Halverson, Long \& Nath}]{Halverson:2017deq}
Halverson J, Long C, Nath P. 2017.
\textit{Phys. Rev.} D96:056025

\bibitem[{Hannuksela et~al.(2019)Hannuksela, Wong, Brito, Berti \&
  Li}]{Hannuksela:2018izj}
Hannuksela OA, Wong KW, Brito R, Berti E, Li TG. 2019.
\textit{Nature Astron.} 3:447--451

\bibitem[{Harari \& Sikivie(1992)}]{Harari:1992ea}
Harari D, Sikivie P. 1992.
\textit{Phys. Lett. B} 289:67--72

\bibitem[{{Harrison} et~al.(2003){Harrison}, {Moroz} \& {Tod}}]{hmt03}
{Harrison} R, {Moroz} I, {Tod} KP. 2003.
\textit{Nonlinearity} 16:101--122

\bibitem[{Helfer et~al.(2019)Helfer, Lim, Garcia \& Amin}]{Helfer:2018vtq}
Helfer T, Lim EA, Garcia MA, Amin MA. 2019.
\textit{Phys. Rev. D} 99:044046

\bibitem[{Helfer et~al.(2017)Helfer, Marsh, Clough, Fairbairn, Lim \&
  Becerril}]{Helfer:2016ljl}
Helfer T, Marsh DJE, Clough K, Fairbairn M, Lim EA, Becerril R. 2017.
\textit{JCAP} 03:055

\bibitem[{Hertzberg \& Schiappacasse(2018)}]{Hertzberg:2018lmt}
Hertzberg MP, Schiappacasse ED. 2018.
\textit{JCAP} 1808:028

\bibitem[{Hezaveh et~al.(2016{\natexlab{a}})Hezaveh, Dalal, Holder, Kisner,
  Kuhlen \& Perreault~Levasseur}]{Hezaveh:2014aoa}
Hezaveh Y, Dalal N, Holder G, Kisner T, Kuhlen M, Perreault~Levasseur L.
  2016{\natexlab{a}}.
\textit{JCAP} 1611:048

\bibitem[{Hezaveh et~al.(2016{\natexlab{b}})}]{Hezaveh:2016ltk}
Hezaveh YD, et~al. 2016{\natexlab{b}}.
\textit{Astrophys. J.} 823:37

\bibitem[{Higaki et~al.(2014)Higaki, Jeong \& Takahashi}]{Higaki:2014ooa}
Higaki T, Jeong KS, Takahashi F. 2014.
\textit{Phys. Lett. B} 734:21--26

\bibitem[{Hills et~al.(2018)Hills, Kulkarni, Meerburg \&
  Puchwein}]{Hills:2018vyr}
Hills R, Kulkarni G, Meerburg PD, Puchwein E. 2018.
\textit{Nature} 564:E32--E34

\bibitem[{Hinshaw et~al.(2013)}]{Hinshaw:2012aka}
Hinshaw G, et~al. 2013.
\textit{Astrophys. J. Suppl.} 208:19

\bibitem[{Hirano et~al.(2018)Hirano, Sullivan \& Bromm}]{Hirano:2017bnu}
Hirano S, Sullivan JM, Bromm V. 2018.
\textit{Mon. Not. Roy. Astron. Soc.} 473:L6--L10

\bibitem[{Hlo\v{z}ek et~al.(2017)Hlo\v{z}ek, Marsh, Grin, Allison, Dunkley \&
  Calabrese}]{Hlozek:2016lzm}
Hlo\v{z}ek R, Marsh DJE, Grin D, Allison R, Dunkley J, Calabrese E. 2017.
\textit{Phys. Rev. D} 95:123511

\bibitem[{Hlozek et~al.(2015)Hlozek, Grin, Marsh \& Ferreira}]{Hlozek:2014lca}
Hlozek R, Grin D, Marsh DJE, Ferreira PG. 2015.
\textit{Phys. Rev.} D91:103512

\bibitem[{Hoekstra et~al.(2004)Hoekstra, Yee \& Gladders}]{Hoekstra:2003pn}
Hoekstra H, Yee HK, Gladders MD. 2004.
\textit{Astrophys. J.} 606:67--77

\bibitem[{Hook(2019)}]{Hook:2018dlk}
Hook A. 2019.
\textit{PoS} TASI2018:004

\bibitem[{Hook et~al.(2018)Hook, Kahn, Safdi \& Sun}]{Hook:2018iia}
Hook A, Kahn Y, Safdi BR, Sun Z. 2018.
\textit{Phys. Rev. Lett.} 121:241102

\bibitem[{Horbatsch \& Burgess(2012)}]{Horbatsch:2011ye}
Horbatsch M, Burgess C. 2012.
\textit{JCAP} 05:010

\bibitem[{Hsueh et~al.(2020)Hsueh, Enzi, Vegetti, Auger, Fassnacht
  et~al.}]{Hsueh:2019ynk}
Hsueh JW, Enzi W, Vegetti S, Auger M, Fassnacht CD, et~al. 2020.
\textit{Mon. Not. Roy. Astron. Soc.} 492:3047--3059

\bibitem[{Hu et~al.(2000)Hu, Barkana \& Gruzinov}]{Hu:2000ke}
Hu W, Barkana R, Gruzinov A. 2000.
\textit{Phys. Rev. Lett.} 85:1158--1161

\bibitem[{Hui(1999)}]{Hui:1998hq}
Hui L. 1999.
\textit{Astrophys. J.} 516:519--526

\bibitem[{Hui \& Gnedin(1997)}]{Hui:1997dp}
Hui L, Gnedin NY. 1997.
\textit{Mon. Not. Roy. Astron. Soc.} 292:27

\bibitem[{Hui et~al.(2020)Hui, Joyce, Landry \& Li}]{Hui:2020hbq}
Hui L, Joyce A, Landry MJ, Li X. 2020

\bibitem[{Hui et~al.(2019)Hui, Kabat, Li, Santoni \& Wong}]{Hui:2019aqm}
Hui L, Kabat D, Li X, Santoni L, Wong SSC. 2019.
\textit{JCAP} 1906:038

\bibitem[{Hui et~al.(2017)Hui, Ostriker, Tremaine \& Witten}]{Hui:2016ltb}
Hui L, Ostriker JP, Tremaine S, Witten E. 2017.
\textit{Phys. Rev.} D95:043541

\bibitem[{Ibata et~al.(2002)Ibata, Lewis \& Irwin}]{Ibata:2001iv}
Ibata R, Lewis G, Irwin M. 2002.
\textit{Mon. Not. Roy. Astron. Soc.} 332:915

\bibitem[{Ibata et~al.(2020)Ibata, Thomas, Famaey, Malhan, Martin \&
  Monari}]{Ibata_2020}
Ibata R, Thomas G, Famaey B, Malhan K, Martin N, Monari G. 2020.
\textit{The Astrophysical Journal} 891:161

\bibitem[{Ibata et~al.(2016)Ibata, Lewis \& Martin}]{Ibata_2016}
Ibata RA, Lewis GF, Martin NF. 2016.
\textit{The Astrophysical Journal} 819:1

\bibitem[{Inoue(2011)}]{Inoue:2009wd}
Inoue S. 2011.
\textit{Mon. Not. Roy. Astron. Soc.} 416:1181--1190

\bibitem[{Irastorza \& Redondo(2018)}]{Irastorza:2018dyq}
Irastorza IG, Redondo J. 2018.
\textit{Prog. Part. Nucl. Phys.} 102:89--159

\bibitem[{Ir\v{s}i\v{c} et~al.(2017)Ir\v{s}i\v{c}, Viel, Haehnelt, Bolton \&
  Becker}]{Irsic:2017yje}
Ir\v{s}i\v{c} V, Viel M, Haehnelt MG, Bolton JS, Becker GD. 2017.
\textit{Phys. Rev. Lett.} 119:031302

\bibitem[{Ir\v{s}i\v{c} et~al.(2020)Ir\v{s}i\v{c}, Xiao \&
  McQuinn}]{Irsic:2019iff}
Ir\v{s}i\v{c} V, Xiao H, McQuinn M. 2020.
\textit{Phys. Rev. D} 101:123518

\bibitem[{Ivanov et~al.(2019)Ivanov, Kovalev, Lister, Panin, Pushkarev
  et~al.}]{Ivanov:2018byi}
Ivanov M, Kovalev Y, Lister M, Panin A, Pushkarev A, et~al. 2019.
\textit{JCAP} 02:059

\bibitem[{Ivezi\'c et~al.(2019)}]{Ivezic:2008fe}
Ivezi\'c v, et~al. 2019.
\textit{Astrophys. J.} 873:111

\bibitem[{Jacobson(1999)}]{Jacobson:1999vr}
Jacobson T. 1999.
\textit{Phys. Rev. Lett.} 83:2699--2702

\bibitem[{Jedamzik(2020)}]{Jedamzik:2020ypm}
Jedamzik K. 2020.
\textit{JCAP} 09:022

\bibitem[{Johnston et~al.(2002)Johnston, Spergel \& Haydn}]{Johnston:2001wh}
Johnston KV, Spergel DN, Haydn C. 2002.
\textit{Astrophys. J.} 570:656

\bibitem[{Kahn et~al.(2016)Kahn, Safdi \& Thaler}]{Kahn:2016aff}
Kahn Y, Safdi BR, Thaler J. 2016.
\textit{Phys. Rev. Lett.} 117:141801

\bibitem[{Kain \& Ling(2010)}]{Kain:2010rb}
Kain B, Ling HY. 2010.
\textit{Phys. Rev.} D82:064042

\bibitem[{Kaplan \& Rattazzi(2016)}]{Kaplan:2015fuy}
Kaplan DE, Rattazzi R. 2016.
\textit{Phys. Rev. D} 93:085007

\bibitem[{Kaplinghat et~al.(2020)Kaplinghat, Ren \& Yu}]{Kaplinghat:2019dhn}
Kaplinghat M, Ren T, Yu HB. 2020.
\textit{JCAP} 06:027

\bibitem[{Kato \& Soda(2020)}]{Kato:2019bqz}
Kato R, Soda J. 2020.
\textit{JCAP} 09:036

\bibitem[{Kaup(1968)}]{Kaup:1968zz}
Kaup DJ. 1968.
\textit{Phys. Rev.} 172:1331--1342

\bibitem[{Keating et~al.(2018)Keating, Puchwein \& Haehnelt}]{Keating:2017lgk}
Keating LC, Puchwein E, Haehnelt MG. 2018.
\textit{Mon. Not. Roy. Astron. Soc.} 477:5501--5516

\bibitem[{Khmelnitsky \& Rubakov(2014)}]{Khmelnitsky:2013lxt}
Khmelnitsky A, Rubakov V. 2014.
\textit{JCAP} 1402:019

\bibitem[{Kim(1979)}]{Kim:1979if}
Kim JE. 1979.
\textit{Phys. Rev. Lett.} 43:103

\bibitem[{Kim \& Marsh(2016)}]{KM}
Kim JE, Marsh D. 2016.
\textit{Phys. Rev.} D93:025027

\bibitem[{Kobayashi et~al.(2017)Kobayashi, Murgia, De~Simone, Ir\v{s}i\v{c} \&
  Viel}]{Kobayashi:2017jcf}
Kobayashi T, Murgia R, De~Simone A, Ir\v{s}i\v{c} V, Viel M. 2017.
\textit{Phys. Rev. D} 96:123514

\bibitem[{Kogut et~al.(2019)Kogut, Abitbol, Chluba, Delabrouille, Fixsen
  et~al.}]{Kogut:2019vqh}
Kogut A, Abitbol M, Chluba J, Delabrouille J, Fixsen D, et~al. 2019.
\textit{arXiv:1907.13195}

\bibitem[{Kolb \& Long(2020)}]{Kolb:2020fwh}
Kolb EW, Long AJ. 2020.
\textit{arXiv:2009.03828}

\bibitem[{Kolb \& Tkachev(1993)}]{Kolb:1993zz}
Kolb EW, Tkachev II. 1993.
\textit{Phys. Rev. Lett.} 71:3051--3054

\bibitem[{Kolb \& Tkachev(1996)}]{Kolb:1995bu}
Kolb EW, Tkachev II. 1996.
\textit{Astrophys. J. Lett.} 460:L25--L28

\bibitem[{Kolb \& Turner(1990)}]{Kolb:1990vq}
Kolb EW, Turner MS. 1990.
\textit{{The Early Universe}}.
vol.~69

\bibitem[{Konoplya \& Zhidenko(2006)}]{Konoplya:2006br}
Konoplya RA, Zhidenko A. 2006.
\textit{Phys. Rev.} D73:124040

\bibitem[{Kuepper et~al.(2010)Kuepper, Kroupa, Baumgardt \&
  Heggie}]{Kuepper:2009sg}
Kuepper A, Kroupa P, Baumgardt H, Heggie D. 2010.
\textit{Mon. Not. Roy. Astron. Soc.} 401:105

\bibitem[{Kulkarni \& Ostriker(2020)}]{Kulkarni:2020pnb}
Kulkarni M, Ostriker JP. 2020.
\textit{arXiv:2011.02116}

\bibitem[{Lancaster et~al.(2020)Lancaster, Giovanetti, Mocz, Kahn, Lisanti \&
  Spergel}]{Lancaster:2019mde}
Lancaster L, Giovanetti C, Mocz P, Kahn Y, Lisanti M, Spergel DN. 2020.
\textit{JCAP} 01:001

\bibitem[{Lentz et~al.(2020)Lentz, Quinn \& Rosenberg}]{Lentz:2018mpf}
Lentz EW, Quinn TR, Rosenberg LJ. 2020.
\textit{Nucl. Phys. B} 952:114937

\bibitem[{Lesgourgues et~al.(2002)Lesgourgues, Arbey \&
  Salati}]{Lesgourgues:2002hk}
Lesgourgues J, Arbey A, Salati P. 2002.
\textit{New Astron. Rev.} 46:791--799

\bibitem[{Levkov et~al.(2018)Levkov, Panin \& Tkachev}]{Levkov:2018kau}
Levkov D, Panin A, Tkachev I. 2018.
\textit{Phys. Rev. Lett.} 121:151301

\bibitem[{Li et~al.(2019)Li, Hui \& Bryan}]{Li:2018kyk}
Li X, Hui L, Bryan GL. 2019.
\textit{Phys. Rev.} D99:063509

\bibitem[{Li et~al.(2021)Li, Hui \& Yavetz}]{Li:2020ryg}
Li X, Hui L, Yavetz TD. 2021.
\textit{Phys. Rev. D} 103:023508

\bibitem[{Li et~al.(2020)Li, Shen \& Schive}]{Li:2020qva}
Li Z, Shen J, Schive HY. 2020.
\textit{arXiv:2001.00318}

\bibitem[{Lidz \& Hui(2018)}]{Lidz:2018fqo}
Lidz A, Hui L. 2018.
\textit{Phys. Rev. D} 98:023011

\bibitem[{Lin et~al.(2018)Lin, Schive, Wong \& Chiueh}]{Lin:2018whl}
Lin SC, Schive HY, Wong SK, Chiueh T. 2018.
\textit{Phys. Rev. D} 97:103523

\bibitem[{Linde(1985)}]{Linde:1985yf}
Linde AD. 1985.
\textit{Phys. Lett. B} 158:375--380

\bibitem[{Liu \& Ng(2017)}]{Liu:2016dcg}
Liu GC, Ng KW. 2017.
\textit{Phys. Dark Univ.} 16:22--25

\bibitem[{Liu et~al.(2019)Liu, Elwood, Evans \& Thaler}]{Liu:2018icu}
Liu H, Elwood BD, Evans M, Thaler J. 2019.
\textit{Phys. Rev.} D100:023548

\bibitem[{Lora et~al.(2012)Lora, Magana, Bernal, Sanchez-Salcedo \&
  Grebel}]{lora12}
Lora V, Magana J, Bernal A, Sanchez-Salcedo FJ, Grebel EK. 2012.
\textit{JCAP} 1202:011

\bibitem[{Lue et~al.(1999)Lue, Wang \& Kamionkowski}]{Lue:1998mq}
Lue A, Wang LM, Kamionkowski M. 1999.
\textit{Phys. Rev. Lett.} 83:1506--1509

\bibitem[{Lund(1991)}]{LUND1991245}
Lund F. 1991.
\textit{Physics Letters A} 159:245 -- 251

\bibitem[{Luscher(1981)}]{Luscher:1980ac}
Luscher M. 1981.
\textit{Nucl. Phys.} B180:317--329

\bibitem[{Luu et~al.(2020)Luu, Tye \& Broadhurst}]{Broadhurst:2018fei}
Luu HN, Tye SHH, Broadhurst T. 2020.
\textit{Phys. Dark Univ.} 30:100636

\bibitem[{Lyth(1990)}]{Lyth:1989pb}
Lyth DH. 1990.
\textit{Phys. Lett. B} 236:408--410

\bibitem[{Macedo et~al.(2013)Macedo, Pani, Cardoso \&
  Crispino}]{Macedo:2013jja}
Macedo CF, Pani P, Cardoso V, Crispino LCB. 2013.
\textit{Phys. Rev. D} 88:064046

\bibitem[{Madelung(1927)}]{Madelung1927}
Madelung E. 1927.
\textit{Zeitschrift f{\"u}r Physik} 40:322--326

\bibitem[{Mao \& Schneider(1998)}]{Mao:1997ek}
Mao Sd, Schneider P. 1998.
\textit{Mon. Not. Roy. Astron. Soc.} 295:587--594

\bibitem[{Marsh(2016)}]{Marsh:2015xka}
Marsh DJE. 2016.
\textit{Phys. Rept.} 643:1--79

\bibitem[{Marsh et~al.(2013)Marsh, Grin, Hlozek \& Ferreira}]{Marsh:2013taa}
Marsh DJE, Grin D, Hlozek R, Ferreira PG. 2013.
\textit{Phys. Rev. D} 87:121701

\bibitem[{Marsh \& Niemeyer(2019)}]{Marsh:2018zyw}
Marsh DJE, Niemeyer JC. 2019.
\textit{Phys. Rev. Lett.} 123:051103

\bibitem[{Marsh \& Silk(2014)}]{Marsh:2013ywa}
Marsh DJE, Silk J. 2014.
\textit{Mon. Not. Roy. Astron. Soc.} 437:2652--2663

\bibitem[{Martynov \& Miao(2020)}]{Martynov:2019azm}
Martynov D, Miao H. 2020.
\textit{Phys. Rev. D} 101:095034

\bibitem[{Mathur et~al.(2020)Mathur, Rajendran \& Tanin}]{Mathur:2020aqv}
Mathur A, Rajendran S, Tanin EH. 2020.
\textit{Phys. Rev. D} 102:055015

\bibitem[{May \& Springel(2021)}]{May:2021wwp}
May S, Springel V. 2021.
\textit{arXiv:2101.01828}

\bibitem[{Mayle et~al.(1988)Mayle, Wilson, Ellis, Olive, Schramm \&
  Steigman}]{Mayle:1987as}
Mayle R, Wilson JR, Ellis JR, Olive KA, Schramm DN, Steigman G. 1988.
\textit{Phys. Lett. B} 203:188--196

\bibitem[{McDonald et~al.(2005{\natexlab{a}})McDonald, Seljak, Cen, Bode \&
  Ostriker}]{McDonald:2004xp}
McDonald P, Seljak U, Cen R, Bode P, Ostriker JP. 2005{\natexlab{a}}.
\textit{Mon. Not. Roy. Astron. Soc.} 360:1471--1482

\bibitem[{McDonald et~al.(2005{\natexlab{b}})}]{McDonald:2004xn}
McDonald P, et~al. 2005{\natexlab{b}}.
\textit{Astrophys. J.} 635:761--783

\bibitem[{{McKee} et~al.(2015){McKee}, {Parravano} \& {Hollenbach}}]{McKee2015}
{McKee} CF, {Parravano} A, {Hollenbach} DJ. 2015.
\textit{\apj} 814:13

\bibitem[{Metcalf \& Madau(2001)}]{Metcalf:2001ap}
Metcalf RB, Madau P. 2001.
\textit{Astrophys. J.} 563:9

\bibitem[{Mirizzi et~al.(2008)Mirizzi, Raffelt \& Serpico}]{Mirizzi:2006zy}
Mirizzi A, Raffelt GG, Serpico PD. 2008.
\textit{Lect. Notes Phys.} 741:115--134

\bibitem[{Mishra-Sharma et~al.(2020)Mishra-Sharma, Van~Tilburg \&
  Weiner}]{Mishra-Sharma:2020ynk}
Mishra-Sharma S, Van~Tilburg K, Weiner N. 2020.
\textit{Phys. Rev. D} 102:023026

\bibitem[{Mocz \& Succi(2015)}]{Mocz:2015sda}
Mocz P, Succi S. 2015.
\textit{Phys. Rev.} E91:053304

\bibitem[{Mocz et~al.(2017)Mocz, Vogelsberger, Robles, Zavala, Boylan-Kolchin
  et~al.}]{Mocz:2017wlg}
Mocz P, Vogelsberger M, Robles VH, Zavala J, Boylan-Kolchin M, et~al. 2017.
\textit{Mon. Not. Roy. Astron. Soc.} 471:4559--4570

\bibitem[{Mocz et~al.(2019)}]{Mocz:2019pyf}
Mocz P, et~al. 2019.
\textit{Phys. Rev. Lett.} 123:141301

\bibitem[{Mondino et~al.(2020)Mondino, Taki, Van~Tilburg \&
  Weiner}]{Mondino:2020rkn}
Mondino C, Taki AM, Van~Tilburg K, Weiner N. 2020

\bibitem[{Nadler et~al.(2020)}]{Nadler:2020prv}
Nadler E, et~al. 2020.
\textit{arXiv:2008.00022}

\bibitem[{Nielsen \& Olesen(1973)}]{Nielsen:1973cs}
Nielsen HB, Olesen P. 1973.
\textit{Nucl. Phys.} B61:45--61

\bibitem[{Niemeyer(2019)}]{Niemeyer:2019aqm}
Niemeyer JC. 2019.
\textit{Prog. Part. Nucl. Phys.} :103787

\bibitem[{Nodland \& Ralston(1997)}]{Nodland:1997cc}
Nodland B, Ralston JP. 1997.
\textit{Phys. Rev. Lett.} 78:3043--3046

\bibitem[{Nori \& Baldi(2018)}]{Nori:2018hud}
Nori M, Baldi M. 2018.
\textit{Mon. Not. Roy. Astron. Soc.} 478:3935--3951

\bibitem[{Nori et~al.(2019)Nori, Murgia, Ir\v{s}i\v{c}, Baldi \&
  Viel}]{Nori:2018pka}
Nori M, Murgia R, Ir\v{s}i\v{c} V, Baldi M, Viel M. 2019.
\textit{Mon. Not. Roy. Astron. Soc.} 482:3227--3243

\bibitem[{O\~norbe et~al.(2019)O\~norbe, Davies, Luki\'c, Hennawi \&
  Sorini}]{Onorbe:2018zoi}
O\~norbe J, Davies F, Luki\'c Z, Hennawi J, Sorini D. 2019.
\textit{Mon. Not. Roy. Astron. Soc.} 486:4075--4097

\bibitem[{{Oh} et~al.(2000){Oh}, {Lin} \& {Richer}}]{oh2000}
{Oh} KS, {Lin} D, {Richer} HB. 2000.
\textit{\apj} 531:727--738

\bibitem[{Oman et~al.(2019)Oman, Marasco, Navarro, Frenk, Schaye \&
  Ben\'\i{}tez-Llambay}]{Oman:2017vkl}
Oman KA, Marasco A, Navarro JF, Frenk CS, Schaye J, Ben\'\i{}tez-Llambay A.
  2019.
\textit{Mon. Not. Roy. Astron. Soc.} 482:821--847

\bibitem[{Oman et~al.(2015)}]{Oman:2015xda}
Oman KA, et~al. 2015.
\textit{Mon. Not. Roy. Astron. Soc.} 452:3650--3665

\bibitem[{{Onsager}(1949)}]{1949NCim....6S.279O}
{Onsager} L. 1949.
\textit{Il Nuovo Cimento} 6:279--287

\bibitem[{{Ostriker} \& {Peebles}(1973)}]{OstrikerPeebles1973}
{Ostriker} JP, {Peebles} PJE. 1973.
\textit{\apj} 186:467--480

\bibitem[{Ouellet et~al.(2019)}]{Ouellet:2018beu}
Ouellet JL, et~al. 2019.
\textit{Phys. Rev. Lett.} 122:121802

\bibitem[{Padmanabhan(2021)}]{Padmanabhan:2020}
Padmanabhan N. 2021.
\textit{in preparation}

\bibitem[{Palanque-Delabrouille et~al.(2013)}]{Palanque-Delabrouille:2013gaa}
Palanque-Delabrouille N, et~al. 2013.
\textit{Astron. Astrophys.} 559:A85

\bibitem[{Palenzuela et~al.(2017)Palenzuela, Pani, Bezares, Cardoso, Lehner \&
  Liebling}]{Palenzuela:2017kcg}
Palenzuela C, Pani P, Bezares M, Cardoso V, Lehner L, Liebling S. 2017.
\textit{Phys. Rev. D} 96:104058

\bibitem[{Peccei \& Quinn(1977)}]{Peccei:1977hh}
Peccei RD, Quinn HR. 1977.
\textit{Phys. Rev. Lett.} 38:1440--1443

\bibitem[{Peebles(2000)}]{Peebles:2000yy}
Peebles PJE. 2000.
\textit{Astrophys. J.} 534:L127

\bibitem[{Porayko et~al.(2018)}]{Porayko:2018sfa}
Porayko NK, et~al. 2018.
\textit{Phys. Rev.} D98:102002

\bibitem[{Pozo et~al.(2020)Pozo, Broadhurst, de~Martino, Chiueh, Smoot
  et~al.}]{Pozo:2020ukk}
Pozo A, Broadhurst T, de~Martino I, Chiueh T, Smoot GF, et~al. 2020.
\textit{arXiv:2010.10337}

\bibitem[{Preskill et~al.(1983)Preskill, Wise \& Wilczek}]{Preskill:1982cy}
Preskill J, Wise MB, Wilczek F. 1983.
\textit{Phys. Lett.} 120B:127--132

\bibitem[{Press et~al.(1990)Press, Ryden \& Spergel}]{Press:1989id}
Press WH, Ryden BS, Spergel DN. 1990.
\textit{Phys. Rev. Lett.} 64:1084

\bibitem[{Press \& Schechter(1974)}]{Press:1973iz}
Press WH, Schechter P. 1974.
\textit{Astrophys. J.} 187:425--438

\bibitem[{Press \& Teukolsky(1972)}]{Press:1972zz}
Press WH, Teukolsky SA. 1972.
\textit{Nature} 238:211--212

\bibitem[{Prusti et~al.(2016)Prusti, de~Bruijne, Brown, Vallenari, Babusiaux
  et~al.}]{GAIA}
Prusti T, de~Bruijne JHJ, Brown AGA, Vallenari A, Babusiaux C, et~al. 2016.
\textit{Astronomy \& Astrophysics} 595:A1

\bibitem[{Pustelny et~al.(2013)}]{Pustelny:2013rza}
Pustelny S, et~al. 2013.
\textit{Annalen Phys.} 525:659--670

\bibitem[{Raffelt \& Seckel(1988)}]{Raffelt:1987yt}
Raffelt G, Seckel D. 1988.
\textit{Phys. Rev. Lett.} 60:1793

\bibitem[{Raffelt \& Stodolsky(1988)}]{Raffelt:1987im}
Raffelt G, Stodolsky L. 1988.
\textit{Phys. Rev. D} 37:1237

\bibitem[{Raffelt(2008)}]{Raffelt:2006cw}
Raffelt GG. 2008.
\textit{Lect. Notes Phys.} 741:51--71

\bibitem[{Raffelt \& Dearborn(1987)}]{Raffelt:1987yu}
Raffelt GG, Dearborn DS. 1987.
\textit{Phys. Rev. D} 36:2211

\bibitem[{Read et~al.(2006)Read, Goerdt, Moore, Pontzen, Stadel \&
  Lake}]{Read:2006fq}
Read JI, Goerdt T, Moore B, Pontzen A, Stadel J, Lake G. 2006.
\textit{Mon. Not. Roy. Astron. Soc.} 373:1451--1460

\bibitem[{Rindler-Daller \& Shapiro(2012)}]{RindlerDaller:2011kx}
Rindler-Daller T, Shapiro PR. 2012.
\textit{Mon. Not. Roy. Astron. Soc.} 422:135--161

\bibitem[{Roberts et~al.(2017)Roberts, Blewitt, Dailey, Murphy, Pospelov
  et~al.}]{Roberts:2017hla}
Roberts BM, Blewitt G, Dailey C, Murphy M, Pospelov M, et~al. 2017.
\textit{Nature Commun.} 8:1195

\bibitem[{Rogers \& Peiris(2020)}]{Rogers:2020ltq}
Rogers KK, Peiris HV. 2020.
\textit{arXiv:2007.12705}

\bibitem[{{Rubin} \& {Ford}(1970)}]{Rubin1970}
{Rubin} VC, {Ford} W.~Kent J. 1970.
\textit{\apj} 159:379

\bibitem[{Ruffini \& Bonazzola(1969)}]{Ruffini:1969qy}
Ruffini R, Bonazzola S. 1969.
\textit{Phys. Rev.} 187:1767--1783

\bibitem[{Safarzadeh et~al.(2018)Safarzadeh, Scannapieco \&
  Babul}]{Safarzadeh:2018hhg}
Safarzadeh M, Scannapieco E, Babul A. 2018.
\textit{Astrophys. J. Lett.} 859:L18

\bibitem[{Safarzadeh \& Spergel(2019)}]{Safarzadeh:2019sre}
Safarzadeh M, Spergel DN. 2019.
\textit{arXiv:1906.11848}

\bibitem[{Sasaki et~al.(2018)Sasaki, Suyama, Tanaka \&
  Yokoyama}]{Sasaki:2018dmp}
Sasaki M, Suyama T, Tanaka T, Yokoyama S. 2018.
\textit{Class. Quant. Grav.} 35:063001

\bibitem[{Savalle et~al.(2019)Savalle, Roberts, Frank, Pottie, McAllister
  et~al.}]{Savalle:2019jsb}
Savalle E, Roberts BM, Frank F, Pottie PE, McAllister BT, et~al. 2019.
\textit{arXiv:1902.07192}

\bibitem[{Schive et~al.(2014{\natexlab{a}})Schive, Chiueh \&
  Broadhurst}]{Schive:2014dra}
Schive HY, Chiueh T, Broadhurst T. 2014{\natexlab{a}}.
\textit{Nature Phys.} 10:496--499

\bibitem[{Schive et~al.(2020)Schive, Chiueh \& Broadhurst}]{Schive:2019rrw}
Schive HY, Chiueh T, Broadhurst T. 2020.
\textit{Phys. Rev. Lett.} 124:201301

\bibitem[{Schive et~al.(2016)Schive, Chiueh, Broadhurst \&
  Huang}]{Schive:2015kza}
Schive HY, Chiueh T, Broadhurst T, Huang KW. 2016.
\textit{Astrophys. J.} 818:89

\bibitem[{Schive et~al.(2014{\natexlab{b}})Schive, Liao, Woo, Wong, Chiueh
  et~al.}]{Schive:2014hza}
Schive HY, Liao MH, Woo TP, Wong SK, Chiueh T, et~al. 2014{\natexlab{b}}.
\textit{Phys. Rev. Lett.} 113:261302

\bibitem[{Schlattl et~al.(1999)Schlattl, Weiss \& Raffelt}]{Schlattl:1998fz}
Schlattl H, Weiss A, Raffelt G. 1999.
\textit{Astropart. Phys.} 10:353--359

\bibitem[{Schmitz \& Yanagida(2018)}]{Schmitz:2018nhb}
Schmitz K, Yanagida TT. 2018.
\textit{Phys. Rev. D} 98:075003

\bibitem[{Schneider(2018)}]{Schneider:2018xba}
Schneider A. 2018.
\textit{Phys. Rev. D} 98:063021

\bibitem[{Schutz(2020)}]{Schutz:2020jox}
Schutz K. 2020.
\textit{Phys. Rev. D} 101:123026

\bibitem[{Schwabe et~al.(2020)Schwabe, Gosenca, Behrens, Niemeyer \&
  Easther}]{Schwabe:2020eac}
Schwabe B, Gosenca M, Behrens C, Niemeyer JC, Easther R. 2020.
\textit{Phys. Rev. D} 102:083518

\bibitem[{Schwabe et~al.(2016)Schwabe, Niemeyer \& Engels}]{Schwabe:2016rze}
Schwabe B, Niemeyer JC, Engels JF. 2016.
\textit{Phys. Rev.} D94:043513

\bibitem[{Seckel \& Turner(1985)}]{Seckel:1985tj}
Seckel D, Turner MS. 1985.
\textit{Phys. Rev. D} 32:3178

\bibitem[{Seidel \& Suen(1994)}]{Seidel:1993zk}
Seidel E, Suen WM. 1994.
\textit{Phys. Rev. Lett.} 72:2516--2519

\bibitem[{Sheth \& Tormen(1999)}]{Sheth:1999mn}
Sheth RK, Tormen G. 1999.
\textit{Mon. Not. Roy. Astron. Soc.} 308:119

\bibitem[{Shifman et~al.(1980)Shifman, Vainshtein \& Zakharov}]{Shifman:1979if}
Shifman MA, Vainshtein AI, Zakharov VI. 1980.
\textit{Nucl. Phys.} B166:493--506

\bibitem[{Sibiryakov et~al.(2020)Sibiryakov, S\o{}rensen \&
  Yu}]{Sibiryakov:2020eir}
Sibiryakov S, S\o{}rensen P, Yu TT. 2020.
\textit{JHEP} 20:075

\bibitem[{Sikivie(1983)}]{Sikivie:1983ip}
Sikivie P. 1983.
\textit{Phys. Rev. Lett.} 51:1415--1417.
[Erratum: Phys. Rev. Lett.\ 52, 695 (1984)]

\bibitem[{Sikivie(2020)}]{Sikivie:2020zpn}
Sikivie P. 2020.
\textit{arXiv:2003.02206}

\bibitem[{Sikivie \& Yang(2009)}]{Sikivie:2009qn}
Sikivie P, Yang Q. 2009.
\textit{Phys. Rev. Lett.} 103:111301

\bibitem[{Silverman \& Mallett(2002)}]{Silverman:2002qx}
Silverman MP, Mallett RL. 2002.
\textit{Gen. Rel. Grav.} 34:633--649

\bibitem[{Sin(1994)}]{Sin:1992bg}
Sin SJ. 1994.
\textit{Phys. Rev.} D50:3650--3654

\bibitem[{Sivertsson et~al.(2018)Sivertsson, Silverwood, Read, Bertone \&
  Steger}]{Sivertsson:2017rkp}
Sivertsson S, Silverwood H, Read J, Bertone G, Steger P. 2018.
\textit{Mon. Not. Roy. Astron. Soc.} 478:1677--1693

\bibitem[{Smith(1936)}]{Smith:1936mlg}
Smith S. 1936.
\textit{Astrophys. J.} 83:23--30

\bibitem[{Spergel \& Steinhardt(2000)}]{Spergel:1999mh}
Spergel DN, Steinhardt PJ. 2000.
\textit{Phys. Rev. Lett.} 84:3760--3763

\bibitem[{Starobinski{\v i}(1973)}]{1973JETP3728S}
Starobinski{\v i} AA. 1973.
\textit{Soviet Journal of Experimental and Theoretical Physics} 37:28

\bibitem[{Stott \& Marsh(2018)}]{Stott:2018opm}
Stott MJ, Marsh DJ. 2018.
\textit{Phys. Rev. D} 98:083006

\bibitem[{Suarez \& Matos(2011)}]{Suarez:2011yf}
Suarez A, Matos T. 2011.
\textit{Mon. Not. Roy. Astron. Soc.} 416:87

\bibitem[{Svrcek \& Witten(2006)}]{Svrcek:2006yi}
Svrcek P, Witten E. 2006.
\textit{JHEP} 06:051

\bibitem[{Terrano et~al.(2015)Terrano, Adelberger, Lee \&
  Heckel}]{Terrano:2015sna}
Terrano W, Adelberger E, Lee J, Heckel B. 2015.
\textit{Phys. Rev. Lett.} 115:201801

\bibitem[{Terrano et~al.(2019)Terrano, Adelberger, Hagedorn \&
  Heckel}]{Terrano:2019clh}
Terrano WA, Adelberger EG, Hagedorn CA, Heckel BR. 2019.
\textit{Phys. Rev. Lett.} 122:231301

\bibitem[{Tremaine \& Gunn(1979)}]{Tremaine:1979we}
Tremaine S, Gunn JE. 1979.
\textit{Phys. Rev. Lett.} 42:407--410

\bibitem[{{Tremaine}(1976)}]{tre76}
{Tremaine} SD. 1976.
\textit{\apj} 203:345--351

\bibitem[{Turner(1983)}]{Turner:1983he}
Turner MS. 1983.
\textit{Phys. Rev. D} 28:1243

\bibitem[{Turner(1988)}]{Turner:1987by}
Turner MS. 1988.
\textit{Phys. Rev. Lett.} 60:1797

\bibitem[{Turner \& Wilczek(1991)}]{Turner:1990uz}
Turner MS, Wilczek F. 1991.
\textit{Phys. Rev. Lett.} 66:5--8

\bibitem[{Uhlemann et~al.(2014)Uhlemann, Kopp \& Haugg}]{CU2014}
Uhlemann C, Kopp M, Haugg T. 2014.
\textit{Phys. Rev.} D90:023517

\bibitem[{Uhlemann et~al.(2019)Uhlemann, Rampf, Gosenca \&
  Hahn}]{Uhlemann:2018gzz}
Uhlemann C, Rampf C, Gosenca M, Hahn O. 2019.
\textit{Phys. Rev. D} 99:083524

\bibitem[{Ullio et~al.(2001)Ullio, Zhao \& Kamionkowski}]{Ullio:2001fb}
Ullio P, Zhao H, Kamionkowski M. 2001.
\textit{Phys. Rev. D} 64:043504

\bibitem[{{Unruh}(1976)}]{unruh1976}
{Unruh} WG. 1976.
\textit{Phys. Rev. D} 14:3251--3259

\bibitem[{Veltmaat \& Niemeyer(2016)}]{Veltmaat:2016rxo}
Veltmaat J, Niemeyer JC. 2016.
\textit{Phys. Rev.} D94:123523

\bibitem[{Veltmaat et~al.(2018)Veltmaat, Niemeyer \&
  Schwabe}]{Veltmaat:2018dfz}
Veltmaat J, Niemeyer JC, Schwabe B. 2018.
\textit{Phys. Rev.} D98:043509

\bibitem[{Vieira et~al.(2014)Vieira, Bezerra \& Muniz}]{Vieira:2014waa}
Vieira HS, Bezerra VB, Muniz CR. 2014.
\textit{Annals Phys.} 350:14--28

\bibitem[{Viel et~al.(2013)Viel, Schaye \& Booth}]{Viel:2012sd}
Viel M, Schaye J, Booth CM. 2013.
\textit{Mon. Not. Roy. Astron. Soc.} 429:1734

\bibitem[{Vlahakis et~al.(2015)Vlahakis, Hunter, Hodge, Pérez, Andreani
  et~al.}]{ALMA}
Vlahakis C, Hunter TR, Hodge JA, Pérez LM, Andreani P, et~al. 2015.
\textit{The Astrophysical Journal} 808:L4

\bibitem[{Wagner et~al.(2012)Wagner, Schlamminger, Gundlach \&
  Adelberger}]{Wagner:2012ui}
Wagner T, Schlamminger S, Gundlach J, Adelberger E. 2012.
\textit{Class. Quant. Grav.} 29:184002

\bibitem[{Walker et~al.(2009)Walker, Mateo, Olszewski, Penarrubia, Evans \&
  Gilmore}]{Walker:2009zp}
Walker MG, Mateo M, Olszewski EW, Penarrubia J, Evans N, Gilmore G. 2009.
\textit{Astrophys. J.} 704:1274--1287.
[Erratum: Astrophys.J. 710, 886--890 (2010)]

\bibitem[{Weinberg et~al.(2015)Weinberg, Bullock, Governato, Kuzio~de Naray \&
  Peter}]{Weinberg:2013aya}
Weinberg DH, Bullock JS, Governato F, Kuzio~de Naray R, Peter AHG. 2015.
\textit{Proc. Nat. Acad. Sci.} 112:12249--12255

\bibitem[{Weinberg(1978)}]{Weinberg:1977ma}
Weinberg S. 1978.
\textit{Phys. Rev. Lett.} 40:223--226

\bibitem[{Weiner(2019)}]{Weiner:2019zrg}
Weiner N. 2019.
\textit{Astrophys. Space Sci. Proc.} 56:153--159

\bibitem[{Weltman et~al.(2020)}]{Bull:2018lat}
Weltman A, et~al. 2020.
\textit{Publ. Astron. Soc. Austral.} 37:e002

\bibitem[{Widdicombe et~al.(2018)Widdicombe, Helfer, Marsh \&
  Lim}]{Widdicombe:2018oeo}
Widdicombe JY, Helfer T, Marsh DJ, Lim EA. 2018.
\textit{JCAP} 10:005

\bibitem[{Widrow \& Kaiser(1993)}]{Widrow:1993qq}
Widrow LM, Kaiser N. 1993.
\textit{Astrophys. J.} 416:L71--L74

\bibitem[{Wilczek(1978)}]{Wilczek:1977pj}
Wilczek F. 1978.
\textit{Phys. Rev. Lett.} 40:279--282

\bibitem[{Wong et~al.(2019)Wong, Davis \& Gregory}]{Wong:2019yoc}
Wong LK, Davis AC, Gregory R. 2019.
\textit{Phys. Rev. D} 100:024010

\bibitem[{Wu et~al.(2019)Wu, McQuinn, Kannan, D'Aloisio, Bird
  et~al.}]{Wu:2019sgk}
Wu X, McQuinn M, Kannan R, D'Aloisio A, Bird S, et~al. 2019.
\textit{Mon. Not. Roy. Astron. Soc.} 490:3177--3195

\bibitem[{Yoshino \& Kodama(2014)}]{Yoshino:2013ofa}
Yoshino H, Kodama H. 2014.
\textit{PTEP} 2014:043E02

\bibitem[{Zel'Dovich(1972)}]{1972JETP351085Z}
Zel'Dovich YB. 1972.
\textit{Soviet Journal of Experimental and Theoretical Physics} 35:1085

\bibitem[{Zhang \& Yang(2020)}]{Zhang:2019eid}
Zhang J, Yang H. 2020.
\textit{Phys. Rev. D} 101:043020

\bibitem[{Zhang \& Chiueh(2017)}]{Zhang:2017dpp}
Zhang UH, Chiueh T. 2017.
\textit{Phys. Rev. D} 96:063522

\bibitem[{Zhitnitsky(1980)}]{Zhitnitsky:1980tq}
Zhitnitsky AR. 1980.
\textit{Sov. J. Nucl. Phys.} 31:260.
[Yad. Fiz.31,497(1980)]

\bibitem[{Zinner(2011)}]{Zinner:2011if}
Zinner NT. 2011.
\textit{Phys. Res. Int.} 2011:734543

\bibitem[{Zwicky(1933)}]{Zwicky:1933gu}
Zwicky F. 1933.
\textit{Helv. Phys. Acta} 6:110--127

\end{thebibliography}

\end{document}